\documentclass[12pt]{JHEP3}
\usepackage{graphicx}

\usepackage{amsmath}
\usepackage[utf8]{inputenc}
\usepackage{amssymb}
\usepackage{amsfonts}
\usepackage{amsbsy}
\usepackage{url}
\usepackage{enumerate}
\usepackage{caption}

\newcommand{\bea}{\begin{eqnarray}}
\newcommand{\eea}{\end{eqnarray}}
\newcommand{\p}{\partial}

\makeatletter
\@addtoreset{equation}{section}
\makeatother

\newcommand{\refer}[1]{(\ref{#1})}

\newcommand{\abs}[1]{\left|#1\right|}

\title{BPS Boojums in ${\cal N}=2$ supersymmetric gauge theories II}
\author{
Masato Arai$^{1}$,
Filip Blaschke$^{2}$ 
and
Minoru Eto$^{3}$ 
\\
{\it Department of Physics, Yamagata University, Kojirakawa-machi 1-4-12, Yamagata,
Yamagata 990-8560, Japan}\\\\
E-mail: $^1$ \email{meto(at)sci.kj.yamagata-u.ac.jp}\\ 
E-mail: $^2$ \email{arai(at)sci.kj.yamagata-u.ac.jp}\\ 
E-mail: $^3$ \email{fblasch(at)sci.kj.yamagata-u.ac.jp}
}

\abstract{
We continue our study of 1/4 Bogomol'nyi-Prasad-Sommerfield (BPS) composite solitons of vortex strings, domain walls and
boojums in ${\cal N}=2$ supersymmetric Abelian gauge theories in four dimensions. In this work, we numerically confirm that 
a boojum appearing at an end point of a string on a thick domain wall behaves as a magnetic monopole with a fractional charge in three dimensions. We introduce a ``magnetic" scalar potential whose gradient gives magnetic fields. Height of the magnetic potential has a geometrical meaning that is shape of the domain wall. We find a semi-local extension of
boojum which has an additional size moduli at an end point of a semi-local string on the domain wall. 
Dyonic solutions are also studied and we numerically confirm that the dyonic domain 
wall becomes an electric capacitor storing opposite electric charges on its skins. At the same time, the boojum becomes fractional
dyon whose charge density is proportional to $\vec E \cdot \vec B$.
We also study dual configurations with an infinite number of boojums and anti-boojums on parallel lines and analyze the ability of domain walls to store magnetic charge as magnetic capacitors. In understanding these phenomena, the magnetic scalar potential plays an important role. We study the composite solitons from the viewpoints of the Nambu-Goto and Dirac-Born-Infeld actions, and
find the semi-local BIon as the counterpart of the semi-local Boojum.
}
\preprint{YGHP-16-01-B}

\begin{document}


\section{Introduction and summary}

Topological solitons, which often appear in physical settings where local or global symmetry is spontaneously broken, are important to various   fields in modern physics such as string theory, field theory, cosmology, nuclear physics and condensed matter physics.
The simplest examples, ordered in increasing number of co-dimension, are domain wall, vortex string \cite{Nielsen:1973cs} and 't Hooft-Polyakov magnetic monopole \cite{tHooft,Polyakov:1974ek}. 
Interestingly, a non-trivial composite of these `elementary' solitons also exists.
 
Among such configurations, composite solitons that vortex strings attach to domain walls have been studied for a long time.
A reason is that such configurations are corresponding objects to the D-branes \cite{Polchinski:1996fm} where F/D strings end in superstring framework. A pioneering analysis was performed in non-supersymmetic (SUSY) field theory in \cite{Carroll:1997pz, Bowick:2003au}. This work was later followed by other studies with SUSY \cite{Witten:1997ep, Kogan:1997dt, Campos:1998db, Portugues, Shifman:2002jm,Shifman:2003uh}.  In a SUSY field theory, it was shown that there exits a Bogomol'nyi-Prasad-Sommefield (BPS) object with negative energy in a junction point where the vortex string attaches on the wall \cite{Sakai2}. This is nothing but the binding energy of the vortex string and the domain wall. This configuration with negative energy is called the boojum, which is originally coined in the context of ${}^3$He superfluid \cite{Mermin,volovik}. Interesting point of such negative binding energy is that there is no corresponding analog in string theory. Further study of the boojum was also performed in ${\cal N}=2$ Abelian gauge theory with two charged matter hypermultiplets \cite{Auzzi}. In \cite{Sakai2, Auzzi} some features of the boojum such as its mass and configuration were investigated. 
However, there were several issues to be confirmed until quite recently.
For instance, in \cite{Sakai2}, although the correct formula for the boojum mass was derived, a certain approximation was used to simplify the calculations. In \cite{Auzzi}, it was also discussed that there is an ambiguity in the definition of the boojum mass given in \cite{Sakai2}. 
Furthermore, no analytic/numerical solutions for the boojums have been obtained and the true shape of boojum was not known.

In order to clarify these issues, we recently studied the boojum in details in ${\cal N}=2$ SUSY QED with $N_F \ge 2$ flavors in the presence of the Fayet-Iliopoulos term in the previous work \cite{Boojum1}. The boojum configuration was numerically/analytically 
obtained by solving the 1/4 BPS equations. Though they are a set of first order differential equations, they amount to a second order 
differential equation called the master equation (see Eq. (\ref{eq:quatermaster})) thanks to the so-called moduli matrix formalism \cite{Isozumi:2004jc, Sakai1, Eto1}. Before \cite{Boojum1}, it was known that this equation can analytically be solved only when the gauge coupling constant is taken to infinity \cite{Isozumi:2004jc} while the finite case is rather difficult. 
In principle, numerical solution can always be obtained if an appropriate boundary condition is given.
However, it is not straightforward task to give it when two or more topological solitons coexist. 
In \cite{Boojum1}, we provided a simple and systematic way to give suitable boundary conditions called the global approximations.
We showed that the global approximation is useful not only to solve numerically the master equation but also to figure out the boojum mass exactly without any ambiguity, such as that discussed in \cite{Auzzi}.
We also derived several exact solutions for 1/4 BPS equations at the finite gauge coupling in models with $N_F=4$ and $N_F=6$ flavors respectively. This was not achieved previously. 
The only composite soliton known exactly was 1/4 BPS junction of domain walls \cite{Kakimoto}. 
 
In our previous work \cite{Boojum1}, we were oriented to solving the master equation and revealing real shape of the boojums.
In contrast, 
in this paper we will focus on physical aspects of the boojums and expand our understanding of composite solitons further 
by using the developments of our previous work \cite{Boojum1}.
First we investigate a composite solution that a (semi-)local string vortex ends on a wall in a weak gauge coupling limit. Note that in our analysis it is possible to take any value of the gauge coupling when we solve the master equation. In the weak coupling limit, the domain wall becomes thick and has 
a fat internal layer where the $U(1)$ gauge symmetry is almost restored. 
In this situation we numerically confirm that the boojums can be identified
with {\it magnetic} point-like sources with a fractional charge from the $3+1$ dimensional viewpoint by taking the thickness of the domain walls into account. This is contrary to the case that the points where vortex-strings terminates on walls are interpreted as {\it electric} point charges in the low energy effective theory in $2+1$ dimensional world volume of the domain walls \cite{Portugues,Shifman:2002jm,Shifman:2003uh}. We show that the two-dimensional distribution of the magnetic flux inside the domain wall can be correctly reproduced by the gradient of a scalar function, which we call the {\it magnetic} scalar potential. Interestingly, the magnetic scalar potential corresponds to the``position'' of the domain wall. Namely, we prove that the shape of the domain wall determines  the magnetic force inside the domain walls. Further insights along this direction are brought by the global approximate solutions. We show that the domain wall's position can be approximately -- but precisely enough -- identified with the solution 
to the Taubes equation \cite{Taubes:1979tm}.

Second we study a numerical solution to a configuration of periodically aligned 
vortex strings attached to the domain wall. The shape of the domain wall exhibits a linear
while the domain wall is bent logarithmically when one vortex string pulls it.
In the setup, as mentioned above,
the point charges are magnetic charges and the magnetic scalar potential corresponds to the domain wall's position/shape.
We consider a configuration, where periodically aligned vortex strings end on the domain wall from one side and
another infinite series of the vortex strings end on the opposite side. As can be easily imagined,
such configuration resembles {\it magnetic} capacitor. We compute the magnetic capacitance per unit length and energy stored there.
When we separate the two parallel lines of endpoints far away, flat but slant domain wall remains in between with
non-zero magnetic flux inside. This is similar to a D-brane with magnetic flux. Putting additional vortex string ending on
the tilt domain wall, the magnetic flux spreading inside the domain wall shows again one-dimensional structure,
which is almost the same as an electric charge placed in an electric capacitor. A similar configuration was already obtained
in the strong gauge coupling limit \cite{Sakai1} and our solution is for the finite gauge coupling case.
This offers a field theoretical D-brane resembling the fundamental string ending on the D-brane with magnetic flux
\cite{Hashimoto:1999xh,Hashimoto:1999zw}.

We also study the  dyonic extension of the 1/4 BPS solutions. Although the BPS equations were derived 
in \cite{Eto2,Lee:2005sv}, no solutions have been obtained in the literature, except for the strong
gauge coupling limit \cite{Portugues}. We first study the 1/2 BPS dyonic domain walls which are
finite gauge coupling version of the Q-kinks \cite{Abraham:1992vb,Abraham:1992qv}.
We confirm that positive and negative electric charges are induced on the skin of the domain wall. As a consequence,
the dyonic domain wall in the weak gauge coupling region is an {\it electric} capacitor.
Then, we numerically solve the master equation for the dyonic 1/4 BPS configuration again with an aid of
the global approximate solutions. When a vortex string attaches to the dyonic domain wall, 
both the magnetic and  electric fluxes coexist inside the domain wall. We show that almost everywhere except for the vicinity of the
junction point, 
the electric flux $\vec B$ and the magnetic flux $\vec B$ 
are perpendicular. $\vec E$ and $\vec B$ becomes parallel 
around the junctions points, and, indeed, we show that
the boojum charge is proportional to $\vec E \cdot \vec B$, which is a CP-violating interaction.

As a novel solution, we find a semi-local boojum which appears
at the endpoint of the semi-local vortex string \cite{Vachaspati:1991dz} on the domain wall in the model with multiple flavors $N_F \ge 3$ with partially degenerate masses for the hypermultiplets.
It has an additional zero mode related to  the
size of the string diameter. We find that the semi-local boojum changes its size unison with the size of the attached semi-local vortex string.

Finally, we  study the 1/4 BPS configuration from the viewpoint of the low energy effective action, the Nambu-Goto action, and the DBI action, for the domain wall. This kind of study 
was already 
performed
for example, in \cite{Portugues,Shifman:2002jm,Shifman:2003uh,Tong2}. 
In these previous works, as the low energy effective action, the DBI action (or its linearization) which is obtained by dualizing the internal moduli of the domain wall to the Abelian gauge field was studied. 
In our paper, we study both the Nambu-Goto action and the DBI action. We first investigate the domain wall and its 
Q-extension (dyonic extension)
in the Nambu-Goto action and find that the energy of those configuration coincides with one in the field theoretical model.
Secondly, we 
study the case
that a point source of a zero size deforms the domain wall to a spike configuration in the Nambu-Goto action. This is precisely counterpart of the Q-lump string ending on the domain wall in the strong gauge coupling limit in the original field theory. 
After that, we study the relation between the Nambu-Goto action and the DBI action. We briefly explain how the Nambu-Goto action is dualized to the DBI action. By using the relations so obtained, we also transform the energy and the BPS equation for the dyonic extension of the spike configuration in terms of the DBI language. We show that the results are the same one as in \cite{Portugues}. By using the DBI action, we
also study a point-like source with a finite size which should be a counterpart of the semi-local boojum. We find the
semi-local BIon which, contrary to local BIon, has the tip of its spike smoothed out with the same order as the size of the source.

This paper is organized as follows. Section 2 serves as a summary of our model and all relevant formulas, such as topological charges and 1/4 BPS equations, which we present both in terms of field and also via moduli matrix method. In that section, we do not repeat derivation of these quantities, which is done in \cite{Boojum1}.
In Section 3 we present the notion of a boojum as a fractional magnetic monopole.
Section 4 is devoted to studying periodically aligned vortex strings. We investigate
the magnetic capacitor there.
In Section 5, we study the dyonic extension of the 1/4 BPS states. We find that the domain wall plays a role
of an electric capacitor and show several numerical solutions.
Section 6 is devoted to analysis from the perspective of Nambu-Goto action together with the analysis in terms of
the DBI action. 
A brief discussion of the future work is given in Section~7. 


\section{The Model}\label{model}

In this section, we write down all relevant formulas such as topological charges, BPS equations and the master equation for 1/4 BPS solitons for convenience. A proper derivation of these quantities is skipped and we refer the reader to look into \cite{Boojum1} for details.

\subsection{Abelian vortex-wall system}

The model we use for our analysis is ${\cal N}=2$ supersymmetric $U(1)$ gauge theory in (3+1)-dimensions with $2N_F$ 
complex scalar fields in the charged hypermultiplets.
The vector multiplet includes the photon $A_\mu$ and 
a real scalar field~$\sigma$. The bosonic Lagrangian is given as
\begin{align}
\label{eq:lag} {\mathcal L} & = -\frac{1}{4 g^2}(F_{\mu\nu})^2+\frac{1}{2g^2}(\partial_{\mu}\sigma)^2+\abs{D_{\mu}H}^2
+\abs{D_{\mu}\hat H^\dagger}^2-V\,,  \\
V & = \frac{g^2}{2}\bigl(v^2-HH^{\dagger} + \hat H^\dagger \hat H\bigr)^2
+ \frac{g^2}{2}\left|H\hat H\right|^2
+\abs{\sigma H-H M}^2
+\abs{\sigma \hat H^\dagger -\hat H^\dagger M}^2\,, 
\end{align}
where $g$ is a gauge coupling constant, $M$ is a real diagonal matrix
\begin{eqnarray}
M = \mbox{diag}(m_1,\ldots, m_{N_F}),
\end{eqnarray}
and $v$ is the Fayet-Illiopoulos D-term.
Without loss of generality we can take $M$ to be traceless, 
namely $\sum_{A=1}^{N_F} m_A = 0$,\footnote{Any overall factor 
$M=m \mathbf{1}_{N_F}+\ldots$ can be absorbed into $\sigma$ by shifting $\sigma\to \sigma-m$}
and align the masses as $m_A > m_{A+1}$. Since $\tilde H$ will play no role,
we will set $\tilde H = 0$ in the rest of this paper.

In the absence of the mass matrix $M$, the Lagrangian \refer{eq:lag} is invariant under $SU(N_F)$ 
flavour transformation of Higgs fields $H\to H U$, $U \in SU(N_F)$. 
The non-degenerate masses in $M$ explicitly break this down to  $U(1)^{N_F-1}$, 
which we from now on assume to be the case unless stated otherwise.

We consider 1/4 BPS solitons, namely the junctions of vortex strings arranged to be parallel to the $x^3$ axis and the domain walls perpendicular to the $x^3$ axis. 
By completing the energy density (see \cite{Boojum1} for details) we obtain  the following Bogomol'nyi bound
\begin{equation}
{\mathcal E} \geq {\cal T}_W + {\cal T}_S + {\cal T}_B + \partial_k j_k\,,
\end{equation}
with
\begin{eqnarray}
{\mathcal T}_W  = \eta v^2  \partial_3 \sigma\,, \quad
{\mathcal T}_S  = - \xi v^2  F_{12}\,, \quad
{\mathcal T}_B = \frac{\eta\, \xi}{g^2} \epsilon_{klm}\partial_k (F_{lm}\sigma) \label{TB} 
\end{eqnarray}
and with the non-topological currents defined as
\begin{eqnarray}
 j_a&=& \frac{\xi i}{2}\epsilon_{ab}(H D_b H^\dagger - D_b H H^\dagger)\,, \quad (a=1,2) \label{ja}\\
 j_3&=&- \eta (\sigma H-HM)H^\dagger\,. \label{jW}
\end{eqnarray}
The above bound is saturated if the following 1/4 BPS equations
\begin{eqnarray}
& \label{eq:bpsws1}D_3 H +\eta \bigl(\sigma H-H M\bigr) = 0\,, & \\
& \label{eq:bpsws2}\bigl(D_1+ i \xi D_2\bigr) H = 0\,,  & \\
& \label{eq:bpsws3} \eta \partial_1 \sigma = \xi F_{23}\,, \hspace{5mm} \eta \partial_2 \sigma = \xi F_{31}\,, & \\ 
& \label{eq:bpsws4} \xi F_{12}-\eta \partial_3 \sigma+g^2\bigl(v^2-\abs{H}^2\bigr) = 0\,. &
\end{eqnarray}
are satisfied. 
Here $\xi = (-1)1$ labels (anti-)vortices and $\eta = (-1)1$ 
denotes (anti-)walls. 

${\mathcal T}_W$ and ${\mathcal T}_S$, the domain wall and the vortex string energy density respectively,
are positive definite. ${\mathcal T}_B$ is the so-called \emph{boojum} energy density, which is interpreted as binding energy 
of vortex string attached to the domain wall, since it is negative irrespective the signs of $\eta$ and $\xi$ \cite{Sakai2, Auzzi}. 
The total energy of $1/4$ BPS soliton is obtained upon the space integration and it consists of three parts
\begin{equation}
E_{1/4} =  T_W A +T_S L + T_B\,, \label{evw}
\end{equation}
where we have denoted the sum of tensions of the domain walls $T_W = \int dx^3\ {\mathcal T}_W$, 
and that of the vortex strings $T_S = \int dx^1dx^2\ {\mathcal T}_S$, respectively.
$A = \int dx^1dx^2$ and $L=\int dx^3$ stand for
the domain wall's area and length of  the vortex string.
Only masses of the boojums $T_B = \int d^3x\ {\mathcal T}_B$ are finite.
Summing up all the elementary domain walls and vortex strings, we have
\begin{eqnarray}
T_W = \sum t_W = v^2 |\Delta m|,\quad T_S = 2\pi v^2 |k|,
\label{eq:TW}
\end{eqnarray}
where we have denoted $\Delta m = \left[\sigma\right]^{x^3 = +\infty}_{x^3 = -\infty}$ and
$k \in \mathbb{Z}$ stands for the number of vortex strings.
In \cite{Boojum1} we directly verify the generic formula 
\begin{eqnarray}
T_B =- \sum  \frac{2\pi}{g^2} |m_{A+1} - m_{A}|,
\end{eqnarray}
where the sum is taken for all the junctions of domain walls and vortex strings in the solution
under consideration.

\subsection{The moduli matrix formalism}
\label{sec:mmf}

The moduli matrix approach \cite{Isozumi:2004jc, Sakai1, Eto1} reduces the set of the equations (\ref{eq:bpsws1})--(\ref{eq:bpsws4}) into a one equation called the master equation. 
The moduli matrix approach is based on the ansatz
\begin{eqnarray}
H = v e^{-\frac{u}{2}}H_0(z) e^{\eta M x^3}\,,\quad
A_1+i\xi A_2 = -i \partial_{\bar z} u\,, \quad
\eta\sigma+i A_3 = \frac{1}{2}\partial_3 u\,, \label{a3}
\end{eqnarray}
where $H_0(z)$ is the so-called the moduli matrix which is holomorphic in a complex coordinate $z\equiv x^1+i\xi x^2$. 
By using the $U(1)$
gauge transformation, we fix 
$u=u(z,\bar{z},x^3)$ to be real. Then we have $A_3 = 0$. It is easy to see that this ansatz solves (\ref{eq:bpsws1})--(\ref{eq:bpsws3}) identically.
The last BPS equation \refer{eq:bpsws4} turns into the master equation
\begin{equation}\label{eq:quatermaster}
\frac{1}{2g^2v^2}\p_k^2 u = 1-\Omega_0 e^{-u}\,, \hspace{5mm} \Omega_0 = H_{0}(z)e^{2\eta M x^3} H_0^{\dagger}(\bar z)\,.
\end{equation}
Now, all fields can be expressed  in terms of $u$ as follows
\begin{eqnarray}
\sigma = \frac{\eta}{2}\p_3 u,\ 
F_{12} = -2\xi \p_z\p_{\bar z} u,\ 
F_{23} = \frac{\xi}{2}\p_3\p_1 u,\ 
F_{31} = \frac{\xi}{2}\p_2\p_3 u.
\label{eq:F_from_u}
\end{eqnarray}
The energy densities are also written as
\begin{eqnarray}
{\mathcal T}_W &=& \frac{v^2}{2}\p_3^2 u, \label{eq:tw}\\
{\mathcal T}_S &=& 2 v^2 \p_z\p_{\bar z}u = \frac{v^2}{2}\left(\p_1^2 + \p_2^2\right)u, \label{eq:ts}\\ 
{\mathcal T}_B &=& \frac{1}{2g^2}\left\{\left(\p_1\p_3 u\right)^2 + \left(\p_2\p_3u\right)^2 - \left(\p_1^2+\p_2^2\right)\! u\  \p_3^2 u\right\}. \label{eq:tb}
\end{eqnarray}
The non-topological current $j_k$ given in Eqs.~(\ref{ja}) and (\ref{jW}) can be rewritten in
the following expression by using the BPS equations
\begin{eqnarray}
j_k = \frac{1}{2}\p_k(HH^\dagger).
\end{eqnarray}
Thus, we also have
\begin{eqnarray}
{\cal T}_4 = \partial_k j_k  = \frac{1}{2}\p^2 (HH^\dagger) = - \frac{1}{4g^2}\p^2 \p^2 u,
\end{eqnarray}
with $\p^2 \equiv \p_k^2$.
Collecting all pieces, the total energy density is given by
\begin{eqnarray}
{\mathcal E} = \frac{v^2}{2}\p_k^2 u 
+ \frac{1}{2g^2}\left\{\left(\p_1\p_3 u\right)^2 + \left(\p_2\p_3u\right)^2 - \left(\p_1^2+\p_2^2\right) u \p_3^2 u\right\}
-\frac{1}{4g^2}(\p_k^2)^2 u. \label{eq:ted}
\end{eqnarray}
Thus, the scalar function $u$ determines everything.

Finally, for further convenience, we will use the following dimensionless coordinates and mass
\begin{eqnarray}
\tilde x^k = \sqrt2 gv x^k,\quad \tilde M = \frac{1}{\sqrt2 gv } M= {\rm diag}\left(\frac{\tilde m}{2},-\frac{\tilde m}{2}\right).
\label{eq:dimless}
\end{eqnarray}
The dimensionless fields are similarly defined by
\begin{eqnarray}
\tilde H  = \frac{H}{v} =  e^{-\frac{u}{2}} H_0 e^{\eta \tilde M \tilde x^3},\quad
\tilde \sigma = \frac{\sigma}{\sqrt{2} gv} = \frac{\eta}{2}\tilde \p_3 u.
\end{eqnarray}
We will also use the dimensionless magnetic fields 
\begin{eqnarray}
\tilde F_{12} &=& \frac{1}{g^2v^2}F_{12} = -\xi \left(\tilde \p_\rho^2 + \frac{1}{\tilde \rho}\tilde \p_\rho\right) u,\ \\
\tilde F_{23} &=& \frac{1}{g^2v^2}F_{23} = \xi \tilde \p_3\tilde \p_\rho u \cos\theta,\ \\
\tilde F_{31} &=& \frac{1}{g^2v^2}F_{31} = \xi \tilde \p_3\tilde \p_\rho u \sin\theta.
\end{eqnarray}
Then, the dimensionless energy density $\tilde {\cal E}$ is defined by
\begin{eqnarray}
{\cal E} = g^2 v^4 \tilde{\cal E} = \tilde {\cal T}_W + \tilde {\cal T}_S + \tilde {\cal T}_B + \tilde {\cal T}_4,
\end{eqnarray}
where
\begin{eqnarray}
\tilde {\cal T}_W  &=& 2 \eta \tilde\p_3 \tilde \sigma = \tilde\p_3^2 u,\\
\tilde {\cal T}_S  &=& - \xi \tilde F_{12} = \left(\tilde\p_\rho^2 + \frac{1}{\tilde\rho}\tilde\p_\rho \right) u,\\
\tilde {\cal T}_B &=& 2\eta\xi \tilde\p_i\left(\epsilon_{ijk} \tilde \sigma \tilde F_{jk}\right) 
=  2 \left[\left(\tilde\p_\rho\tilde\p_3 u\right)^2 - \left(\tilde\p_\rho^2 + \frac{1}{\tilde\rho}\tilde\p_\rho\right)\! u\ \tilde\p_3^2 u\right],\\
\tilde {\cal T}_4 &=& - \left(\tilde\p_\rho^2 + \frac{1}{\tilde\rho}\tilde\p_\rho + \tilde\p_3^2\right)^2 u.
\end{eqnarray}
The relations to the original values are given as
\begin{eqnarray}
T_W 
&=&  \int dx^3\ {\cal T}_W 
= \frac{gv^3}{\sqrt 2}  \int d\tilde x^3\ \tilde{\cal T}_W =  \frac{gv^3}{\sqrt 2}  \tilde T_W \label{eq:tw2}
\\
T_S &=& \int d^2x\ {\cal T}_S = \frac{v^2}{2}\int d^2\tilde x\ \tilde{\cal T}_S = \frac{v^2}{2}\tilde T_S,\\
T_B &=& \int d^3x\ {\cal T}_B = \frac{v}{2\sqrt2 g}  \int d^3\tilde x\ \tilde{\cal T}_B = \frac{v}{2\sqrt2 g} \tilde T_B.
\end{eqnarray}

In what follows, we will not distinguish $x^k$ and $\tilde x^k$, unless stated otherwise. 
An exception is the mass: we will use the notation $\tilde m$ in order not to forget that we are using the 
dimensionless variables.


\section{Boojum as a fractional magnetic monopole and monostick}
\label{sec:fractional_monopole}

\subsection{Weak coupling regime}
\label{sec:weak}

\begin{figure}[t]
\begin{center}
\includegraphics[width=12cm]{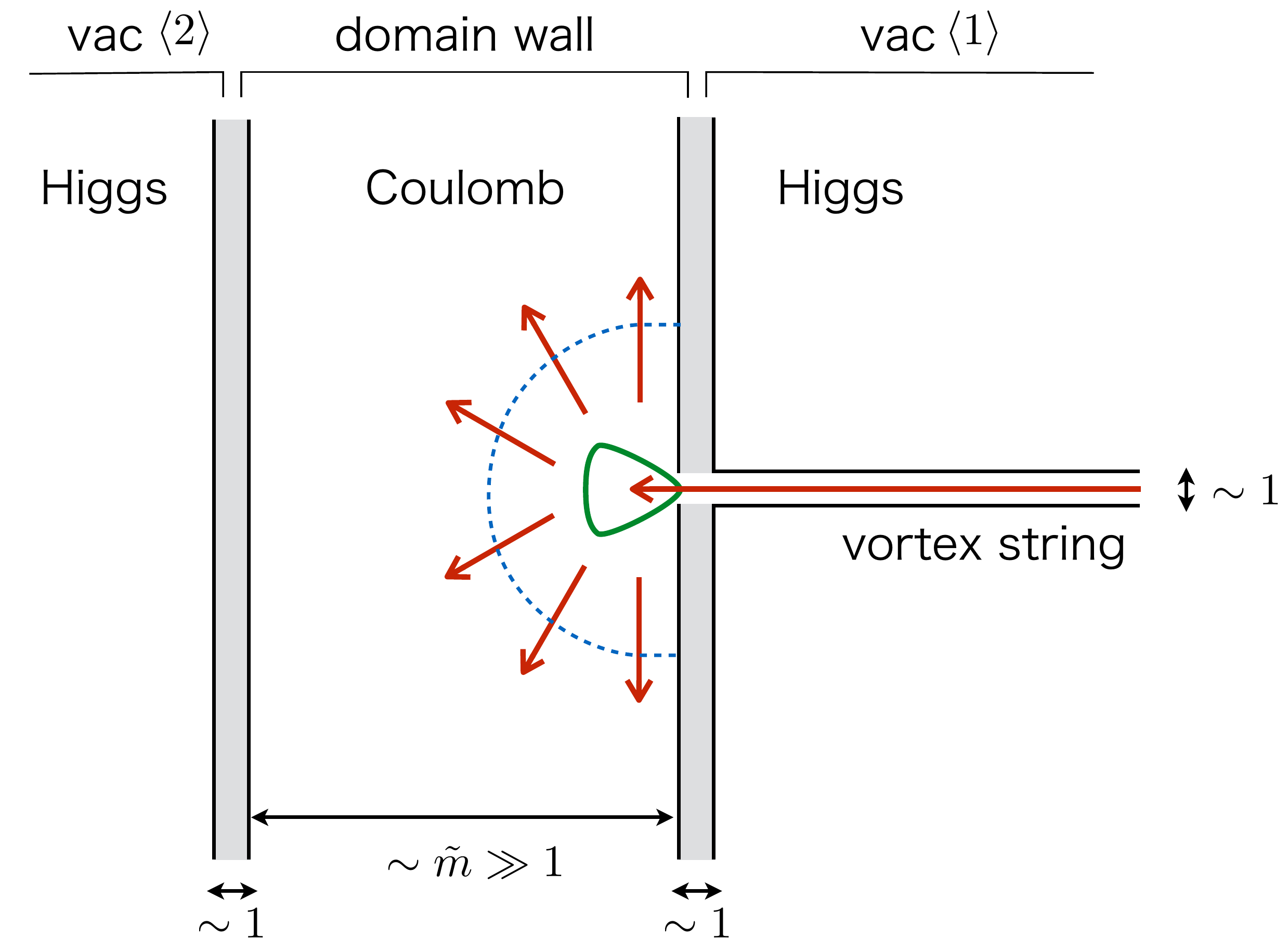}
\caption{A sketch of a boojum inside the domain wall in the weak coupling region. The boojum is
naturally identified with the magnetic source with a fractional magnetic charge ($2\pi/g$) 
which sticks  to the inner boundary of the domain wall.}
\label{fig:schematic_boojum}
\end{center}
\end{figure}

There are no magnetic sources, namely no magnetic monopoles, in our $U(1)$ gauge theory.
Indeed, Bianchi identity  $\epsilon_{ijk} \p_i F_{jk} = 0$ always holds.
The non-zero boojum charge  $T_B$ seems to yield non-zero magnetic charge, but it is not true.
The boojum has $\epsilon_{ijk} \p_i \hat F_{jk} \neq 0$ with $\hat F_{ij} = \sigma F_{ij}$ while
it has $\epsilon_{ijk} \p_i F_{jk} = 0$.
Of course, $\hat F_{ij} = \sigma F_{ij}$ is not a genuine magnetic field.
Nevertheless, there is a case that the boojum can be naturally identified as a magnetic source in
the week gauge coupling region $\tilde m \gg 1$.
In this region, the domain wall has a fat inner layer of the width $\sim \tilde m$, where the 
$U(1)$ gauge symmetry is almost recovered.
When the magnetic flux is injected into the domain wall through the vortex string, the magnetic flux
almost freely spreads out inside the domain wall, see Fig.~\ref{fig:schematic_boojum}.
Therefore, for one living inside  the domain wall, who is blind to outside world, the boojum is really a magnetic
source. It is a point-like source, so one may call it the magnetic monopole. The difference from 
the ordinary magnetic monopole, be it Dirac or 't Hooft-Polyakov monopole, is that the boojum sticks to
the boundary of the semi-compact  world, where he lives.

\begin{figure}[t]
\begin{center}
\includegraphics[width=15cm]{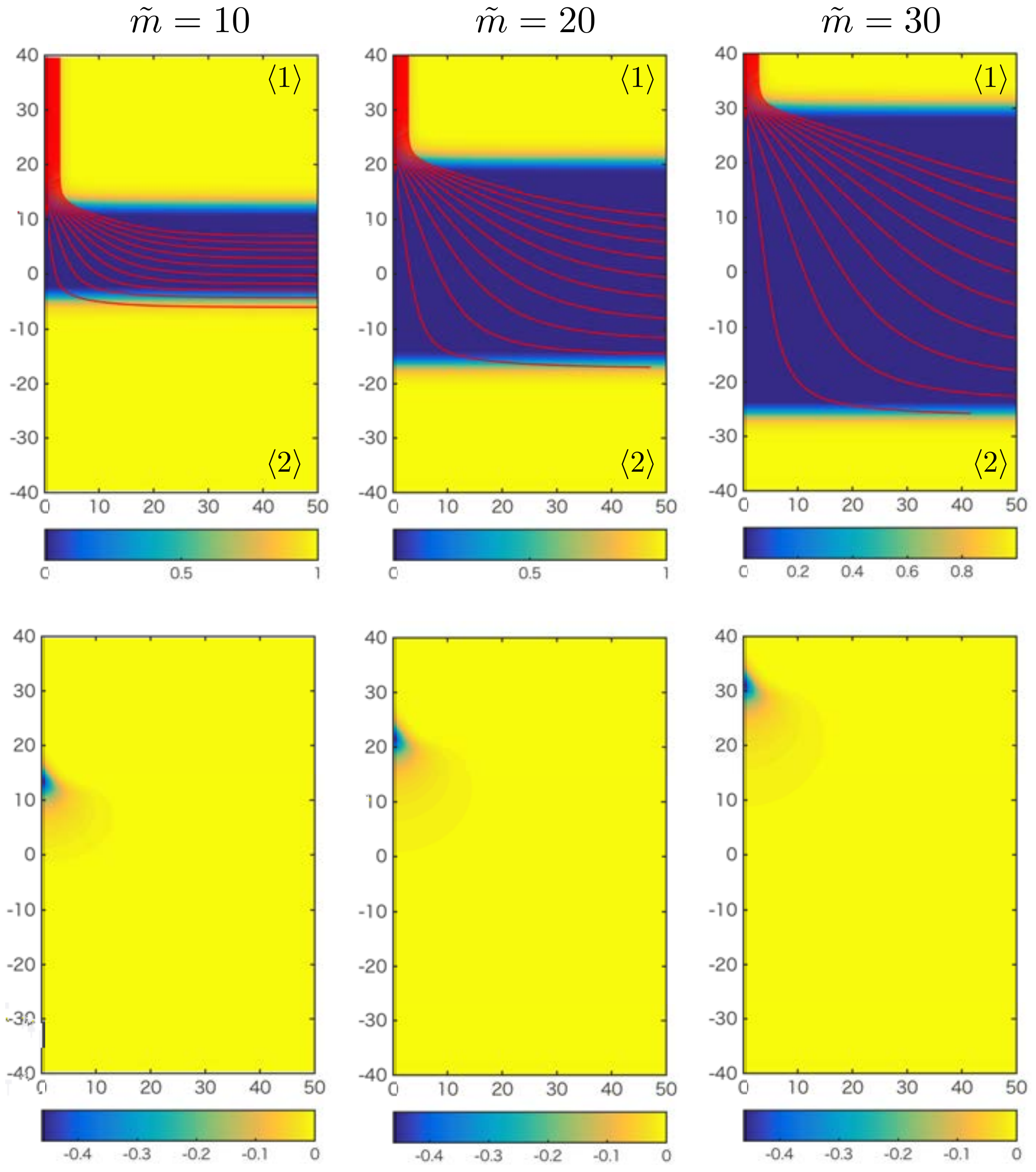}
\caption{Fat domain walls with $\tilde m = 10,20,30$. The 1st row shows
the effective photon mass $m_{\rm v}^2$ and the magnetic force lines. The 2nd row shows the boojum
charge density.}
\label{fig:boojum_monopole}
\end{center}
\end{figure}

Note that, here, we are trying to identify the boojum as
the magnetic monopole in the semi-compact space $d = 2+1$ where $2$ corresponds to the 2 dimensional
infinite plane and $1$ to the finite segment of width $\sim \tilde m$.
This is different from the well-known arguments that the endpoint of the vortex string can be identified with
an {\it electric} charge in the $1+2$ dimensional ($1$ is time and $2$ space) effective theory of the domain wall.
To this end, one needs to integrate out the normal direction to the domain wall ($x^3$) and then to dualize
the $S^1$ internal moduli parameter to the dual $U(1)$ gauge field in $1+2$ dimensional spacetime {\it \`a la} Polyakov.
Here, we do not do this and instead are dealing with the original $U(1)$ gauge field.

In Fig.~\ref{fig:boojum_monopole} we see that the magnetic flux inside the fat domain wall ($\tilde m \geq 10$)
linearly spreads for a while until it encounters the boundary. Characteristic length of this linear spreading is proportional
to the width of the inner layer. 

An observer inside the domain wall can measure the magnetic charge of the boojum by counting the magnetic flux flowing
through a hemisphere enclosing the boojum (blue dotted line in Fig.~\ref{fig:schematic_boojum}) as
\begin{eqnarray}
\tilde \Phi_{\rm boojum} = \int_{\rm hemisphere} dS_i\ \frac{1}{2}\epsilon_{ijk} \tilde F_{jk} = 4\pi.
\end{eqnarray}
Here we integrated only on the hemisphere with excluding the boundary of the wall
from the surface integral. We have $4\pi$ simply due to the flux conservation, 
$\tilde \Phi_{\rm boojum} = - \tilde \Phi_{\rm in}$.
From the Gauss's law (the integration is taken only on the hemisphere) we conclude that the 
magnetic charge of the boojum is
\begin{eqnarray}
\tilde {\cal M}_B = 4\pi.
\end{eqnarray}
Note that this is calculated with the dimensionless variables.
In terms of the original variables
and with respect to the usual notation $A_\mu \to g A_\mu$,
the magnetic charge of the boojum in a conventional notation is given by
\begin{eqnarray}
{\cal M}_B = \frac{2\pi}{g}.
\end{eqnarray}
This is a half of the magnetic charge ${\cal M}_{\rm TP} = 4\pi/g$ of 't Hoof-Polyakov type magnetic monopole.
Thus, the boojum can be identified with a fractional magnetic monopole from the point of view of wall-bound observers.

\begin{figure}[t]
\begin{center}
\includegraphics[width=12cm]{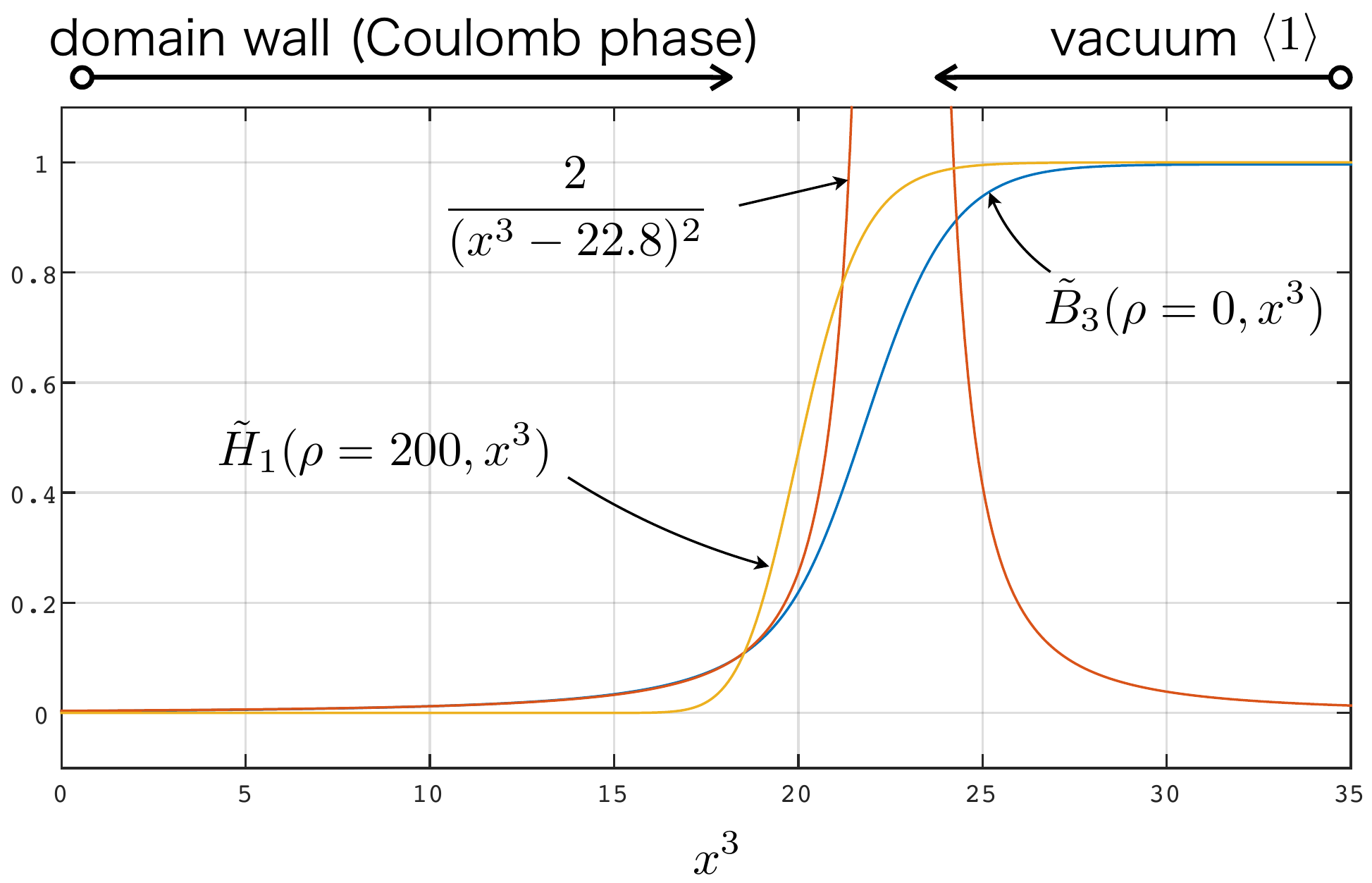}
\caption{
The magnitude of the magnetic field $\tilde B_3$ on the string axis (the $x^3$ axis) for $\tilde m=20$ is shown
in the blue curve. The red curve shows the Coulomb force with an appropriate shift $x^3 \to x^3 - 22.8$.
The yellow curve corresponds to $\tilde H_1$ on the $x^3$ axis. The region $\tilde H_1 = 0$ is in
the Coulomb phase.
$\tilde B_3 \to 1$ at the vacuum $\left<1\right>$ corresponds to the magnitude of the magnetic field
at the center of the vortex string.
}
\label{fig:coulomb_law}
\end{center}
\end{figure}

Since we have solved the equation of motion  for the finite gauge coupling constant, we can check 
this statement numerically.
In the Coulomb phase, the magnetic field of the magnetic charge $\tilde {\cal M}$ 
stuck on a wall at $x^3 = X$
should obey the normal $1+3$ dimensional Coulomb's law
\begin{eqnarray}
\tilde B_i \equiv 
\frac{1}{2}\epsilon_{ijk} \tilde F_{jk} = \frac{\tilde {\cal M}}{2\pi  r^2} \frac{ x^i-\delta^{i3}X}{ r},\quad
 x^3 <  X,
\end{eqnarray}
with $r = \sqrt{ (x^1)^2 + (x^2)^2 + (x^3-X)^2}$.
Note that not $4\pi r^2$ but $2\pi r^2$ (the area of hemisphere surrounding the boojum) 
appears in the denominator reflecting the fact that the magnetic flux
spread for one side ($ x^3 <  X$) of the right boundary of the domain wall at $ x^3 =  X$.
Thus magnitude of the magnetic field from the boojum with $ \tilde {\cal M}_B = 4\pi$ is
given by
\begin{eqnarray}
\left|\vec{\tilde B}\right| = \frac{2}{ (x^1)^2 + (x^2)^2 + (x^3-X)^2}.
\label{eq:Coulomb_boojum}
\end{eqnarray}
We show the magnetic flux $\tilde B_3(\rho=0,x^3)$ on the $x^3$ axis for $\tilde m = 20$
in Fig.~\ref{fig:coulomb_law}
and find that it asymptotically approaches to the Coulomb law as
\begin{eqnarray}
\tilde B_3(\rho=0,x^3) \simeq \frac{-2}{(x^3-22.8)^2},\qquad \left(x^3 \lesssim20\right).
\end{eqnarray}
Thus, the boojum is identical to the magnetic point particle put on
$X =22.8$ with the magnetic charge $\tilde {\cal M}_B = 4\pi$ if it is observed sufficiently far away.
Since the boundary of the inner layer at the vortex string side 
is about $x^3 \sim 22$, it is quite natural that the above
approximation works well.

\begin{figure}[t]
\begin{center}
\includegraphics[width=10cm]{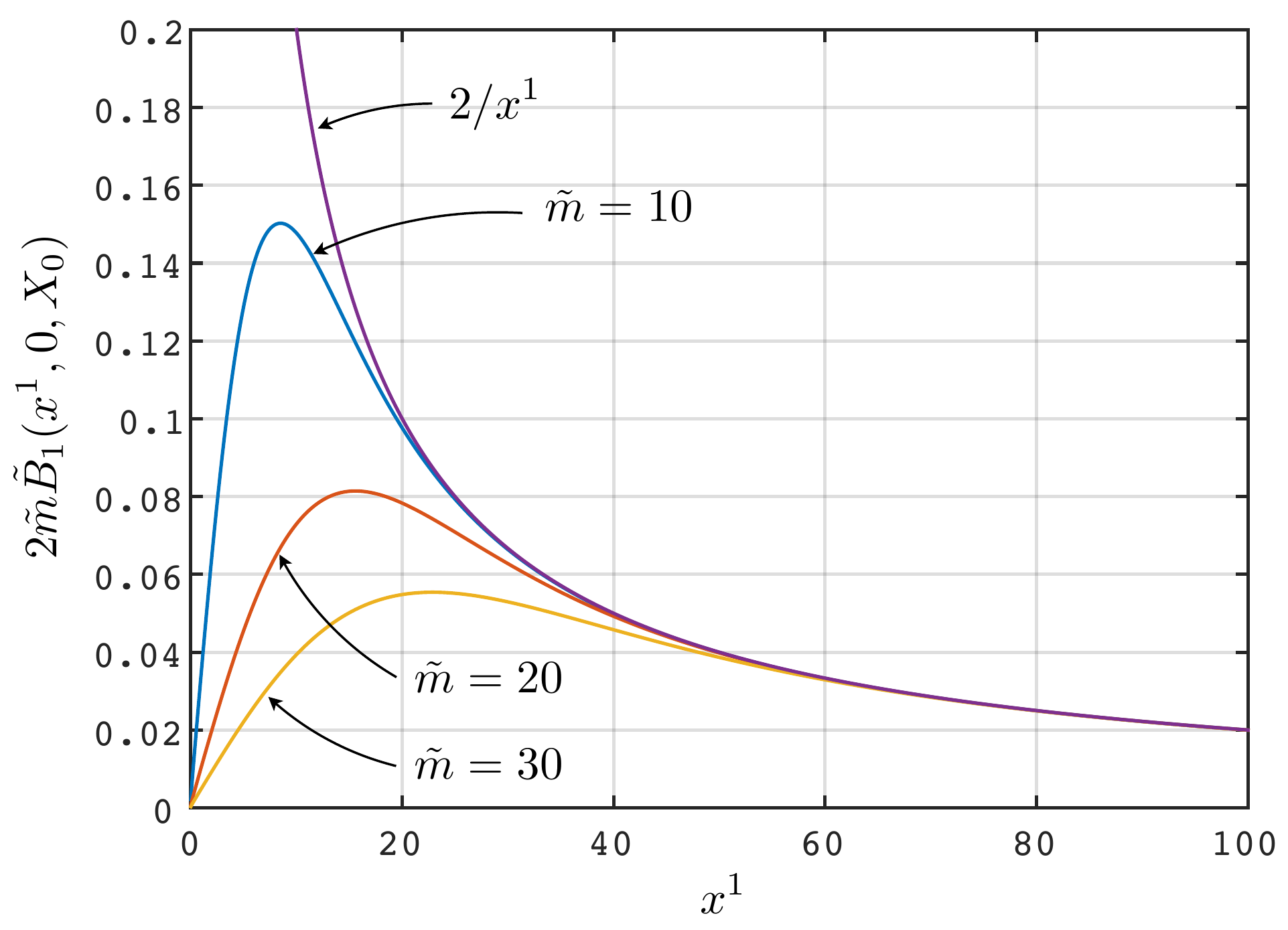}
\caption{A numerical verification for the formula (3.8). We plot $2\tilde m\tilde B_1(x^1,0,X_0)$ for 
$\tilde m = 10,20,30$ which is numerically obtained. All functions  approach to $\tilde{\cal M}_B/(2\pi x^1) = 2/ x^1$ 
for $x^1 > 2\tilde m$.}
\label{fig:B1}
\end{center}
\end{figure}

Let us now leave the boojum and travel inside the domain wall along $x^1-x^2$ plane. When we reach the distance
much farther than the domain wall width $\sim 2\tilde m$, the magnetic flux expands as if 
in the 2-dimensional plane. Therefore, the magnetic field should behave as $|\tilde B| \propto 1/\rho$.
Thus, one may expect for $\rho \gg 2 \tilde m$
\begin{eqnarray}
\tilde B_a = \frac{\tilde{\cal M}_B}{2\pi \rho} \frac{x^a}{\rho},\quad (a=1,2),
\label{eq:magnetic_field_inside_dw}
\end{eqnarray}
where $2\pi \rho$ in the denominator is the circumference of a circle surrounding the boojum.
However, this is too naive. We should not forget that the inner layer is not a 2-dimensional plane,
but it has the thickness $\tilde d_W = 2\tilde m$. Therefore, the magnetic field lines are parallelly distributed along the $x^3$
direction and, effectively, the magnetic charge is weakened by $1/2\tilde m$. Thus, the correct
asymptotic behavior of the magnetic field for $\rho \gg 2 \tilde m$ should be
\begin{eqnarray}
\tilde B_a = \frac{\tilde{\cal M}_B}{2\pi \tilde d_W\rho} \frac{x^a}{\rho},\qquad
B_a = \frac{2\pi g^2v^2}{m}\frac{x^a}{2\pi \rho^2}.
\label{eq:asym_B1}
\end{eqnarray}
In order to verify this, we plot $\tilde B_1(x^1,0,x^3=X_0)$, where $x^3=X_0$ is the center of the domain wall
for $\tilde m = 10,20,30$ in Fig.~\ref{fig:B1}. We read $x^3=X_0$ at which $\sigma$ becomes zero, and find
$X_0 = 2.07, 2.26 ,2.36$ for $\tilde m = 10, 20, 30$, respectively.
As seen from Fig.~\ref{fig:B1}, the numerical solution perfectly supports the formula
 Eq.~(\ref{eq:asym_B1}).
Thus, when seen at a distance, the boojum is suitable to be called {\it magnetic monostick} of height $2\tilde m$.

\subsection{Collinear vortex strings from both sides}
\label{sec:colinear}

Let us next consider collinear vortex strings ending on the domain wall from both sides.
Since the tensions of the vortex strings are the same, the domain wall remains flat.
For $N_F=2$ case with $M = (\tilde m/2,-\tilde m/2)$, 
such configuration is given by the moduli matrix $H_0 = (z,z)$.
The collinear vortex strings sit on the $x^3$ axis and the flat domain wall is at $x^3=0$.
The corresponding master equation is
\begin{eqnarray}\label{eq:mastercoll}
\left(\p_3^2 + \p_\rho^2 + \frac{1}{\rho}\p_\rho\right)u - 1 
+ \rho^2 \left( e^{\tilde m x^3} + e^{-\tilde mx^3}\right)e^{-u} = 0.
\end{eqnarray}
The correct global approximation for the solution of this equations reads (see the discussion below Eq.~(5.9) in our previous paper \cite{Boojum1})
\begin{equation}
{\mathcal U} = u_{W}(x^3)+u_S(\rho)\,,
\end{equation}
where $u_W(x^3)$ is a solution to a single domain wall master equation, which is obtained by putting $\partial_\rho = 1$ and $\rho = 1$ in Eq.~\refer{eq:mastercoll} and where $u_S(\rho)$ is a solution to the single vortex master equation, which is obtained by setting $\partial_3 = 0$ and $x^3 = -\log (2)/\tilde m$ in Eq.~\refer{eq:mastercoll}.

\clearpage
\begin{figure}
\begin{center}
\includegraphics[width=13.5cm]{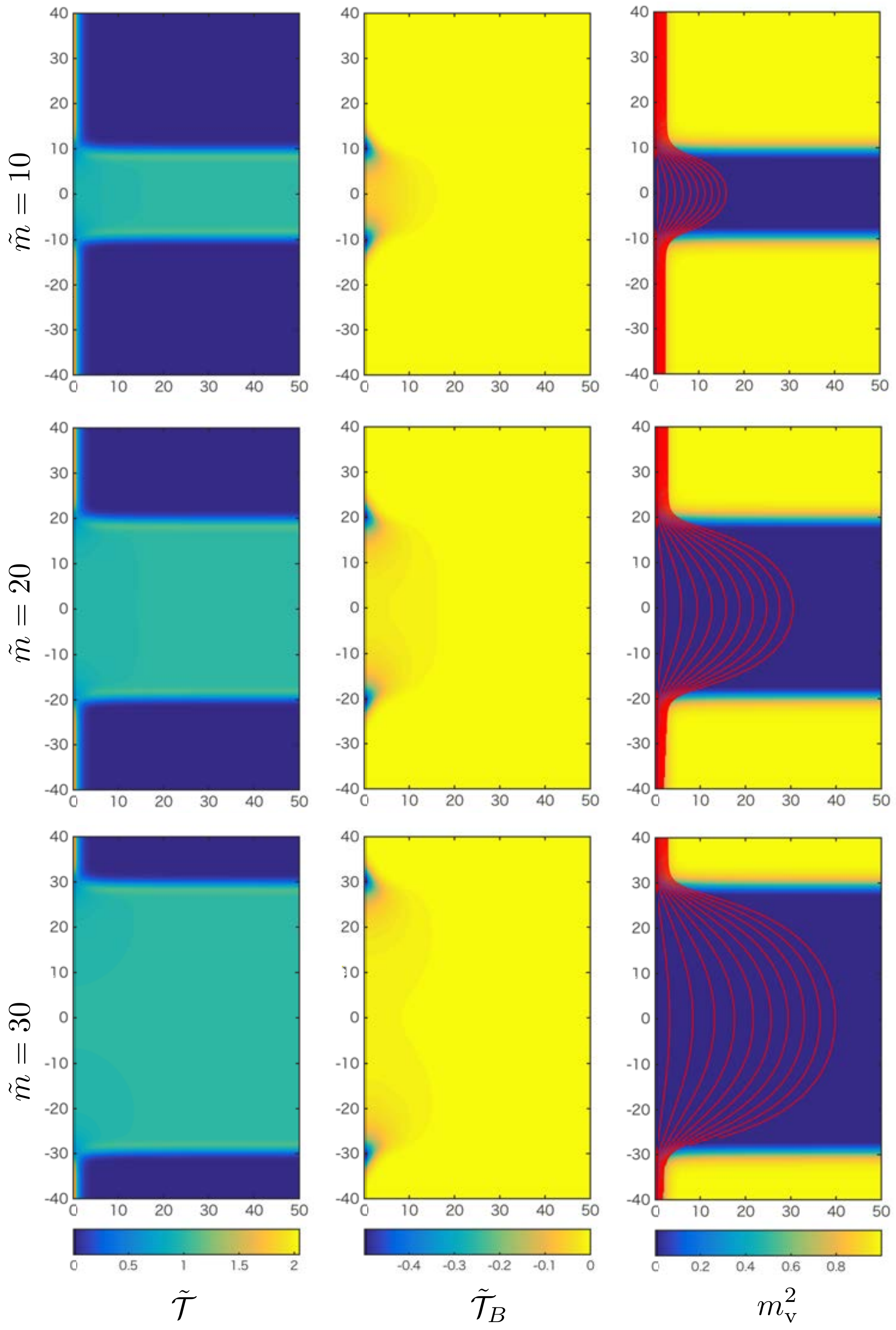}
\caption{Collinear vortex strings ending of the thick flat domain wall for the weak coupling region
$(\tilde m=10,20,30)$. The panels in left, middle, and right columns  show density plots of
the total energy density $\tilde {\cal T}$, the boojum charge density $\tilde{\cal T}_B$, and the  photon mass square $m_{\rm v}^2$, respectively. The red lines in the right column stand
for the magnetic force lines incoming from the upper vortex string and outgoing to the lower one.
The horizontal axis is $\rho$ and the vertical one is $x^3$.}
\label{fig:coaxial}
\end{center}
\end{figure}

\clearpage

\noindent

To help us solve the gradient flow equation
\begin{eqnarray}
\left(\p_3^2 + \p_\rho^2 + \frac{1}{\rho}\p_\rho\right)U - 1 
+ \rho^2 \left( e^{\tilde m x^3} + e^{-\tilde mx^3}\right)e^{-U} = \p_tU,
\end{eqnarray}
we use the above global approximation as the initial condition
\begin{eqnarray}
U(\rho,x^3,t=0) = u_W(x^3) + u_S(\rho).
\end{eqnarray}

We show three typical configurations at weak gauge coupling 
with $\tilde m = 10,20,30$ in Fig.~\ref{fig:coaxial}.
The solution is symmetric  under the reflection through the $x^1$-$x^2$ plane.
Since the domain walls have quite wide inner layers of the Coulomb phase, the upper and lower
boojums are well isolated, see the panels in the middle column of Fig.~\ref{fig:coaxial}.
Incoming magnetic fluxes from the upper vortex string freely spread inside the domain walls,
and then they are swallowed by the lower vortex string. There are no magnetic force lines going
to infinity along the domain walls ($\rho = \infty$) due to the flux conservation.
The expanse of the magnetic flux inside the domain wall is of the same order as the
width of the domain wall, as expected in Ref.~\cite{Sakai2}.
We numerically integrate the boojum charge density and get 
 $-\tilde T_B/8\pi\tilde m =1.95, 1.97, 1.98$ for $\tilde m = 10, 20, 30$, respectively.
These numbers are in good agreement with the analytical result $-\tilde T_B/8\pi\tilde m = 2$.

\begin{figure}[t]
\begin{center}
\includegraphics[width=15cm]{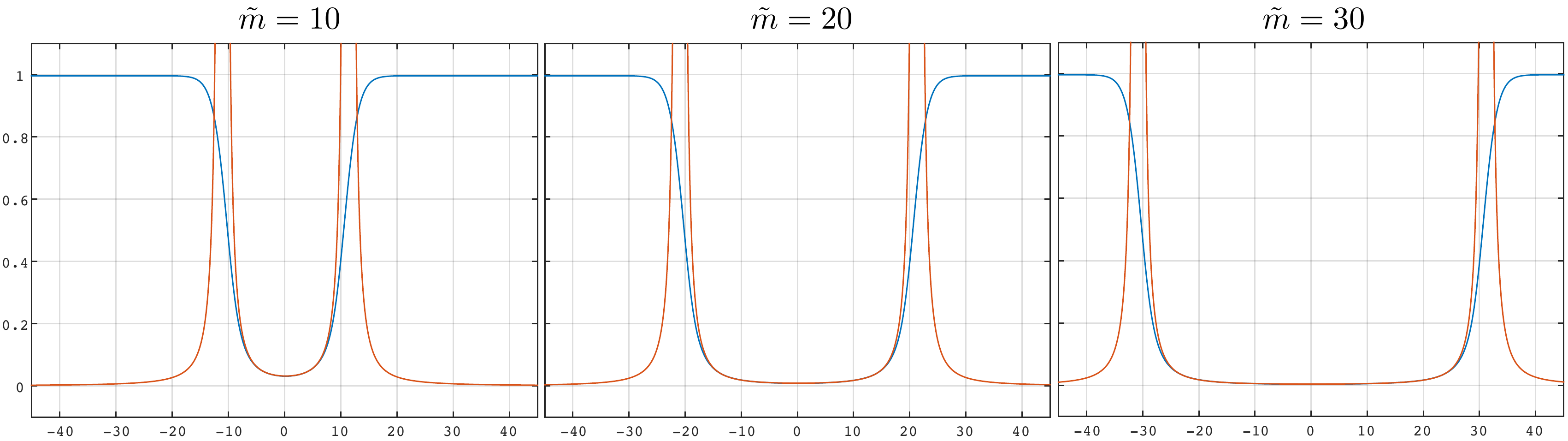}
\caption{The blue lines show the magnitude of $\tilde B_3$ on the string axis ($\rho=0$), which are numerically
obtained. The red lines correspond to the magnetic fields from the point-like monopole approximations 
given in Eq.~(3.69) with
$X = 11.2, 21.1, 31.0$ for $\tilde m = 10, 20, 30$. The horizontal axis is $x^3$.}
\label{fig:coulomb2}
\end{center}
\end{figure}

For observers sitting near the origin, the upper boojum is the fractional magnetic monopole
with the magnetic charge $\tilde {\cal M}_B = 4\pi$ whereas the lower boojum is the fractional anti-magnetic
monopole with $\tilde {\cal M}_{\bar B} = -4\pi$. The magnetic field observed by them should be
a simple superposition 
\begin{eqnarray}
\tilde B_i = \frac{\tilde {\cal M}_B}{2\pi  r_+^2} \frac{ x^i-\delta^{i3}X}{ r_+} + 
 \frac{\tilde {\cal M}_{\bar B}}{2\pi  r_-^2} \frac{ x^i+\delta^{i3}X}{ r_-},
\end{eqnarray}
with $r_\pm = \sqrt{ (x^1)^2 + (x^2)^2 + (x^3 \mp X)^2}$ and $X\ge 0$.
Especially, the third element of the $x^3$ axis for $|x^3| \ll X$ is 
\begin{eqnarray}
\tilde B_3(\rho = 0,x^3) &=& \frac{\tilde {\cal M}_B}{2\pi  (x^3-X)^2} \frac{ x^3-X}{ |x^3-X|} + 
\frac{\tilde {\cal M}_{\bar B}}{2\pi  (x^3+X)^2} \frac{ x^3+X}{|x^3+X|} \nonumber\\
&=& - \frac{2}{(x^3-X)^2} - \frac{2}{(x^3+X)^2}.
\label{eq:B3_bab}
\end{eqnarray}
The parameter $X$ should be tuned to fit the numerically obtained solutions for each $\tilde m$.
For example, we find $X = 11.2, 21.1, 31.0$ for $\tilde m = 10, 20, 30$ respectively, see Fig.~\ref{fig:coulomb2}.
The right and the left boundary of the inner layer are at $x^3 = \pm \tilde m$. 
So the boojums, fractional magnetic monopoles, are really stuck on the boundaries.

\subsection{Semi-local Boojums, semi-local magnetic monostick}
\label{sec:semilocal}

So far, we have mostly considered $N_F=2$ model which has three kinds of topological objects,
the domain wall, the vortex string and the boojum. In the models with a higher number of flavors $N_F>2$,
other kinds of topological objects enter the game. Namely, the semi-local vortex strings \cite{Vachaspati:1991dz}
and the semi-local
boojums. They appear when some of the masses are degenerate. The minimal model is $N_F =3$
with $\tilde M = {\rm diag}(\tilde m/2,\tilde m/2,-\tilde m/2)$.
The model has a non-Abelian flavor symmetry $SU(2) \times U(1)$, and two isolated vacua:
the first vacuum
$\left<1\right>$ is determined by $|\tilde H_1|^2 + |\tilde H_2|^2 = 1$, $\tilde H_3 = 0$
and $\tilde \sigma = \tilde m/2$, while
the second vacuum $\left<2\right>$ is determined by $\tilde H_1 = \tilde H_2 = 0$, $\tilde H_3 = 1$ and 
$\tilde \sigma = - \tilde m/2$.
Thus, the vacuum manifold for $\left<1\right>$ is $\mathbb{C}P^1$, while that for $\left<2\right>$ is
a point. The vortex string put in the degenerate vacuum $\left<1\right>$ is the so-called
semi-local vortex string \cite{Vachaspati:1991dz}. 
The semi-local vortex string can change its transverse size with its tension preserved. Namely,
it has a size moduli.
To prevent confusion, 
vortex string in the non-degenerate vacuum $\left<2\right>$ is sometimes called the local vortex string.
Size zero limit of the semi-local vortex string corresponds to the local vortex string.

Here we consider 1/4 BPS configuration of the semi-local vortex string ending on the domain wall.
Naturally, the boojum at the junction point changes its size with the semi-local string, therefore
we may call it a {\it semi-local Boojum}.
The simplest configuration is generated by the moduli matrix
\begin{eqnarray}
H_0 = (z,a,1),
\end{eqnarray}
where $a$ is a complex constant. We can assume that $a$ is a positive real number without loss of generality.
This yields an axially symmetric configuration with the master equation
\begin{eqnarray}
\left(\p_3^2 + \p_\rho^2 + \frac{1}{\rho}\p_\rho\right) u - 1 + \left[
(\rho^2 + a^2)e^{\tilde m x^3} + e^{-\tilde m x^3}\right] e^{-u} = 0.
\label{eq:bps_semilocal_boojum}
\end{eqnarray}
An appropriate initial configuration for the gradient flow equation for Eq.~(\ref{eq:bps_semilocal_boojum}) is based on the  suitable global approximation (see Eq.~(5.12) in \cite{Boojum1} for details)
\begin{eqnarray}
U(x^3,\rho,t=0) = u_W\left(x^3 + \frac{1}{2\tilde m} u_{SLS}\right) + \frac{1}{2} u_{SLS},
\label{eq:ini_sls}
\end{eqnarray}

\begin{figure}[t]
\begin{center}
\includegraphics[width=15cm]{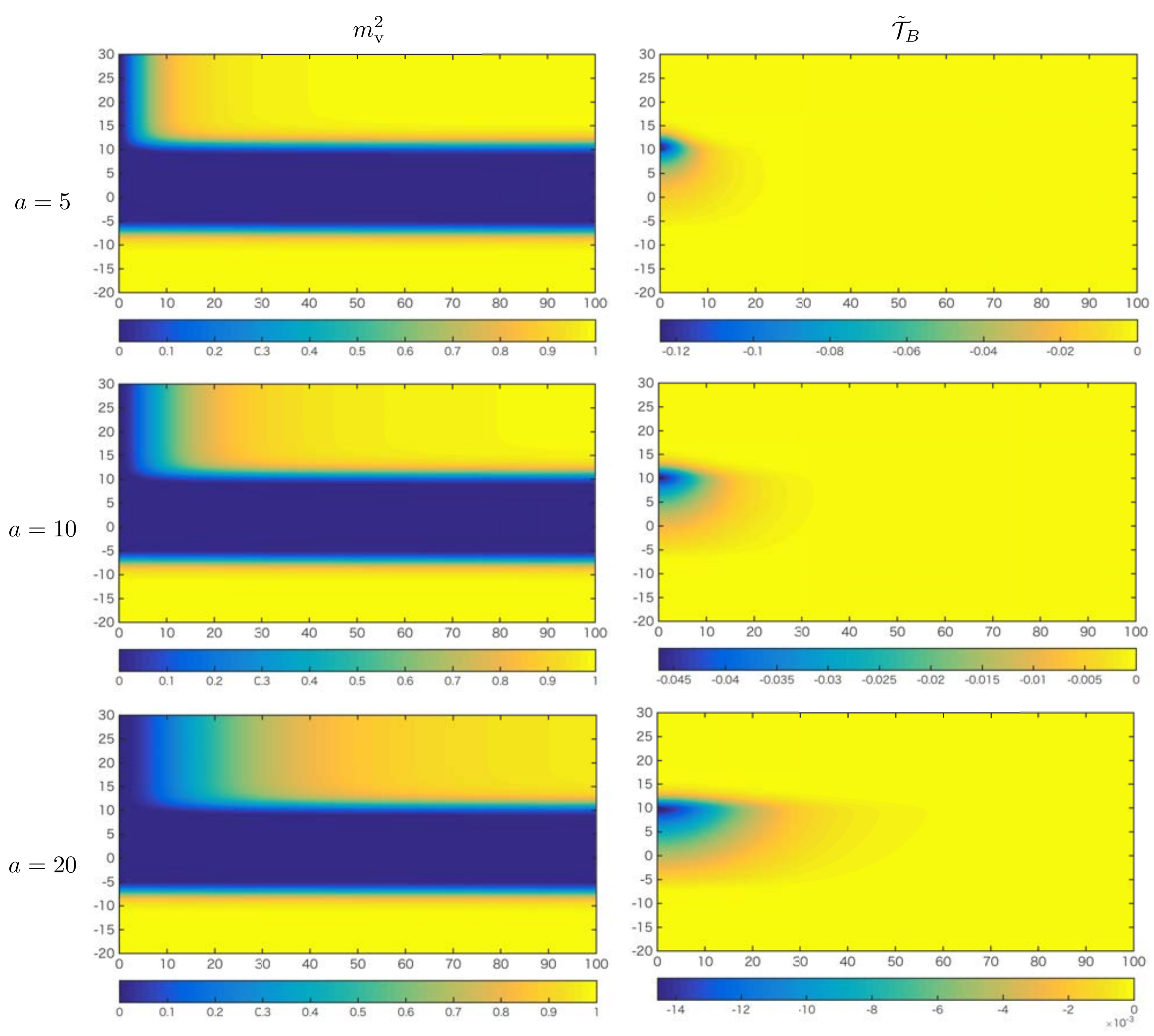}
\caption{Semi-local vortex string and semi-local boojum for $\tilde m = 10$ and $a=5,10,20$ (top, middle
, bottom). The left panels show $m_{\rm v}^2 = |\tilde H_1|^2 + |\tilde H_3|^2$ and the boojum charge density is plotted 
in the right panels.}
\label{fig:semilocal_boojum}
\end{center}
\end{figure}

We show the numerical solutions with $\tilde m = 10$ (in the weak coupling region) and
$a = 5, 10, 20$ in Fig.~\ref{fig:semilocal_boojum}.
This should be compared with the panels in the left column of Fig.~\ref{fig:boojum_monopole}, which
shows $\tilde m=10$ with $a=0$ (the local vortex string and local boojum).
Increasing the size of the parameter $a$, the cross-section of the semi-local vortex string grows linearly. At the same time,
the semi-local boojum is inflated.  Fig.~\ref{fig:semilocal_boojum} clearly shows that
the transverse size in the $x^1$-$x^2$ plane follows the semi-local vortex string but
the vertical size along the $x^3$ axis is limited by the domain wall size.

\begin{figure}[t]
\begin{center}
\includegraphics[width=15cm]{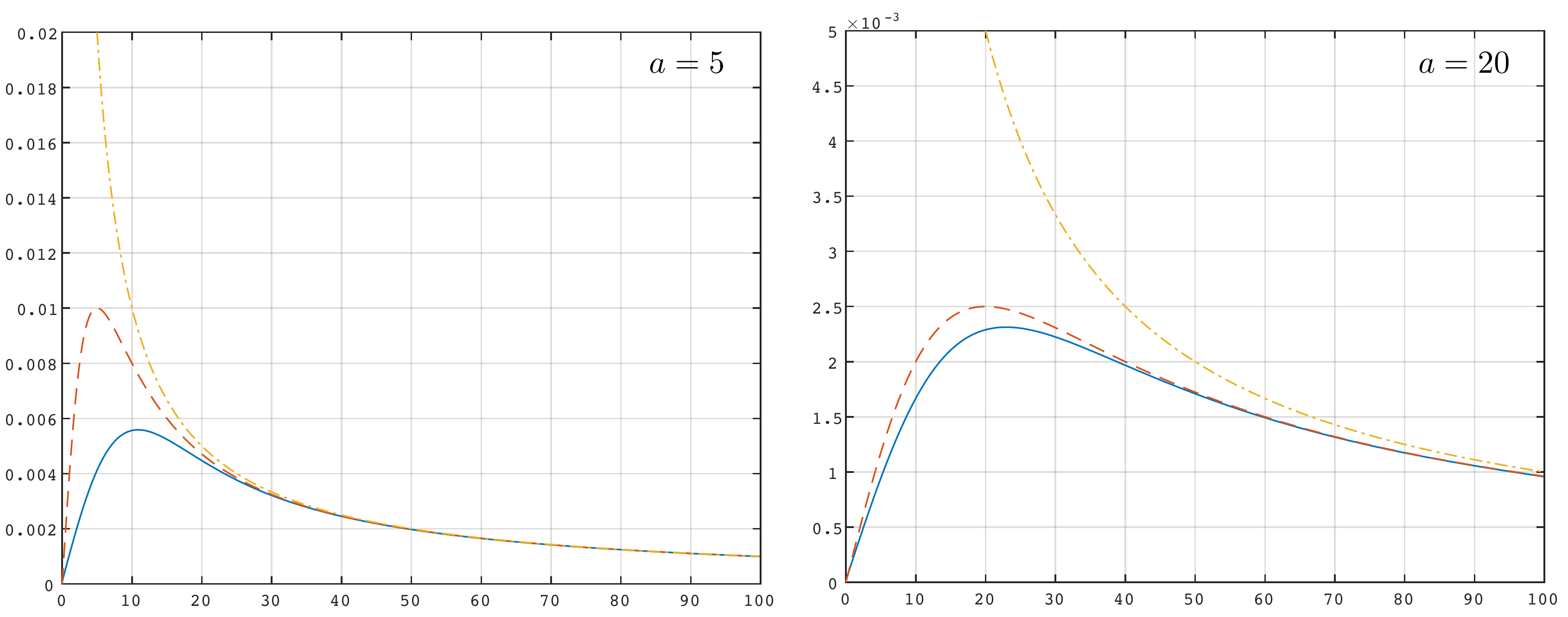}
\caption{The modified Coulomb's law $\tilde B_1(x^1,0,x^3 = 2.07)$ 
for $\tilde m = 10$ given in Eq.~(3.73) is plotted (red dashed line). The blue solid line
shows the numerical result and the yellow dot-dashed line is the normal Coulomb's law.
The left (right) panel corresponds to the case with $a=5$ ($a=20$). The horizontal axis is $x^1$. }
\label{fig:semilocal_monopole}
\end{center}
\end{figure}

Thanks to the moduli matrix formalism and the fact that all physical quantities can be expressed as a function of $u$, we can immediately conclude that the mass of the semi-local boojum is the same as that of the local boojum, namely 
\begin{eqnarray}
\tilde T_{\rm SLB} = - 8\pi \tilde m,\quad T_{\rm SLB} = - \frac{2\pi m}{g^2}.
\end{eqnarray}
This is because the change in the master equation involves only the replacement $\rho \to \sqrt{\rho^2+a^2}$, which does not affect the asymptotic behavior at the boundary, where the boojum mass is calculated (compare with the discussion in \cite{Boojum1}). 
Similarly, due to the flux conservation, the magnetic charge of the semi-local boojum is
\begin{eqnarray}
\tilde{\cal M}_{\rm SLB} = 4\pi,\quad {\cal M}_{\rm SLB} = \frac{2\pi}{g}.
\end{eqnarray}
Thus, the quantum numbers of the semi-local boojum are  the same as those for the local boojum.
Where can we find the effect of the size moduli? 
It appears in the Coulomb's law. When the boojum is seen far from the string axis, the magnetic 
field should spread according to the $1+2$ dimensional Coulomb's law as in Eq.~(\ref{eq:asym_B1})
for the local boojum. In the semi-local case, we find the following modified Coulomb's law
\begin{eqnarray}
\tilde B_b = \frac{\tilde{\cal M}_{\rm SLB}}{2\pi \tilde d_W} \frac{x^b}{\rho^2+a^2},
\quad
B_b = \frac{2\pi g^2v^2}{m}\frac{x^b}{2\pi(\rho^2+a^2)},
\quad \text{for}
\quad \rho \gg a,
\label{eq:asym_B1_SLB}
\end{eqnarray}
with $b=1,2$ and $\tilde d_W = 2\tilde m$. 
We compare it to the numerically obtained magnetic field for $\tilde m=10$ and 
$a = 5, 20$ in Fig.~\ref{fig:semilocal_monopole}
by plotting $\tilde B_1(x^1,0,x^3 = X_0)$ with $X_0=2.07$ ($\tilde\sigma(x^3=X_0) = 0$ as before).
We also show the magnetic field with $a=0$ (the ordinary Coulomb's law). 
As clearly seen in Fig.~\ref{fig:semilocal_monopole},
the modified Coulomb's law reproduces the numerical result much better than the normal Coulomb's law.

If we extrapolate  the magnetic field in Eq.~(\ref{eq:asym_B1_SLB}) to $\rho = 0$, it implies the 
following $1+2$ dimensional Gauss's law for the magnetic field
\begin{eqnarray}
\p_a \tilde B_a = \tilde f,\quad
\tilde f \simeq  \frac{\tilde{\cal M}_{\rm SLB}}{4\pi \tilde m} \frac{2 a^2 }{\left(\rho^2+a^2\right)^2}.
\end{eqnarray}
Therefore, the semi-local boojum is not point-like.
Roughly speaking, the magnetic charge is distributed into a cylinder of the height $2\tilde m$ and the radius
$\rho = a$. 
Thus, it is suitable to call the semi-local boojum as {\it semi-local  magnetic monostick}.

\begin{figure}[t]
\begin{center}
\includegraphics[width=15cm]{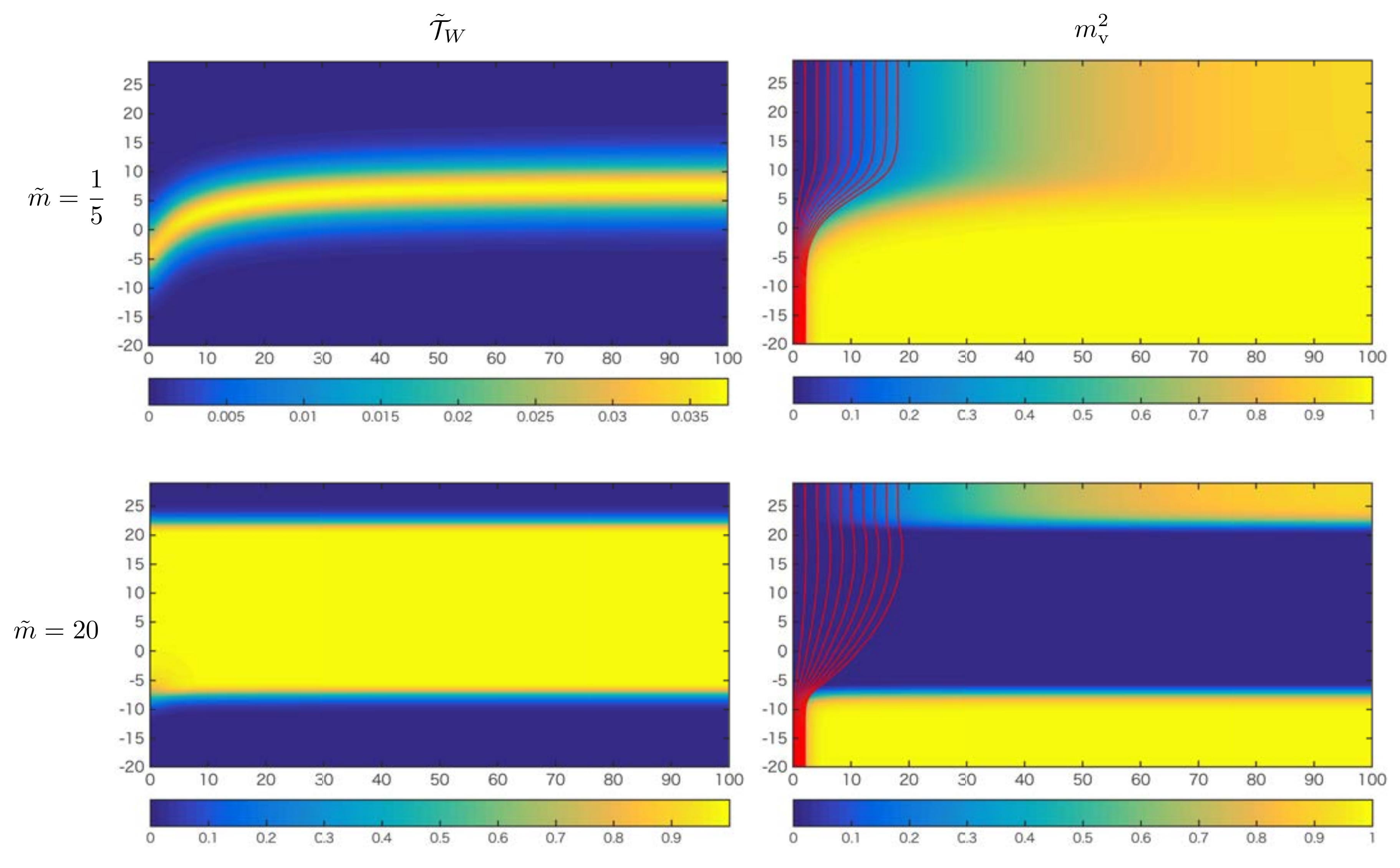}
\caption{The collinear semi-local vortex strings with $a_2=0$ from blow and $a_1=30$ from top end on
the asymptotically flat domain wall. The domain wall energy density $\tilde{\cal T}_W$ and $m_{\rm v}^2 = 
|\tilde H_1|^2 + |\tilde H_3|^2$ are shown in the left and right panels, respectively. The horizontal axis is $\rho$
and the vertical axis is $x^3$.}
\label{fig:asymmetric_colinear}
\end{center}
\end{figure}

Finally, we show collinear semi-local vortex strings with different sizes ending on the domain wall.
A minimal model for this is $N_F= 4$ with 
$\tilde M = {\rm diag}(\frac{\tilde m}{2}, \frac{\tilde m}{2}, -\frac{\tilde m}{2}, -\frac{\tilde m}{2})$.
The moduli matrix is $H_0 = (z,a_1, z, a_2)$. A suitable initial function for the gradient flow equation
in this case is
\begin{eqnarray}
U(\rho,x^3,t=0) = u_W\left(x^3 + \frac{u_{SLS}(a_1) - u_{SLS}(a_2)}{2\tilde m}\right)
+ \frac{u_{SLS}(a_1) + u_{SLS}(a_2)}{2}.
\end{eqnarray}
The domain wall is asymptotically flat, but it can be logarithmically bent around the junction point when
the sizes of two strings are very different. Such local bending is visible in the strong gauge 
coupling limit $\tilde m \ll 1$. In Fig.~\ref{fig:asymmetric_colinear}, we show two typical configurations
that have two collinear strings, the single local vortex string ($a_2=0$) from $x^3 < 0$ side and the single very fat 
semilocal vortex string with $a_1=30$ from $x^3>0$ side, ending of the domain wall for $\tilde m = 1/5$ (
strong gauge coupling) and $\tilde m = 20$ (weak gauge coupling). The domain wall steeply bends near
the collinear string axis for $\tilde m = 1/5$, but it asymptotically becomes flat at large $\rho$ due to 
the balance of the tensions of two vortex strings. On the other hand, the domain wall tension becomes 
sufficiently large for $\tilde m = 20$, so that the local curving structure near the string
axis is almost invisible. The well-squeezed magnetic flux tube from the local vortex string is 
magnified, when it goes into the semi-local vortex string as is shown in the right panels 
of Fig.~\ref{fig:asymmetric_colinear}. This is a lens effect for the magnetic force lines.

\subsection{Strong coupling regime}\label{scl}

Let us next consider the strong gauge coupling limit\footnote{
In this subsection, we will use the original variables $x^\mu$, $m$ and so on.} where the kinetic term of the gauge field disappears 
in the Lagrangian (\ref{eq:lag}), and 
the Higgs fields are restricted to satisfy $HH^\dagger = v^2$. Because of this, the domain wall has no internal
structure, namely both inside and outside the domain wall are in the Higgs phase.
The gauge fields are infinitely heavy and no longer dynamical.
Indeed, their equations of motion give
\begin{eqnarray}
A_\mu = \frac{-i}{2v^2}\left(H\p_\mu H^\dagger - \p_\mu HH^\dagger\right).
\end{eqnarray}
As a result, the Abelian-Higgs model with $N_F$ flavor reduces to the massive $\mathbb{C}P^{N_F-1}$ 
nonlinear sigma model.
One can introduce fictitious electromagnetic fields by $F_{\mu\nu} = \p_\mu A_\nu - \p_\nu A_\nu$ from
the gauge fields given above as
\begin{eqnarray}
F_{\mu\nu} &=& \frac{-i}{v^2} \left(\p_\mu H \p_\nu H^\dagger - \p_\nu H \p_\mu H^\dagger\right).
\label{eq:F_strong}
\end{eqnarray}

The BPS equations and the energy formulae obtained in Sec.~\ref{model} remain unchanged except for
dropping the terms proportional to $1/g^2$. Furthermore, the moduli matrix formalism explained in
Sec.~\ref{sec:mmf} still works without  any changes. One advantage is that the master equation is exactly solvable 
\begin{eqnarray}
u_{g\to \infty} = \log \Omega_0 = \log H_0 e^{2\eta Mx^3} H_0^\dagger.
\end{eqnarray}
In the following we will set $\eta = \xi = +1$.
Let us take the simplest example of a singular lump string (singular at a spatial infinity) ending on the domain wall,
which is generated by the moduli matrix $H_0=(z,1)$ in $N_F=2$ model with $M=(m/2,-m/2)$. The exact solution
is given by
\begin{eqnarray}
u_{g\to \infty} = \log \left( \rho^2 e^{mx^3} + e^{-mx^3}\right).
\end{eqnarray}
The domain wall's position can be read from the condition $\rho^2 e^{mx^3} = e^{-mx^3}$, namely, it is given by
\begin{eqnarray}
x^3 = - \frac{1}{m}\log \rho.
\label{eq:dw_posi_strong}
\end{eqnarray}
The fictitious magnetic flux given in Eq.~(\ref{eq:F_strong}) can be easily calculated by making use of the
formulae (\ref{eq:F_from_u}) as
\begin{eqnarray}
B_a = \frac{2mx^a}{\left(\rho^2 e^{mx^3}+e^{-mx^3}\right)^2},\quad
B_3 = - \frac{2}{\left(\rho^2 e^{mx^3}+e^{-mx^3}\right)^2}.
\end{eqnarray}
At the domain wall, the $a=1,2$ components becomes
\begin{eqnarray}
B_a\big|_{x^3 = - \frac{1}{m}\log \rho} = \frac{m}{2} \frac{x^a}{\rho^2}.
\label{eq:Ba_strong}
\end{eqnarray}

Similarly, the configuration with one regular lump string of the size $a$ ending on the domain wall
given by $H_0 = (z,a,1)$ in $N_F=3$ model with $M = (m/2,m/2,-m/2)$ can be obtained by just 
replacing $\rho \to \sqrt{\rho^2 + a^2}$ in the above results. Therefore, the $a=1,2$ components of the
magnetic flux at the domain wall is given by
\begin{eqnarray}
B_a\big|_{x^3 = - \frac{1}{m}\log \sqrt{\rho^2+a^2}} = \frac{m}{2} \frac{x^a}{\rho^2+a^2}.
\label{eq:Ba_strong_lump}
\end{eqnarray}

\subsection{The magnetic scalar potential}\label{msp}

As observed in the previous subsections,
the boojum, precisely speaking the ending point of the vortex string on the domain wall, 
can be regarded as the magnetic source inside the domain wall.
In order to pursue the identification, let us  
introduce the {\it magnetic} scalar potential,  whose gradient gives the $a=1,2$ component
of the magnetic fields:
\begin{eqnarray}
B_a = - \p_a \varphi(x^b),\quad (a=1,2).
\label{eq:def_msp}
\end{eqnarray}
In this subsection, we will use the original variables $x^\mu$, $m$ and so on, 
 and we will concentrate on $B_{a=1,2}$ only, while ignoring the third component $B_3$.

\subsubsection{Strong coupling limit} \label{sec:mag-charge}

Let us first consider the strong gauge coupling limit where the magnetic fields are given as in Eq.~(\ref{eq:Ba_strong}).
The magnetic scalar potential for this is given by
\begin{eqnarray}
\varphi = -\frac{q_B}{2\pi}\log \rho,\quad q_B =m \pi.
\label{eq:mag_pot_strong}
\end{eqnarray}
Because of $\p_a^2 (\log \rho)/2\pi = \delta^{(2)}(x^a)$, we see that the singular lump string
can be thought of as a point magnetic source with the charge $q_B$:
\begin{eqnarray}
\p_a B_a = - \p_a^2 \varphi = q_B \delta^{(2)}(x^a).
\end{eqnarray}
This identification of the lump string to the point magnetic source is consistent with the fact 
that the lump string is asymptotically singular far away from the domain wall.

The magnetic charge $q_B = m \pi$ can be understood as follows. The total magnetic flux coming from the
lump string is $2\pi$. Furthermore, (as explained in Sec.~3.1.1 of \cite{Boojum1}) the width of the domain wall in the strong
gauge coupling is given by $d_W = 2/m$. Thus, the mean value of the total magnetic flux going through the center of the domain 
wall corresponds to the magnetic charge
\begin{eqnarray}
\frac{2\pi}{d_W} = m\pi = q_B.
\label{eq:q_B_strong}
\end{eqnarray}

Now we come across an interesting coincidence: the domain wall curve given in Eq.~(\ref{eq:dw_posi_strong}) and
the magnetic scalar potential introduced in Eq.~(\ref{eq:mag_pot_strong}) are related as
\begin{eqnarray}
\varphi = \frac{m^2}{2} x^3 = \frac{mx^3}{d_W}.
\label{eq:mag_pot_x^3}
\end{eqnarray}
Factor $m^2$ is needed for consistency of the mass dimension.
If we integrate all the magnetic flux going through the domain wall, we have the total magnetic scalar potential
\begin{eqnarray}
\Phi = d_W \varphi = m x^3.
\end{eqnarray}
This coincidence tells us that the wall curve function $x^3(\rho)$ gives the magnetic scalar potential.

This is quite similar to another identification of an endpoint of the singular lump string on
the domain wall in the massive $\mathbb{C}P^1$
nonlinear sigma model to an {\it electric} point source of a {\it dual} electromagnetic field on
the $2+1$ dimensional domain wall world volume theory \cite{Portugues}, as will be studied in Sec.~\ref{sec:DBI}. 
In this section, however, we do not take the dual viewpoint and we deal with the magnetic field of the
original $U(1)$ gauge field.

The endpoint of the finite size lump string on the domain wall can be similarly regarded as
a magnetic source but as a source with finite size magnetic density. The magnetic scalar
potential leading to Eq.~(\ref{eq:Ba_strong_lump}) is given by
\begin{eqnarray}
\varphi = -\frac{q_B}{4\pi}\log (\rho^2 + a^2),
\end{eqnarray}
As in the case of the singular lump string, the same relation (\ref{eq:mag_pot_x^3})
between the magnetic scalar potential and the wall-curve function holds.

\subsubsection{Weak coupling regime}

Let us next consider the finite gauge coupling limit in which the vortex string has the finite size of order $1/gv$.
The boojum also has a finite size, so that it should be identified with a magnetic source with finite size distribution
in $2+1$ dimensions. In the finite gauge coupling, the domain wall's position in terms of the original variables is given as (compare with Eq.~(3.38) in \cite{Boojum1})
\begin{eqnarray}
x^3(\rho) = - \frac{1}{2m} u_S(\rho).
\label{eq:rela_x3_u_S}
\end{eqnarray}
Now we identify this wall curve function with the magnetic scalar potential by Eq.~(\ref{eq:mag_pot_x^3}).
Before doing this, let us remember that the width of the domain wall in the weak gauge coupling region
is $d_W = m/g^2 v^2$.
Thus the magnetic scalar potential in the weak gauge coupling region is given by
\begin{eqnarray}
\varphi = \frac{mx^3}{d_W} = - \frac{g^2v^2}{2m}u_S.
\label{eq:mag_pot_finite}
\end{eqnarray}
Since $u_S$ is asymptotically $\log \rho^2$, we read the magnetic charge as
\begin{eqnarray}
q_B = \frac{2\pi g^2 v^2}{m} = \frac{2\pi}{d_W}.
\end{eqnarray}
This is consistent with the observation in the strong gauge coupling limit given in Eq.~(\ref{eq:q_B_strong}).

Let us verify if the magnetic scalar potential correctly reproduces the numerical 
results explained in Sec.~\ref{sec:fractional_monopole}. The corresponding magnetic field obtained
from the magnetic scalar potential (\ref{eq:mag_pot_finite}) is
\begin{eqnarray}
B_a = \frac{q_B}{4\pi} \p_a u_S.
\end{eqnarray}
This asymptotically behaves as $\rho \to \infty$
\begin{eqnarray}
B_a \to \frac{q_B}{4\pi} \p_a \log \rho^2 = \frac{2\pi g^2 v^2}{m} \frac{x^a}{2\pi \rho^2},
\end{eqnarray}
which perfectly agrees with the previous result given in Eq.~(\ref{eq:asym_B1}).

Distribution of the magnetic charge density can be found as
\begin{eqnarray}
- \p_a^2 \varphi = \frac{q_B}{4\pi}\p_a^2 u_S = - q_B\frac{F_{12}}{2\pi}.
\end{eqnarray}
Thus, we are lead to a quite reasonable magnetic density $F_{12}/2\pi$, which corresponds to the 
magnetic field made by the vortex string.

\begin{figure}[t]
\begin{center}
\includegraphics[width=15cm]{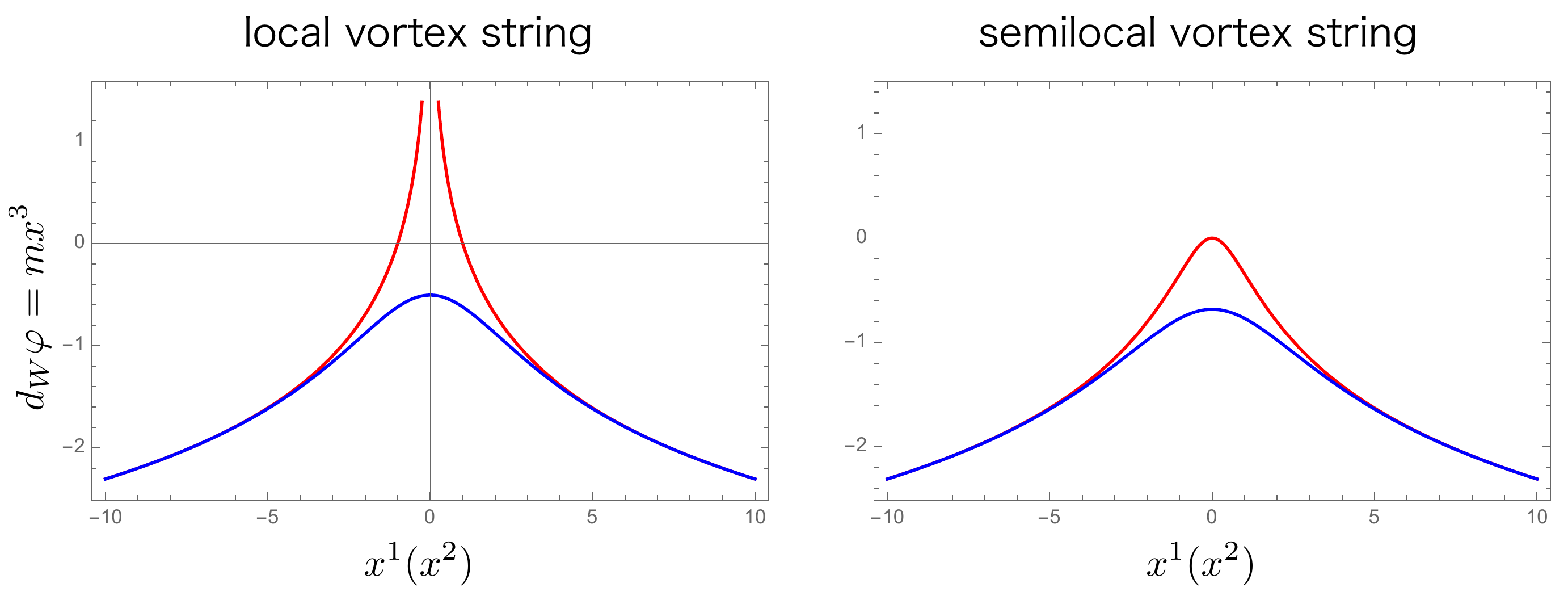}
\caption{The magnetic scalar potentials corresponding to the local vortex string (singular lump string) in the
left panel and the semi-local vortex string (regular lump string) with the size moduli $a=1$ in the right panel
for the strong gauge coupling limit (red) and the finite gauge coupling (blue). The horizontal axes are in the unit
of $1/\sqrt 2 gv$.}
\label{fig:mag_scalar_pot}
\end{center}
\end{figure}

The same  can be said for the semi-local boojum studied in Sec.~\ref{sec:semilocal}.
The domain wall's position can be read from Eq.~(\ref{eq:ini_sls}), which leads to the magnetic scalar potential
\begin{eqnarray}
\varphi = - \frac{q_{SLB}}{4\pi} u_{SLS} 
\to  - \frac{q_{SLB}}{4\pi} \log(\rho^2 + a^2),\quad
(\rho \to \infty),
\end{eqnarray}
with $q_{SLB} =  2\pi g^2 v^2/m$.
From this, one can compute the asymptotic magnetic field as
\begin{eqnarray}
B_b = - \p_b \varphi \to  \frac{q_{SLB}}{4\pi} \frac{2x^b}{\rho^2 + a^2} = \frac{2\pi g^2 v^2}{m} \frac
{x^b}{2\pi(\rho^2 + a^2)}.
\end{eqnarray}
Again, this perfectly agrees with the previous result given in Eq.~(\ref{eq:asym_B1_SLB}).

We plot the magnetic scalar potentials in Fig.~\ref{fig:mag_scalar_pot} for the strong gauge coupling limit and 
the finite gauge coupling case. In the left panel, the red curve shows that the potential made by the
point magnetic source at the origin which corresponds to the endpoint of the singular lump string in the
strong gauge coupling limit. When the gauge coupling is finite, the string size becomes finite of order $1/gv$,
and the charge distribution gets fat with the same size. Then the magnetic potential written in the blue curve
in the left panel becomes regular at the origin. In the right panel, we show the similar potentials for the semi-local
configurations with the moduli matrix $H_0=(z,a,1)$ and the mass matrix $M=(m/2,m/2,-m/2)$. We set $a=1$ so
that the semi-local strings are nonsingular even in the strong gauge coupling limit.

The magnetic scalar potential can be explained in a different way at a more technical level as follows.  
The exact formula for the magnetic field reads 
\begin{eqnarray}
B_a = \p_a\left(\frac{1}{2}\p_3u\right),
\end{eqnarray}
where $u$ is a solution to the master equation for the full 1/4 BPS equations.
Therefore, we should extract the magnetic scalar potential $\varphi$ from $\p_3 u/2$.
How can we do it?
A hint is in the approximate solution
\begin{eqnarray}
{\mathcal U}(x^k)= u_W\left(x^3 + \frac{1}{2m}u_S(x^a)\right) + u_S(x^a).
\end{eqnarray}
Let us evaluate $\p_a\p_3 {\mathcal U}$ on the domain wall's position $x^3 = - u_S/2m$. We find
\begin{eqnarray}
\p_3\p_a {\mathcal U} \big|_{x^3 = -\frac{u_S}{2m}} &=& \p_3 \left[\left(\frac{\p_a u_S}{2m}\right) u'_W\left(x^3 + \frac{1}{2m}u_S\right)\right]_{x^3 = -\frac{u_S}{2m}}  \nonumber\\
&=& \p_a\left(\frac{u''_W(0)}{2m} u_S\right),
\end{eqnarray}
where the prime stands for a $x^3$ derivative.
From Eq.~(\ref{eq:F_from_u}), we have $\sigma = \p_3 u_W/2$, so that $u''_W(0)/2$ corresponds to
$\sigma'(0)$, namely the derivative of $\sigma$ at the center of the domain wall. Furthermore,
$\sigma$ transits from $-m/2$ to $m/2$ inside the domain wall of the thickness $d_W$ \cite{Boojum1}. Thus, we have
$u''_W(0)/2 = \sigma'(0) = m/d_W$. Combining all the pieces, we reach the desired result
\begin{eqnarray}
B_a\big|_{x^3 = -\frac{u_S}{2m}} = -\p_a\varphi,\quad \varphi = -\frac{u_S}{2d_W} = \frac{mx^3}{d_W}.
\end{eqnarray}

In summary, we found that the solution $u_S$ to the master equation for the vortex string gives 
the magnetic scalar potential for $B_{a=1,2}$.


\section{A magnetic capacitor}

In this section, we will study the 1/4 BPS solutions which have multiple vortex strings aligned in a 
line attached to one or both sides of the domain wall in the model with $N_F=2$ and $M = (\tilde m/2,-\tilde m/2)$.
We have already studied similar configurations in Sec.~4 in our previous paper \cite{Boojum1}. For completeness, let us repeat the corresponding master equation
\begin{eqnarray}
\p_k^2 u = 1 - \left(|P_{n_1}|^2 e^{\tilde m x^3}  + |P_{n_2}|^2 e^{-\tilde mx^3}\right) e^{-u},
\label{eq:gf_two_vor}
\end{eqnarray}
for which an appropriate global approximation is given as
\begin{eqnarray}
{\mathcal U}(x^k) = u_W\left(x^3+\frac{u_S^{(n_1)}-u_S^{(n_2)}}{2\tilde m}\right)+\frac{u_S^{(n_1)}+u_S^{(n_2)}}{2},
\label{eq:U0_two_vor}
\end{eqnarray}
where $u_W(x^3)$ is the domain wall solution and $u_S^{(n)}(x^1,x^2)$ is the $n$ vortex string solution to the master equation
\begin{eqnarray}
(\p_1^2+\p_2^2) u_S^{(n)} - 1 + |P_n|^2 e^{-u_S^{(n)}} = 0.
\label{eq:master_dimless_vortex}
\end{eqnarray}

\subsection{Linearly aligned vortex strings ending on a domain wall from one side}

First, we align $n_1$ vortex strings in the vacuum $\left<1\right>$ on a line,
while we set no vortex strings in the opposite vacuum $\left<2\right>$.
More precisely, we will consider the moduli matrix $H_0 = (P_{n_1}(z),\ P_{n_2}(z))$ with
\begin{eqnarray}
P_{n_1} (z) = e^{-\tilde m X/2}\prod_{k=-(n_1-1)/2}^{(n_1-1)/2} (z - L k),\quad P_{n_2=0}(z) = e^{\tilde m X/2},
\label{eq:multiple_vor_mm}
\end{eqnarray}
where $L$ and $X$ are real constants.
For this moduli matrix, the $n_1$ string axes in the vacuum $\left<1\right>$ are aligned on the $x^1$ axis with
the separation $L$. The other real parameter $X$ is introduced to shift the configuration along the $x^3$ direction.

\begin{figure}[t]
\begin{center}
\includegraphics[width=14cm]{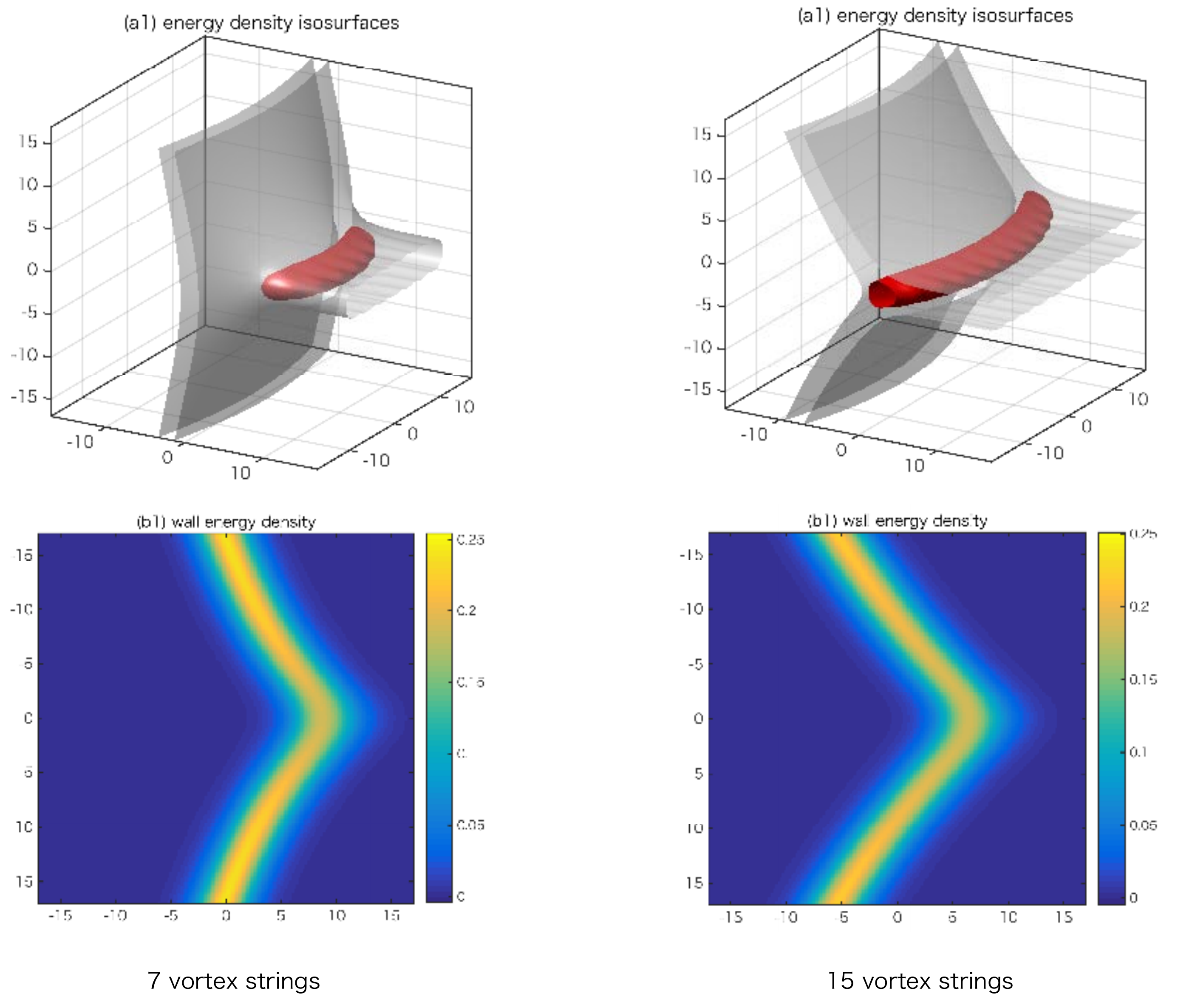}
\caption{The $n_1 = 7$ (left) and $n_1=15$ (right) vortex strings ending on the domain wall from one side.
The separation of neighboring strings is $L=3$. The gray surfaces in the panels (a1) are energy density isosurfaces
and the red ones correspond to a boojum charge density isosurfaces. The panels (b1) show the domain wall energy
density on the cross-section at $x^2 = 0$.}
\label{fig:7_15_vor}
\end{center}
\end{figure}

In Fig.~\ref{fig:7_15_vor}, we show two examples for $n_1=7$ and $n_1=15$. We set $X=20$ for $n_1=7$
and $X=40$ for $n_1 = 15$, and  $L=3$ for both solutions.
Since the vortex strings are degenerate if they are seen very  far from the junction points, 
the asymptotic bending of the domain wall  is logarithmic $\sim \log \rho^{2n_1}$.
However, the structure near the junction points is not logarithmic.
The panels (b1) of Fig.~\ref{fig:7_15_vor} show the domain wall energy density on the cross-section 
at $x^1=0$. Increasing the number of aligned vortex strings, the domain wall at the vicinity of junction points
becomes locally flat. Area of the flat region increases if we put more and more vortex strings on the line.

The emergence of the flat part can be understood as follows.
As before, the domain wall's position can be read from the master equation as
\begin{eqnarray}
x^3(x^1,x^2) = -\frac{1}{2\tilde m} u_S^{(n_1)}(x^1,x^2) + X,
\end{eqnarray}
where $u_S^{(n_1)}$ is a solution of the vortex master equation
\begin{eqnarray}
\p_a^2 u_S^{(n_1)} - 1 + \prod_k \left|z - kL\right|^2 e^{-u_S^{(n_1)}} = 0.
\label{eq:vor_master_n1}
\end{eqnarray}
If the separation $L$ is sufficiently larger than 1, the solution $u_S^{(n_1)}$ to the vortex master equation
can be well approximated by a simple superposition of $u_S^{(1)}$ as
\begin{eqnarray}
u_S^{(n_1)} (x^1,x^2) \simeq \hat u_S^{(n_1)} (x^1,x^2) \equiv \sum_k u_S^{(1);k}(x^1,x^2),\quad (L \gg 1),
\end{eqnarray}
where $u_S^{(1);k}(x^1,x^2)$ is the single vortex string at $z = kL$, namely solution to the master equation
\begin{eqnarray}
\p_a^2 u_S^{(1);k} - 1 + \left|z - kL\right|^2 e^{-u_S^{(1);k}} = 0.
\end{eqnarray}
In the region around $z = kL$, we have $u_S^{(1);k'} \simeq \log |z-k'L|^2$ for $k' \neq k$. Therefore, the
approximate solution there becomes
\begin{eqnarray}
\hat u_S^{(n_1)} (x^1,x^2) = u_S^{(1);k} + \sum_{k'(\neq k)} \log |z-k'L|^2,\quad (|z - kL|\ll 1).
\end{eqnarray}
Plugging this into Eq.~(\ref{eq:vor_master_n1}), one can confirm that the approximation works well.

\begin{figure}[t]
\begin{center}
\includegraphics[height=5cm]{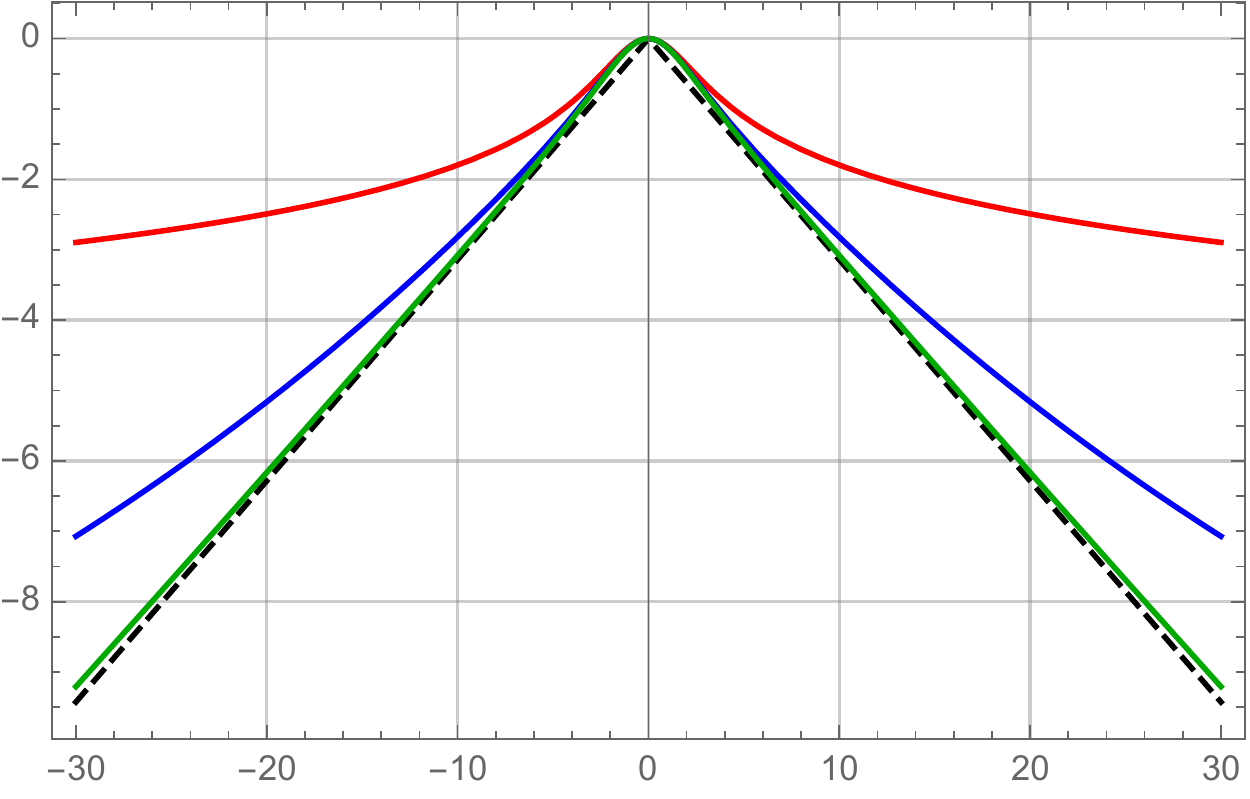}
\caption{The approximate solution $-\frac{1}{2}\left(\hat u_S^{(n_1)}(x^1=0,x^2)-\hat u_S^{(n_1)}(0,0)\right)$ is shown 
for $n_1 = 1$ (red),
$n_1=7$ (blue), and $n_1=101$ (green). The black dashed line is $- \frac{\pi}{L}|x^2|$. $L$ is set to be $10$.}
\label{fig:wall_bending}
\end{center}
\end{figure}

Thus, the domain wall's position in the $x^1=0$ plane is approximated by
\begin{eqnarray}
x^3(0,x^2) &\simeq& - \frac{1}{2\tilde m} \sum_k \hat u_S^{(1);k}(0,x^2) + X ,\nonumber\\
&\sim& - \frac{1}{2\tilde m}\sum_k \log \left((x^2)^2 + k^2L^2\right) + X.
\end{eqnarray}
We choose $X = (\sum_k \log kL)/\tilde m$, so that the junction point at $z\simeq 0$ is independent of $k$.
In Fig.~\ref{fig:wall_bending} we show $x^3(0,x^2)$ for $n_1 = 1,7,101$ for $L=10$ and $\tilde m = 1$.
The domain wall becomes linear as $n_1$ is increased, and it gets close to the linear function at $n_1 \to \infty$ limit
\begin{eqnarray}
\lim_{n_1 \to \infty} \tilde m x^3(0,x^2) 
&=&  - \frac{1}{2}\sum_{k=-\infty}^\infty \log \left((x^2)^2 + k^2L^2\right) + \sum_{k=1}^\infty \log (kL) \nonumber\\
&=& - \log|x^2| -  \sum_{k=1}^\infty\log\left(1 + \frac{(x^2)^2}{k^2L^2}\right) \nonumber\\
&=& -  \log |x^2| -  \log \frac{\sinh\frac{\pi |x^2|}{L}}{\frac{\pi |x^2|}{L}} \nonumber\\ 
&=& - \log \sinh\frac{\pi |x^2|}{L} - \log \frac{L}{2\pi} \nonumber \\ 
&\to& - \frac{\pi}{L}|x^2| - \log \frac{L}{2\pi},\quad (\pi |x^2| \gg L),
\label{eq:x^3_asym}
\end{eqnarray}
where we have used the relation $\prod_{k=1}^\infty\left(1+\frac{\alpha^2}{k^2}\right) = \frac{\sinh\pi\alpha}{\pi\alpha}$.

\begin{figure}[t]
 \begin{minipage}{0.5\hsize}
  \begin{center}
   \includegraphics[width=7cm]{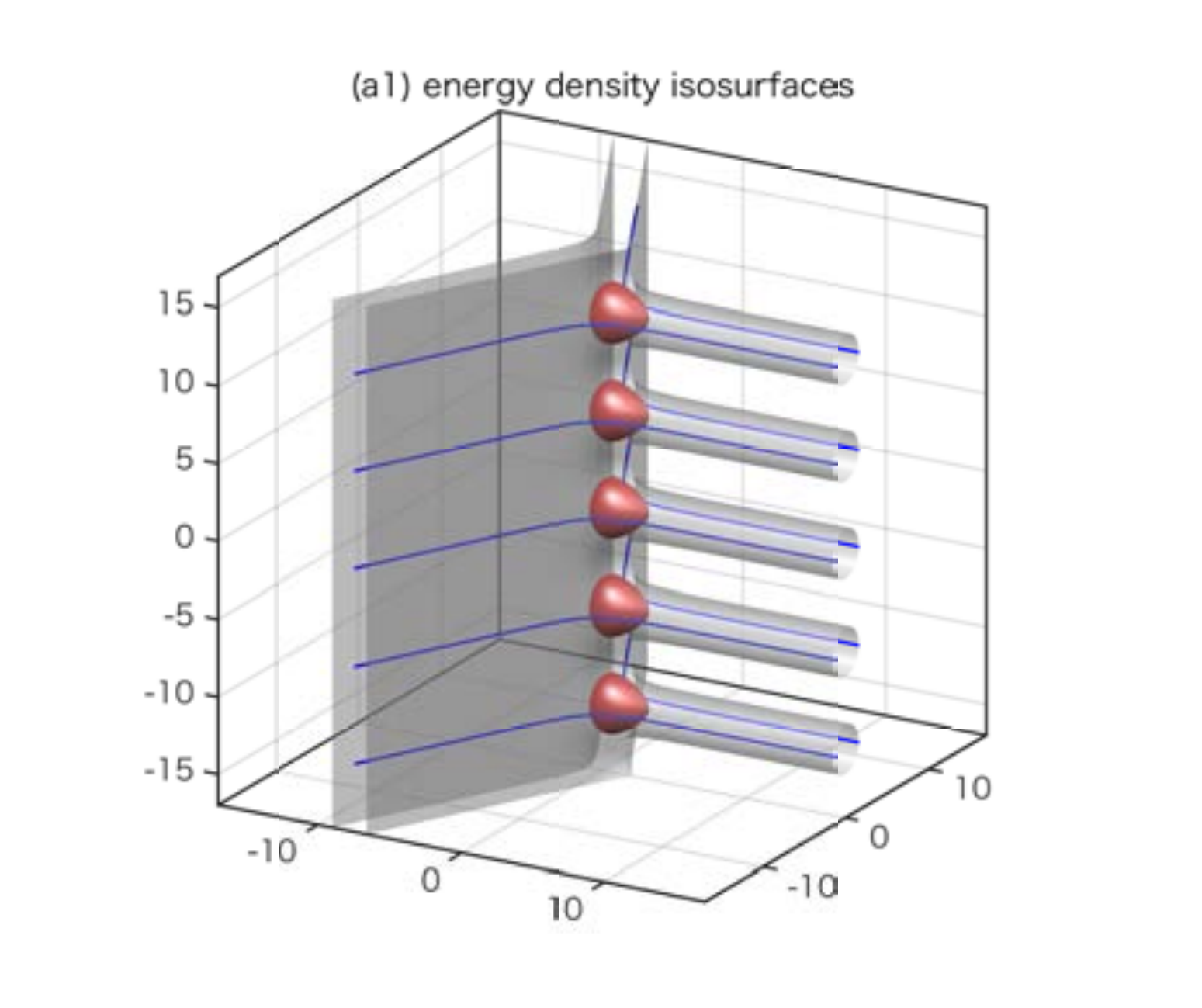}
  \end{center}
 \end{minipage}
  \begin{minipage}{0.5\hsize}
  \begin{center}
   \includegraphics[width=7cm]{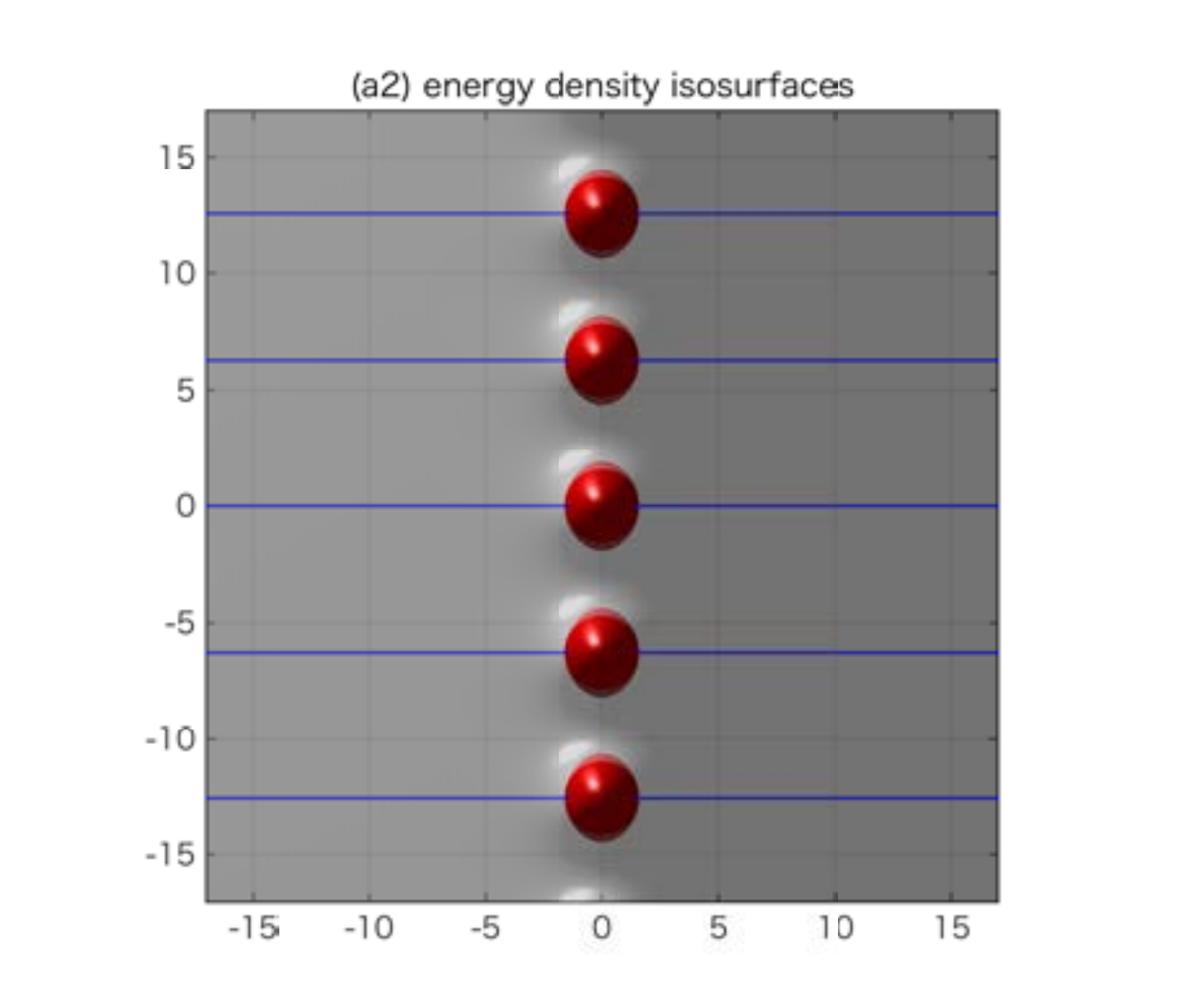}
  \end{center}
 \end{minipage}\\
 \begin{minipage}{0.5\hsize}
  \begin{center}
   \includegraphics[width=7cm]{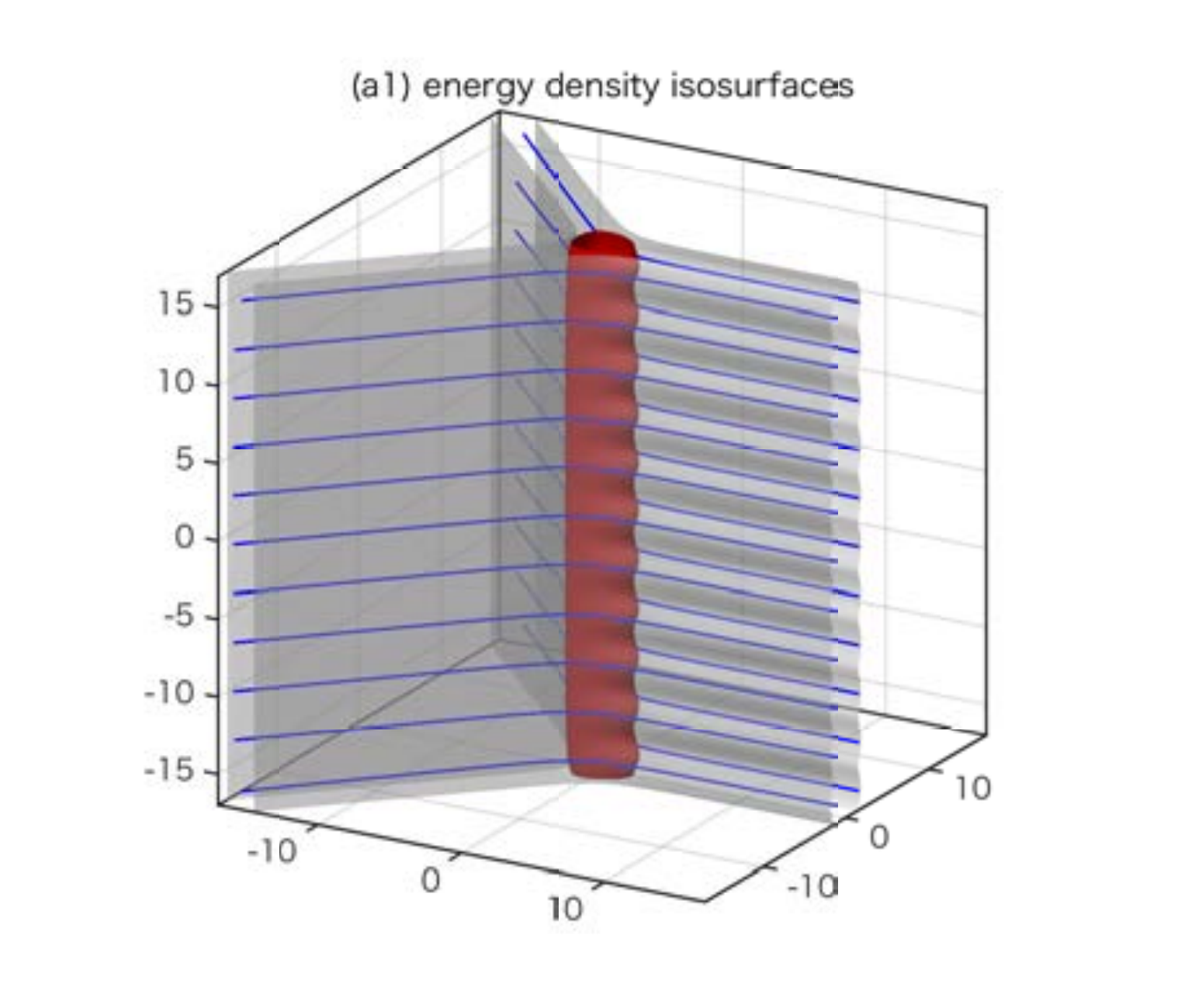}
  \end{center}
 \end{minipage}
 \begin{minipage}{0.5\hsize}
  \begin{center}
   \includegraphics[width=7cm]{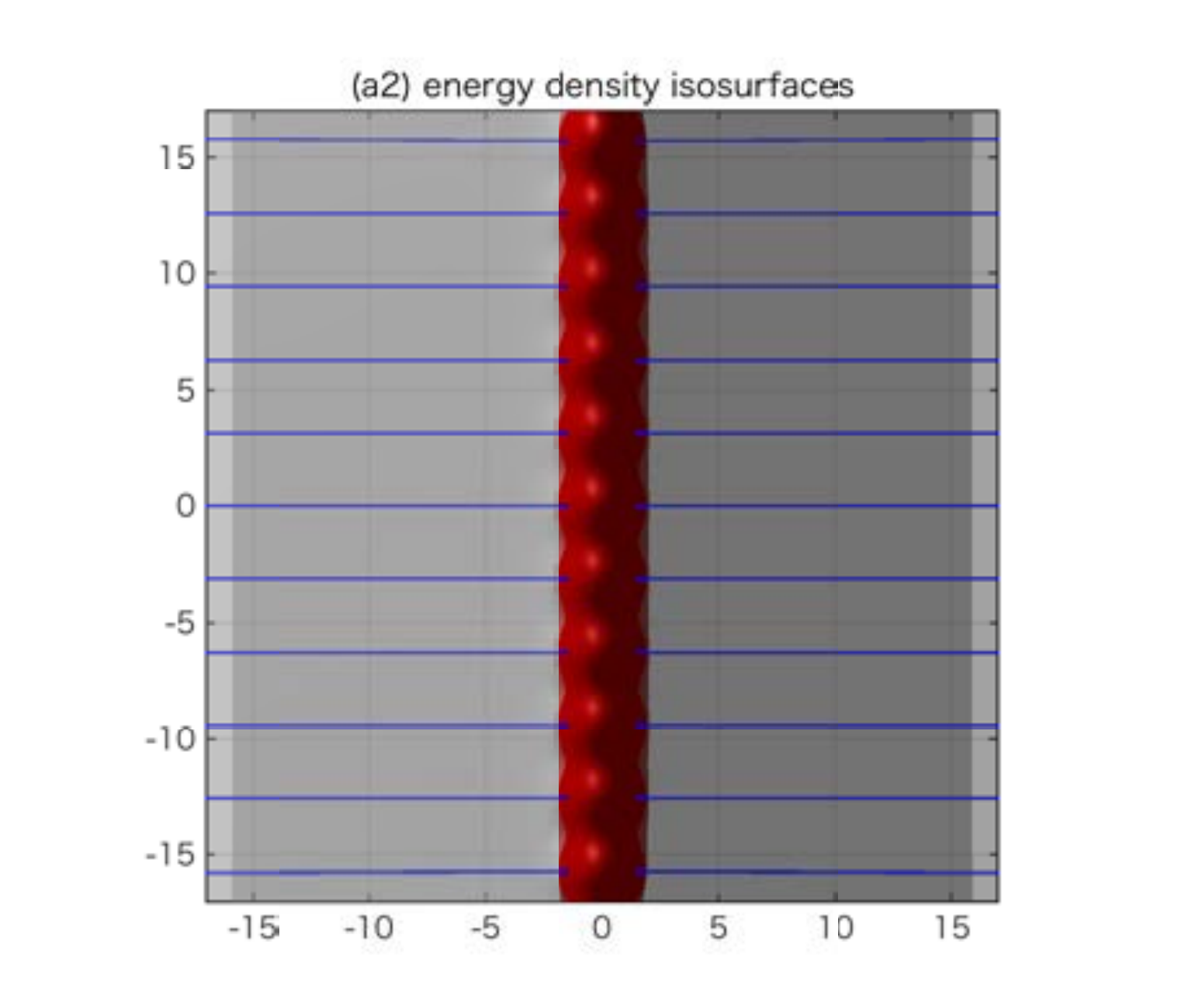}
  \end{center}
 \end{minipage}
 \caption{Periodically aligned vortex strings with period $2\pi$ (upper panels) and $\pi$ (lower panels)
 are shown. The gray surfaces show the energy density isosurfaces, the red surfaces show the boojum density 
 isosurfaces and the blue lines show several magnetic force lines.}
 \label{fig:sin_oneside}
\end{figure}

There is nice physical observation which explains the appearance of the factor $\pi/L$
in the asymptotic angle of the flat domain wall.
From the viewpoint of the domain wall, the endpoints of vortex strings are interpreted 
as the magnetic sources in $2+1$ dimensional sense. Consider the magnetic scalar potential defined by
\begin{eqnarray}
\tilde \varphi(x^1,x^2) = \frac{\tilde m }{\tilde d_W} x^3(x^1,x^2)= \frac{-1}{2\tilde d_W} u_S^{(n_1)}(x^1,x^2).
\label{eq:mag_pot_L2}
\end{eqnarray}
Suppose that we have the infinite point-like magnetic sources on a line, say the $x^1$ axis, with the period $L$. 
Due to the symmetry of this source arrangement, the magnetic force lines far from the $x^1$ axis become
parallel to the $x^2$ axis. 
There is one magnetic source of the magnetic charge $\tilde q_B = 2\pi/\tilde d_W$ (see the discussions
in Sec.~\ref{msp})
at every finite segment 
$x^1 \in [x_0,x_0+L]$ for an arbitrary $x_0$. Therefore, seen far from
the sources, the magnetic charge density is $\tilde q_B/L$. The magnetic force lines from these sources equally 
expand both to $x^2>0$ and $x^2<0$ regions. Thus, we should have
\begin{eqnarray}
\tilde \varphi \simeq -\frac{\tilde q_B}{2L}|x^2| = - \frac{\pi}{\tilde d_W L }|x^2|.
\end{eqnarray}
Combining this with Eq.~(\ref{eq:mag_pot_L2}), we correctly find the asymptotic behavior 
given in Eq.~(\ref{eq:x^3_asym}).

\begin{figure}[t]
  \begin{center}
   \includegraphics[width=14cm]{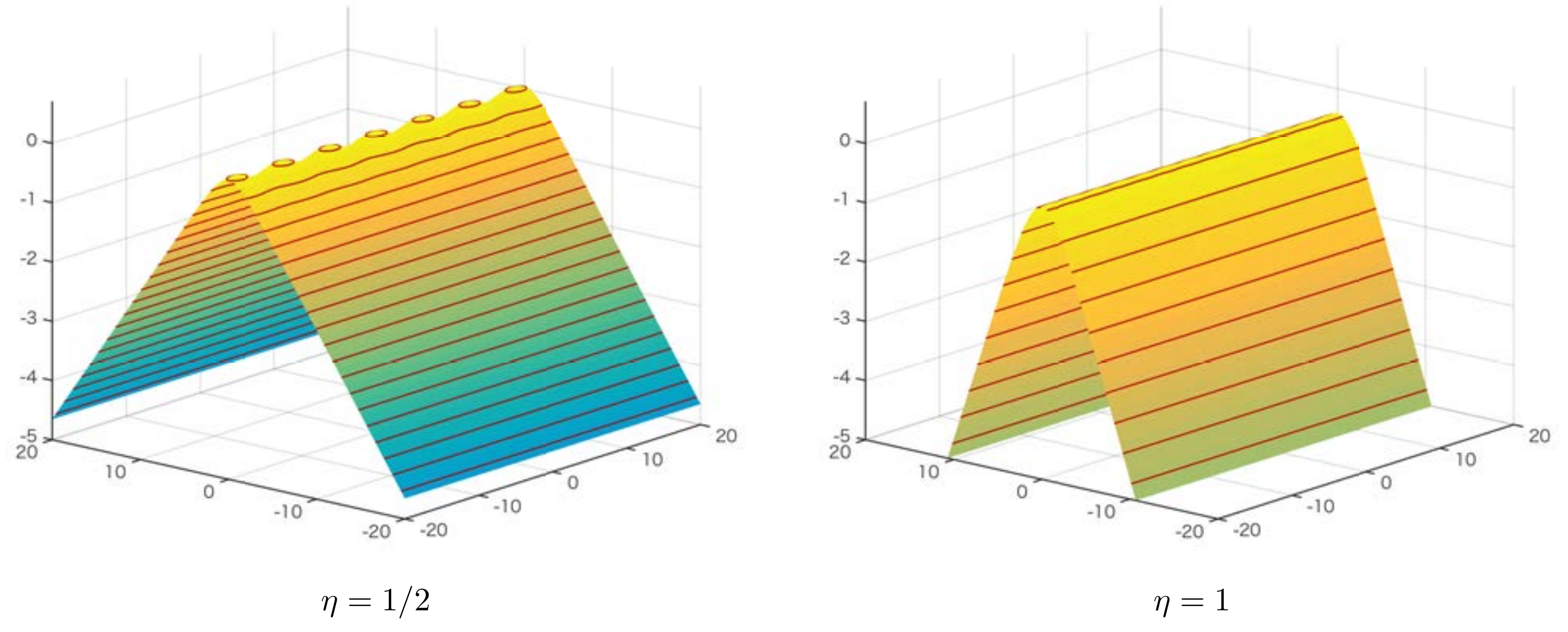}
 \caption{The magnetic scalar potentials for the periodically aligned vortex strings with periods $2\pi$ ($\eta=1/2$)
 and $\pi$ ($\eta=1$).}
 \label{mag_pot_sine}
\end{center}
\end{figure}

In order to get vortex strings periodically aligned on a line, it is better to use holomorphic trigonometric functions
rather than polynomial functions \cite{Eto:2004rz}. 
For example, we choose the following moduli matrix
\begin{eqnarray}
H_0 = \left(\sin i\eta z,\, 1\right).
\end{eqnarray}
The positions of the vortex strings correspond to the zeros of the first elements, namely
$z = (\pi i/\eta) n$ ($n\in \mathbb Z$).
An advantage of using the trigonometric function is that one does not need to shift the domain wall's position by
adjusting the $X$ parameter for the polynomial case in Eq.~(\ref{eq:multiple_vor_mm}). 
In Fig.~\ref{fig:sin_oneside}, we show two examples with sparsely aligned ($\eta = 1/2$: the period is $2\pi$)  
and densely alined ($\eta = 1$: the period $\pi$)  vortex strings ending on the domain wall from one side.
Since the moduli matrix includes the infinite number of vortex strings, the domain wall becomes
asymptotically exactly flat with the slanting angle $\eta$. Namely, the domain wall's position is estimated by
$e^{-2\tilde mx^3} \simeq |\sin i\eta z|^2$,
\begin{eqnarray}
x^3 &=& - \frac{1}{2\tilde m} \log\left(\cosh^2\eta x^1 \sin^2 \eta x^2 + \sinh^2\eta x^1\cos^2\eta x^2\right) \nonumber\\
&\to&  - \frac{\eta}{\tilde m} |x^1| + \frac{1}{\tilde m} \log 2,\quad (|x^1| \to \infty).
\label{eq:angle_slant_wall}
\end{eqnarray}

The magnetic scalar potentials $\tilde \varphi = - u_{S}/2\tilde d_W$ for 
$\eta = 1/2, 1$ are shown in Fig.~\ref{mag_pot_sine} for $\tilde m = 1$ ($\tilde d_W = 2$).
As expected, 
it is clear that the potentials are asymptotically exactly linear in $|x^1|$, reflecting the asymptotic flatness of
domain wall.

\subsection{Linearly aligned vortex strings ending on a domain wall from two sides}

\begin{figure}[h]
\begin{center}
\includegraphics[width=14cm]{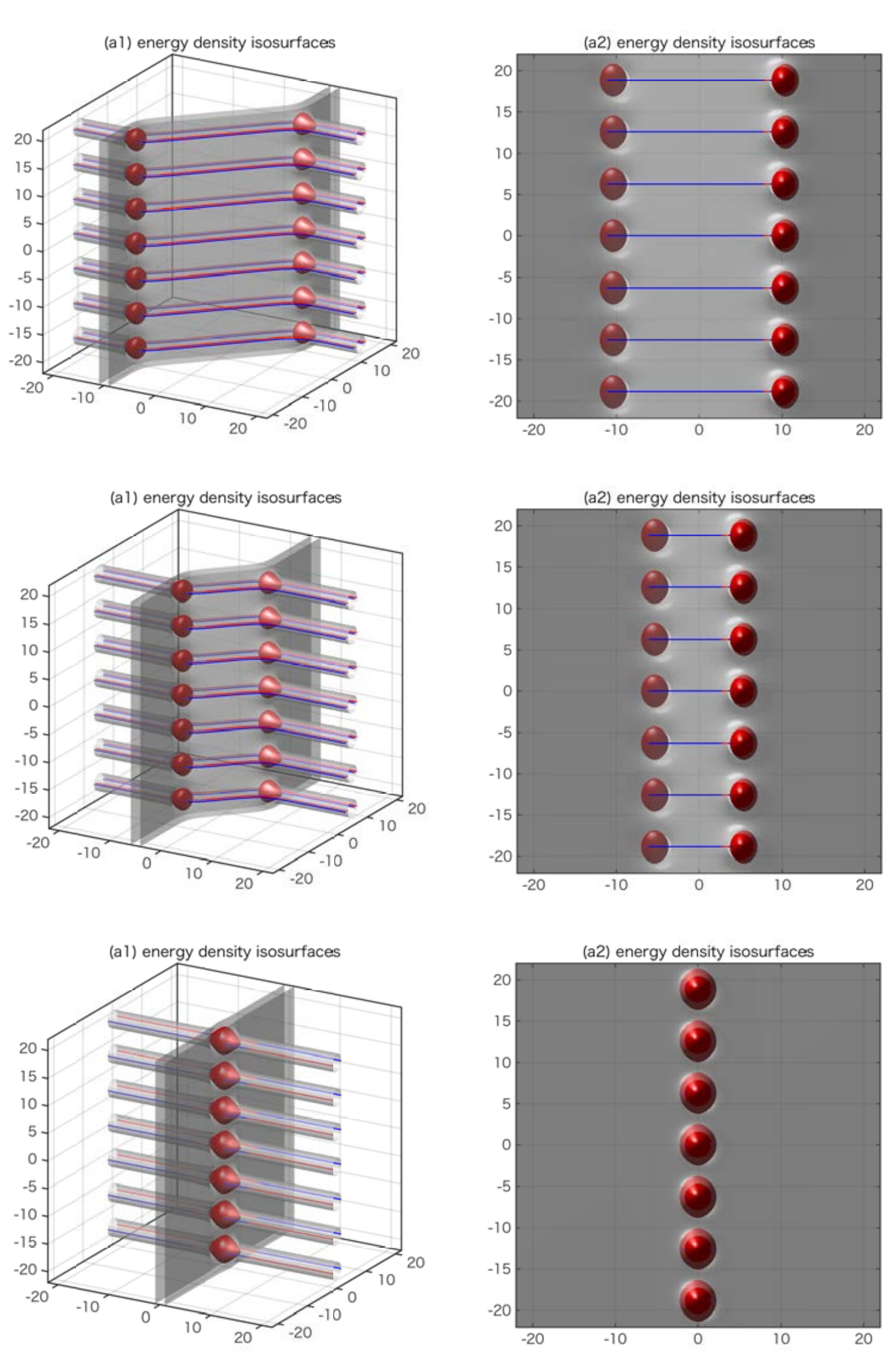}
 \caption{The plots of the energy density isosurfaces of periodic vortices ending on 1 wall from two sides. The distances between vortices are $Z=10$ (top), $Z=5$ (middle), and $Z=0$ (bottom).}
 \label{fig:sin_2side_a}
 \end{center}
 \end{figure}

\begin{figure}[h]
\begin{center}
\includegraphics[width=14cm]{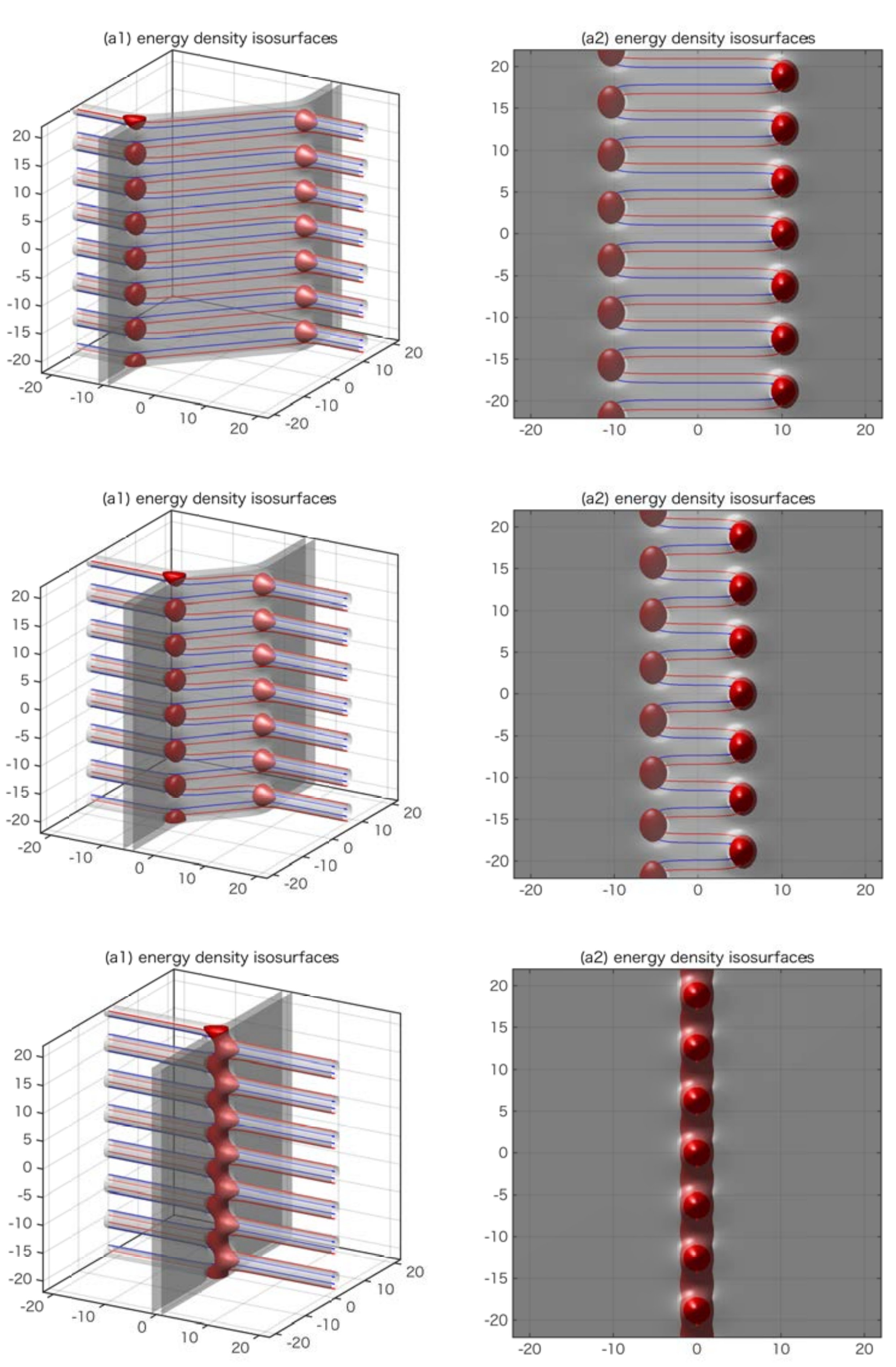}
  \caption{The plots of the energy density isosurfaces of periodic vortices aligned alternately ending on 1 wall from two sides. The distances between vortices are $Z=10$ (top), $Z=5$ (middle), and $Z=0$ (bottom).}
\label{fig:sin_2side_b}
\end{center}
 \end{figure}

Let us next consider configurations with periodically aligned infinite vortex strings ending on 
the domain wall from both sides.
The corresponding moduli matrix is given by
\begin{eqnarray}
H_0 = \left(\sin i\eta_1 (z-Z_1),\ \sin i \eta_2 (z-Z_2) \right),
\label{eq:mm_sin_2sides}
\end{eqnarray}
with $Z_{1,2}$ being complex constants. We set $\eta_1 = \eta_2 = \eta$ to be real constants, so that the vortex strings are aligned
on the lines $x^1 = {\rm Re}(Z_{1,2})$ parallel to the $x^2$ axis with period $\pi/\eta$.
We show two examples of this kind in Figs.~\ref{fig:sin_2side_a} and \ref{fig:sin_2side_b} with $\tilde m=1$.
In the former figure, we take $Z_1 = - Z_2 = 10,5,0$ with $\eta=1/2$ (the period is $2\pi$).
In the latter figure, we shift the vortex strings at the negative $x^3$ side by $\delta x^2 = \pi$. Namely,
we take $Z_1 = - Z_2 + i\pi = 10,5,0$. Far from the vortex strings, the domain wall is flat 
and perpendicular to the $x^1-x^2$ plane. On the other hand, between the vortex strings, the domain wall is flat but
slanted as $x^3 \simeq 2\eta x^1$, which is twice steeper than for the domain wall with the vortex strings just on the one side, see Eq.~(\ref{eq:angle_slant_wall}).
This is, of course, because we have two lines of vortex strings. The shape of the domain wall
is determined by superposition.
For example, the domain wall's position can be estimated for real positive
$Z > 0$ as follows,
\begin{eqnarray}
\tilde m x^3 \simeq - \eta |x_1 - Z| + \eta |x_1 + Z|  = \left\{
\begin{array}{ccc}
2 \eta Z & &  2Z < x^1\\
2\eta x^1 & \quad\text{for}\quad & -2 Z< x^1 < 2Z\\
-2\eta Z & & x^1 < -2Z
\end{array}
\right..
\label{eq:superpose_two_slant}
\end{eqnarray}

Comparing Figs.~\ref{fig:sin_2side_a} and \ref{fig:sin_2side_b} we can see that the shift $\delta x^2 = \pi$ in Fig.~\ref{fig:sin_2side_b} did not affect the resulting configuration very much. 
Only the local structure around the endpoints received small deformation but the asymptotic structure is not changed.
The formula (\ref{eq:superpose_two_slant}) also remains correct.

\begin{figure}[h]
\begin{center}
\includegraphics[height=7cm]{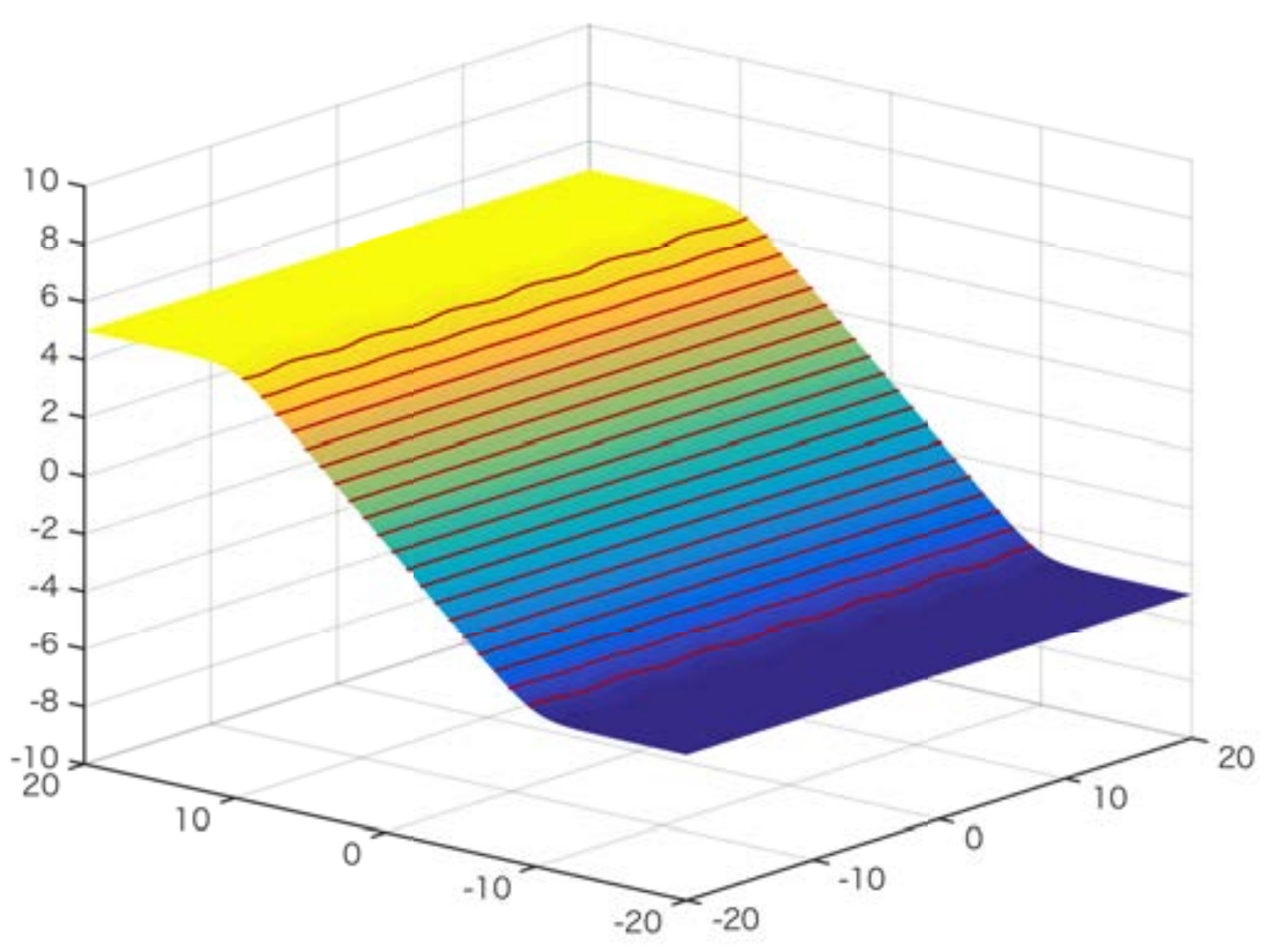}
  \caption{The magnetic scalar potential (5.15)  for the moduli matrix given in 
  Eq.~(5.13) with $Z = Z_1=-Z_2=10$,  $\eta = 1/2$ and $\tilde m=1$.}
\label{fig:mag_pot_sin_2side}
\end{center}
 \end{figure}

Let us interpret the above 1/4 BPS configurations from the viewpoint of $2+1$ dimensions. 
The corresponding magnetic scalar potential is given as follows
\begin{eqnarray}
\tilde \varphi 
&=&-\frac{1}{2\tilde d_W} u_S\big|_{+Z} + \frac{1}{2\tilde d_W} u_S\big|_{-Z} \nonumber\\
&=& \frac{\tilde m}{\tilde d_W}\left(x^3 \big|_{+Z}+ x^3\big|_{-Z}\right)\nonumber\\
&\simeq& \frac{\eta}{\tilde d_W} \left(-  |x_1 - Z| +  |x_1 + Z|\right).
\label{eq:magnetic_capacitor}
\end{eqnarray}
We plot the magnetic scalar potential $\tilde \varphi$ in Fig.~\ref{fig:mag_pot_sin_2side} which reproduces
the correct structure of the kinky domain wall.

\clearpage

The last expression in Eq.~(\ref{eq:magnetic_capacitor}) is reminiscent of the electric scalar potential for 
an electric capacitor. Hence, we may call the configuration with the magnetic scalar potential
given in Eq.~(\ref{eq:magnetic_capacitor}) a {\it magnetic capacitor} in $2+1$ dimensions.
Let us define a density of magnetic capacitance $\tilde c_M$ by
\begin{eqnarray}
\tilde c_M \delta \tilde V_B = \frac{\tilde Q_B}{\pi/\eta} ,
\end{eqnarray}
where $\delta\tilde  V_B = \tilde d_W \tilde \varphi\big|_{x^2=Z} - \tilde d_W \tilde \varphi\big|_{x^2=Z}$ 
stands for the difference of the magnetic potential and $\tilde Q_B = \tilde q_B \tilde d_W$ is 
the magnetic charge per unit length.
We have $\delta \tilde \varphi = 4 \eta Z/\tilde d_W$ and $\tilde q_B = 2\pi/\tilde d_W$, thus we conclude that
the domain wall has the magnetic capacitance 
\begin{eqnarray}
\tilde c_M = \frac{1}{2Z}.
\end{eqnarray}
As an ordinary electric capacitance of a flat capacitor, the capacitance is inversely proportional to the distance of
the charges.

\begin{figure}[t]
\begin{center}
\includegraphics[width=14cm]{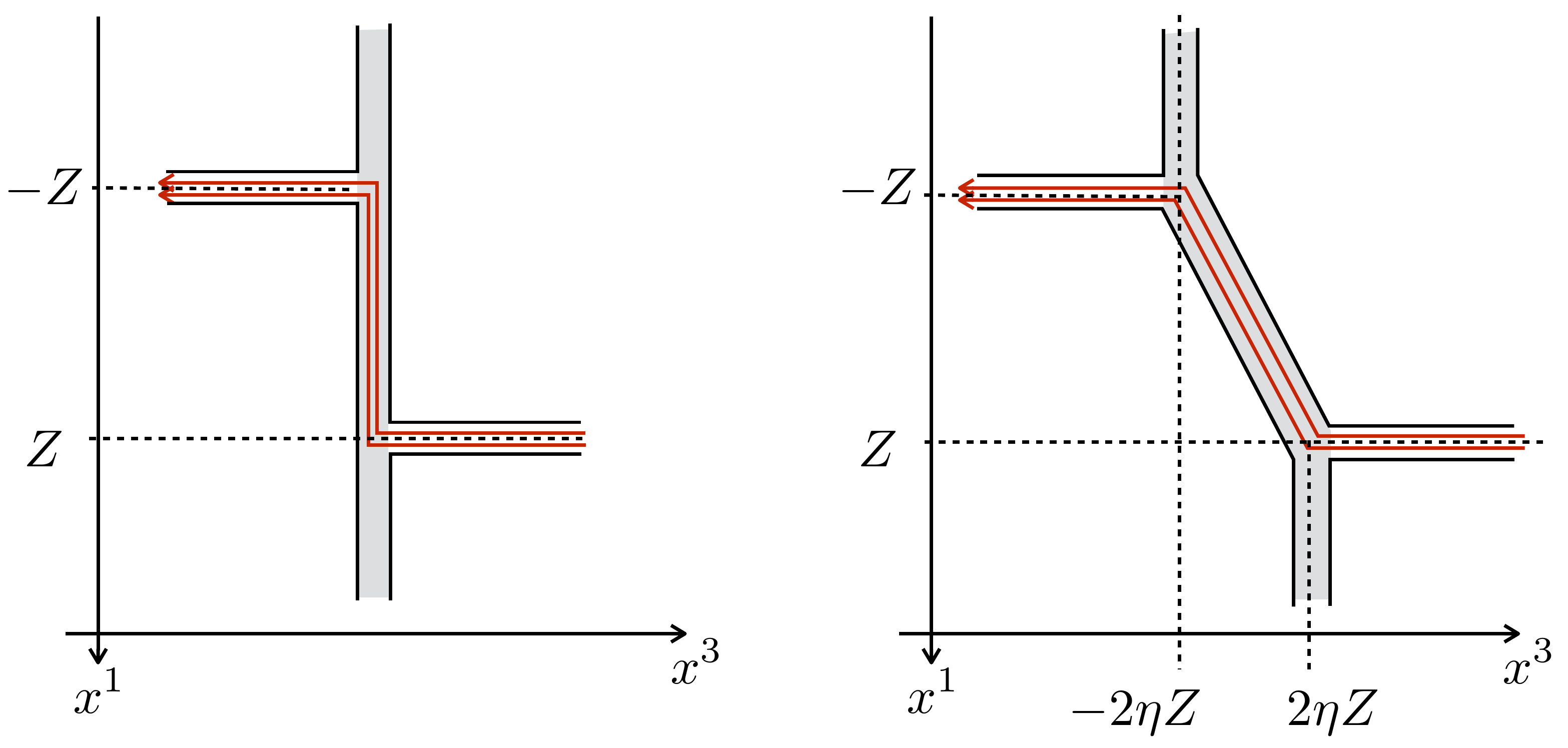}
\caption{Schematic picture for the ideal (real) domain wall (gray region) with periodically aligned vortex strings from both sides
is shown in the left (right) panel. Red lines stand for the incoming and outgoing magnetic fluxes from the vortex strings.}
\label{fig:sche_magnetic_capacitor}
\end{center}
\end{figure}

The energy stored in the magnetic capacitor is given by
\begin{eqnarray}
\delta \tilde {\cal E}_M = \frac{1}{2}\tilde c_M \delta \tilde V_M^2 = 4\eta^2 Z.
\label{eq:ene_mag_capacitor}
\end{eqnarray}
This can be accounted by the following geometric consideration about the domain wall and the vortex strings.
Suppose the domain wall did not bend by the vortex strings. Then the energy for the part between two linearly 
aligned vortex strings, namely $x^1 \in [-Z,Z]$ are proportional to the distance $2Z$ as is depicted in the left
panel of Fig.~\ref{fig:sche_magnetic_capacitor}. In reality, of course, the domain wall linearly bends as is shown in the right 
panel of Fig.~\ref{fig:sche_magnetic_capacitor}. The bent domain wall is longer than flat one by
\begin{eqnarray}
\delta \tilde L = 2Z\sqrt{1 + 4\eta^2} - 2 Z\simeq 4\eta^2Z,
\end{eqnarray}
for $\eta \ll 1$. This coincides  with the energy stored in the magnetic capacitor given in Eq.~(\ref{eq:ene_mag_capacitor}).

\subsection{Vortex strings ending on a slanting domain wall}

\begin{figure}[t]
\begin{center}
\includegraphics[width=14cm]{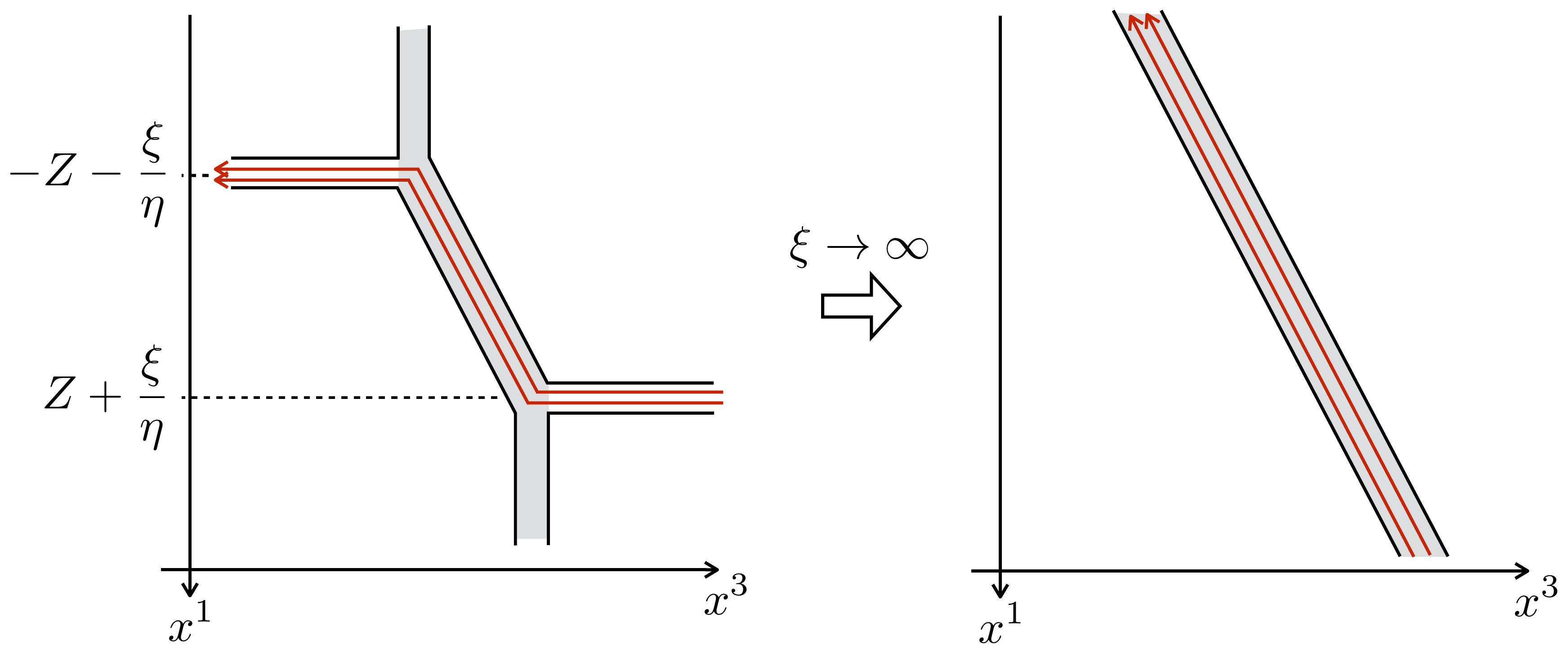}
\caption{The slant domain wall as a limit of sending the vortex strings toward the spatial infinities.}
\label{fig:sche_magnetic_capacitor2}
\end{center}
\end{figure}

Next, we consider the following moduli matrix which is slightly different from
the one given in Eq.~(\ref{eq:mm_sin_2sides})
\begin{eqnarray}
H_0 &=& \left(
\frac{e^{-\eta(z-Z)}-e^{\eta(z-Z)-2\xi}}{2i}
,\ 
\frac{e^{-\eta(z+Z)-2\xi}-e^{\eta(z+Z)}}{2i}\right) \nonumber\\
&=& e^{-\xi}\left(\sin i\eta\left(z-Z-\frac{\xi}{\eta}\right),\ 
\sin i\eta\left(z+Z+\frac{\xi}{\eta}\right)\right).
\end{eqnarray}
As is discussed in the previous subsection, this moduli matrix generates the configuration
with the linearly alined vortex strings at $z =i\frac{\pi}{\eta} n \pm\left( Z + \frac{\xi}{\eta}\right)$ with
$n \in \mathbb{Z}$.
Now, we send all the vortex strings to the spatial infinity by taking the limit $\xi \to \infty$.
We are left with
\begin{eqnarray}
H_0 = \left(
\frac{e^{-\eta(z-Z)}}{2i}
,\ 
\frac{-e^{\eta(z+Z)}}{2i}\right)
\simeq \left(1,\ -e^{2\eta z}\right),
\label{eq:mm_slant_wall}
\end{eqnarray}
where we used the so-called $V$-transformation that transforms the moduli matrix 
as $H_0 \to V(z) H_0$ and $u \to 2\log V(z)$ with arbitrary invertible holomorphic function 
$V(z)$ \cite{Isozumi:2004jc, Sakai1, Eto1}. The $V$-transformation does not change any physics.
Since we just shifted the vortex strings to the spatial infinities, the domain wall shape is given by
the Eq.~(\ref{eq:superpose_two_slant}) with replacement $Z$ by $Z + \xi/\eta$.
Especially, the domain wall between the lines
of the vortex-strings keep being slant with the same angle, see Fig.~\ref{fig:sche_magnetic_capacitor2}.
This holds even in the limit $\xi \to \infty$.
Furthermore, we know the existence of the vortex strings behind the spatial boundaries $x^1 = \pm \infty$, 
which provides a background magnetic fluxes $4\pi/(\pi/\eta) = 2\eta$ per unit length. In short, the flat domain wall
slants when  a background magnetic field is turned on \cite{Sakai1}.
The 1/4 BPS master equation for the moduli matrix (\ref{eq:mm_slant_wall}) is given by
\begin{eqnarray}
\p_k^2 u - 1 + \left(e^{\tilde m x^3} + e^{4\eta x^1}e^{-\tilde mx^3}\right) e^{-u} = 0.
\end{eqnarray} 
This can be rewritten as
\begin{eqnarray}
\p_k^2 u - 1 + \left(e^{\tilde m x^3-2\eta x^1} + e^{-\tilde mx^3 + 2\eta x^1}\right) e^{-u+2\eta x^1} = 0.
\end{eqnarray} 
Introducing the new coordinate by
\begin{eqnarray}
\left(
\begin{array}{c}
y_3\\
y_1
\end{array}
\right) = \left(\begin{array}{cc}
\cos \alpha & - \sin \alpha\\
\sin\alpha & \cos \alpha
\end{array}
\right)
\left(
\begin{array}{c}
x_3\\
x_1
\end{array}
\right),\quad
\tan\alpha = \frac{2\eta}{\tilde m},
\end{eqnarray}
and define a function by
\begin{eqnarray}
\hat u = u - 2\eta x^1,
\end{eqnarray}
we find that $\hat u$ is the solution to
\begin{eqnarray}
\left(\frac{\p^2}{\p y_3^2} + \frac{\p^2}{\p y_2^2}\right)\hat u - 1 + \left(
e^{\hat m y^3} + e^{-\hat m y^3}\right)e^{-\hat u} = 0,
\end{eqnarray}
where we defined $\hat m \equiv \sqrt{\tilde m^2 + 4\eta^2}$. Clearly, $\hat u$ does not depend on $y^1$, so that
we identify that $\hat u$ is identical to the domain wall solution written in the rotated coordinate $y^3$ with 
the mass parameter $\hat m$.
In the original coordinates, the solution is given by
\begin{eqnarray}
u(x^k) =u_W(x^3   - x^1\tan \alpha ) + 2\eta x^1.
\end{eqnarray}
The position of domain wall is determined by the condition $x^3 \cos\alpha  - x^1\sin \alpha=0$, namely it is
\begin{eqnarray}
x^3 = (\tan\alpha) x^1 = \frac{2\eta}{\tilde m}x^1.
\end{eqnarray}
This is consistent with the previous result given in Eq.~(\ref{eq:superpose_two_slant}).

\begin{figure}[h]
 \begin{minipage}{0.5\hsize}
  \begin{center}
   \includegraphics[width=7cm]{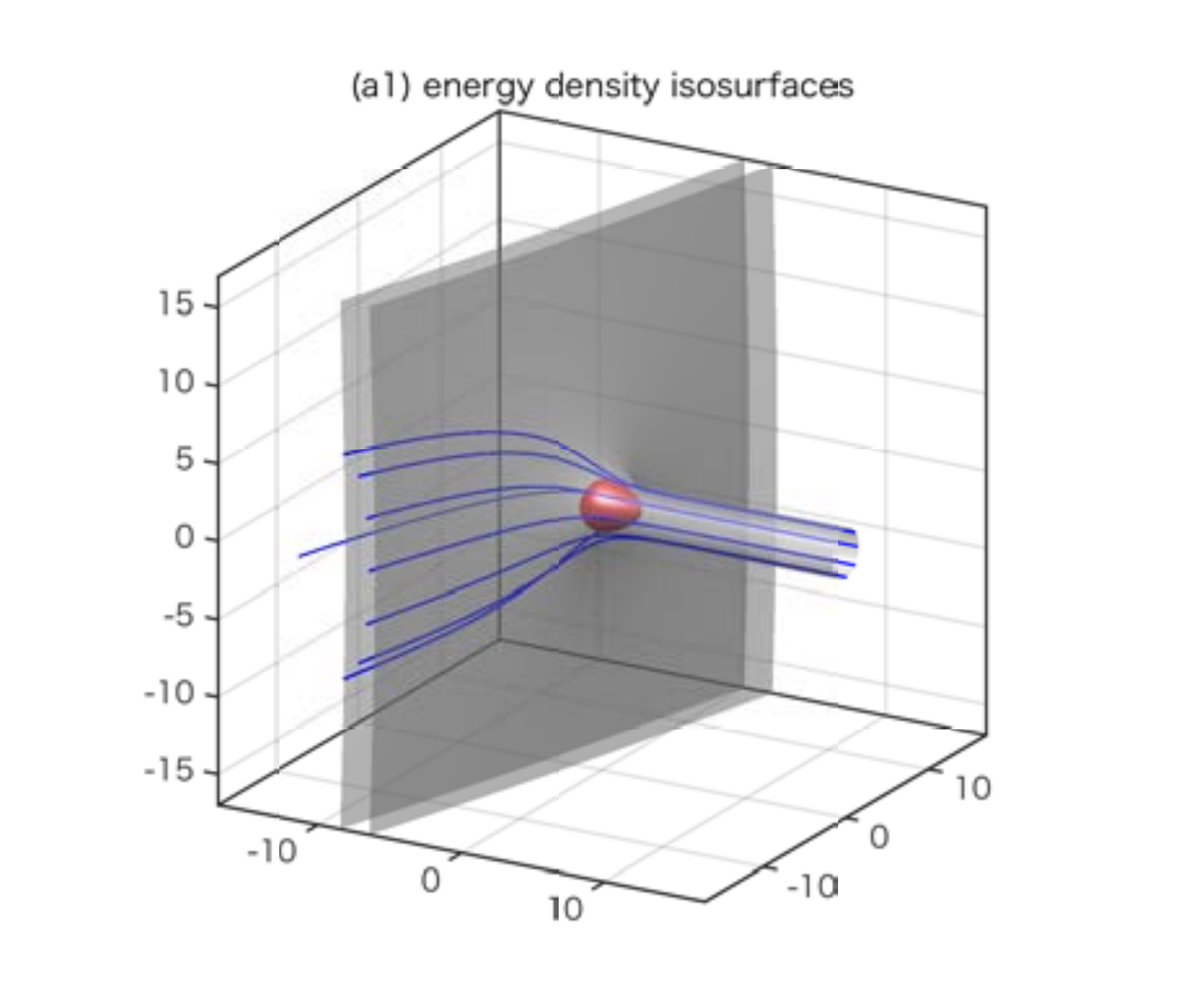}
  \end{center}
 \end{minipage}
  \begin{minipage}{0.5\hsize}
  \begin{center}
   \includegraphics[width=7cm]{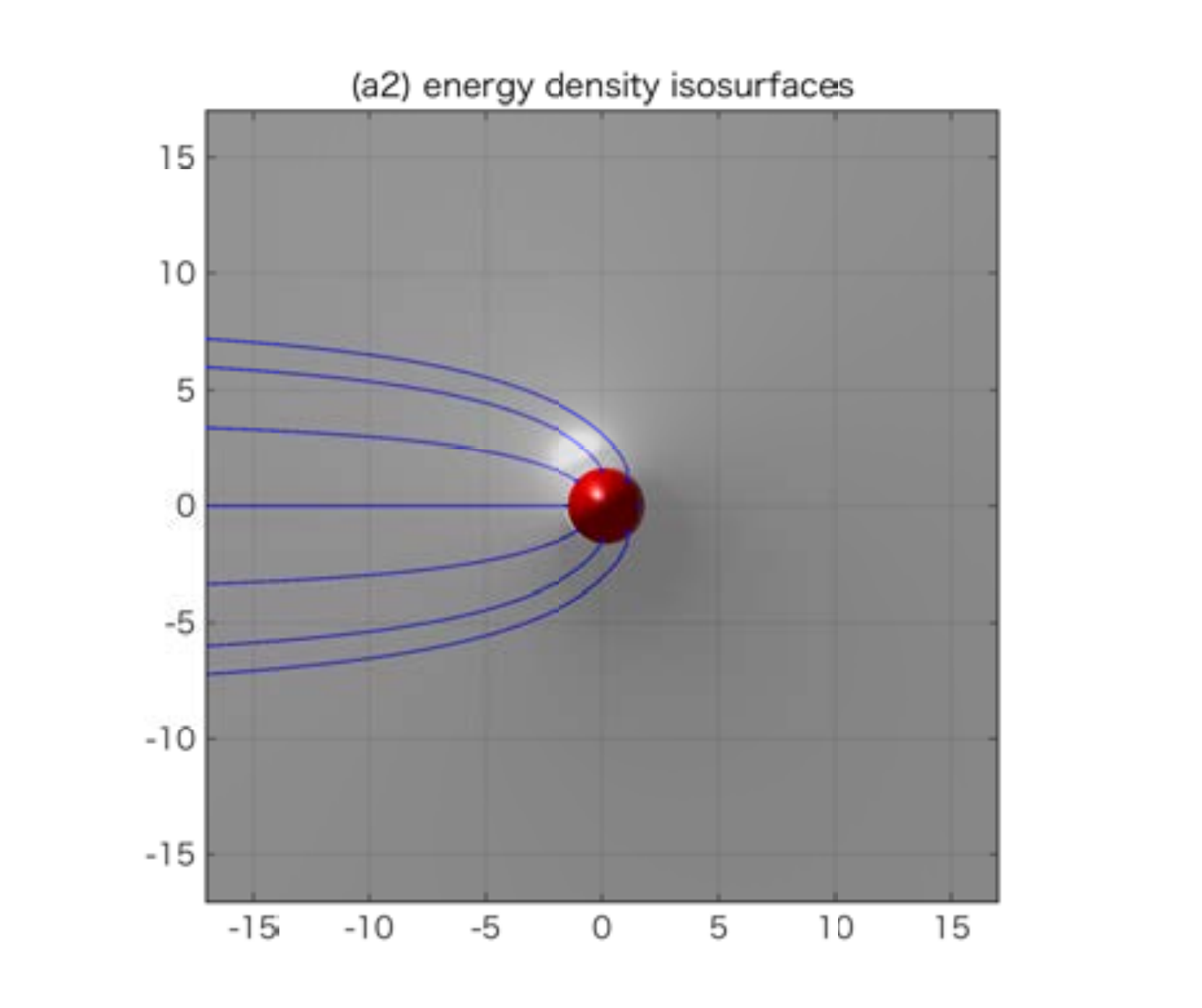}
  \end{center}
 \end{minipage}\\
 \begin{minipage}{0.5\hsize}
  \begin{center}
   \includegraphics[width=7cm]{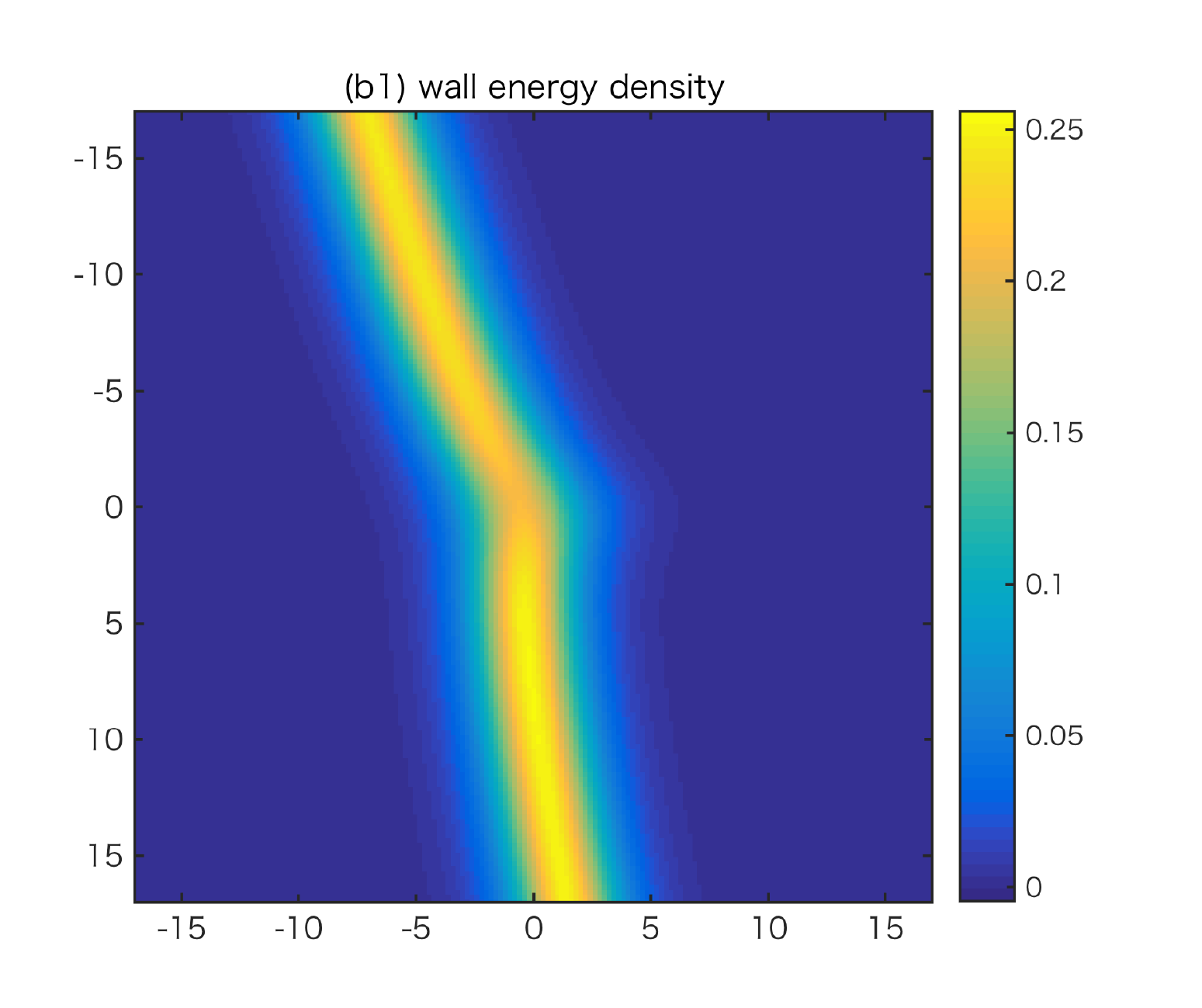}
  \end{center}
 \end{minipage}
 \begin{minipage}{0.5\hsize}
  \begin{center}
   \includegraphics[width=7cm]{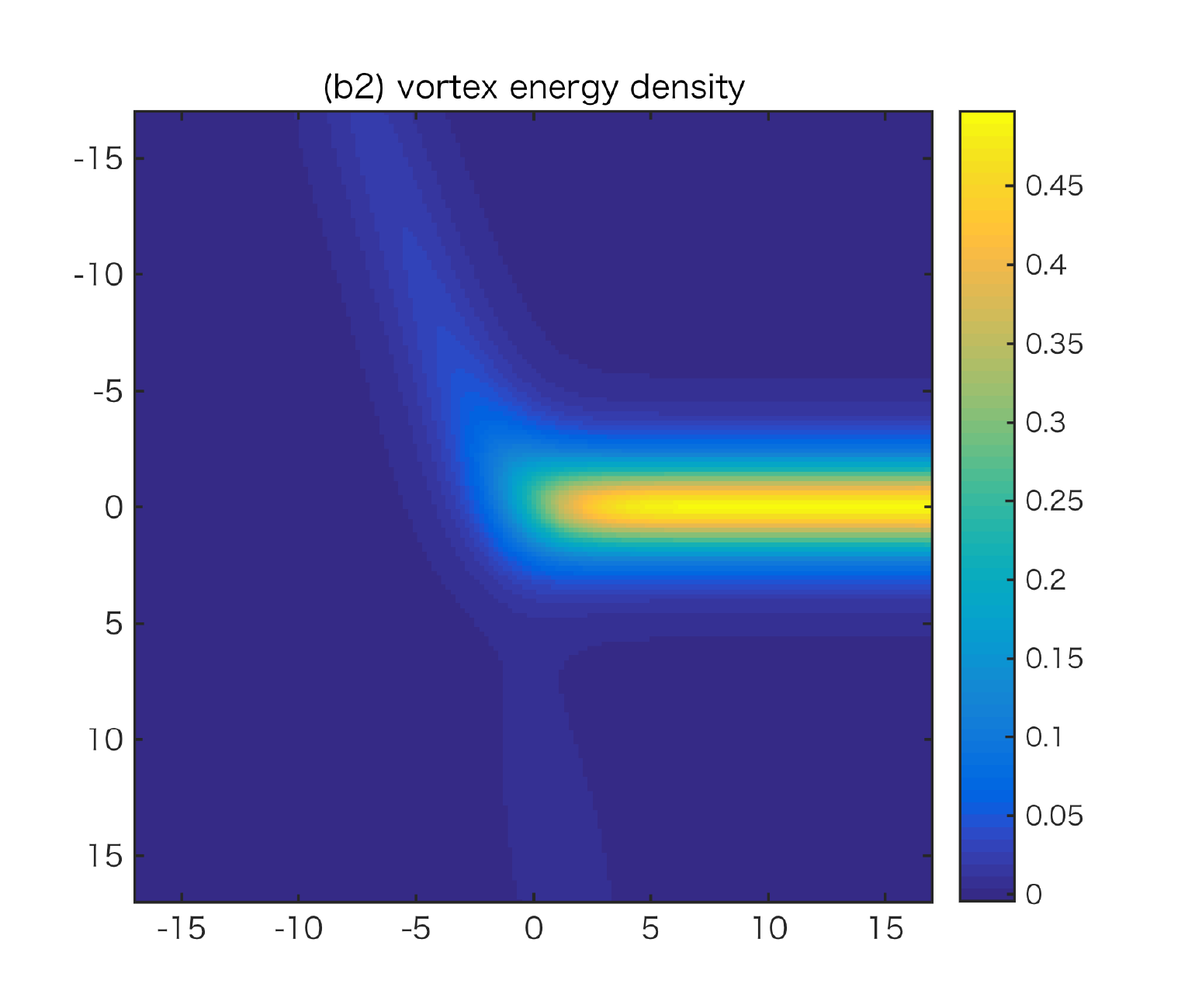}
  \end{center}
 \end{minipage}\\
  \begin{minipage}{0.5\hsize}
  \begin{center}
   \includegraphics[width=7cm]{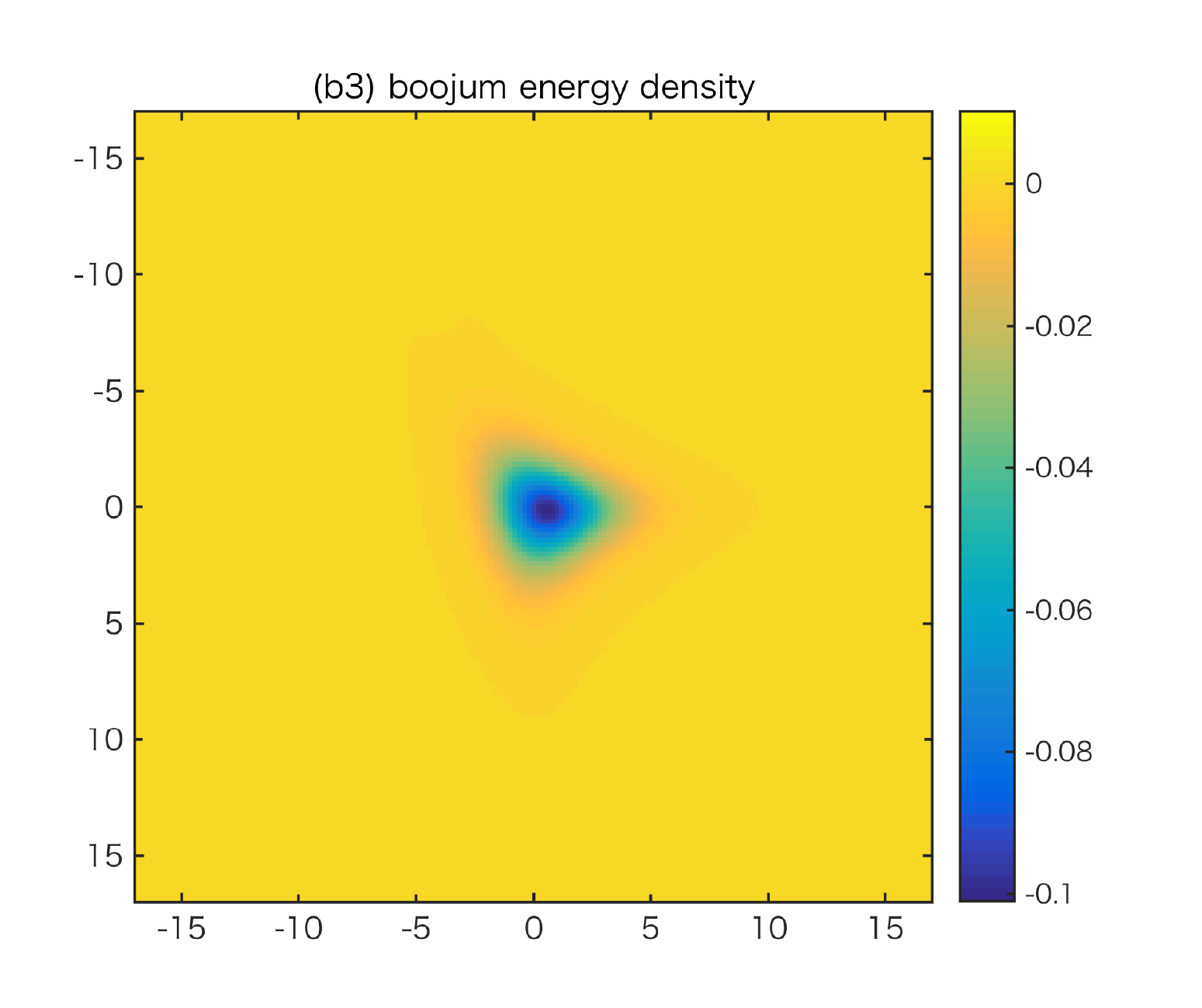}
  \end{center}
 \end{minipage}
 \begin{minipage}{0.5\hsize}
  \begin{center}
   \includegraphics[width=7cm]{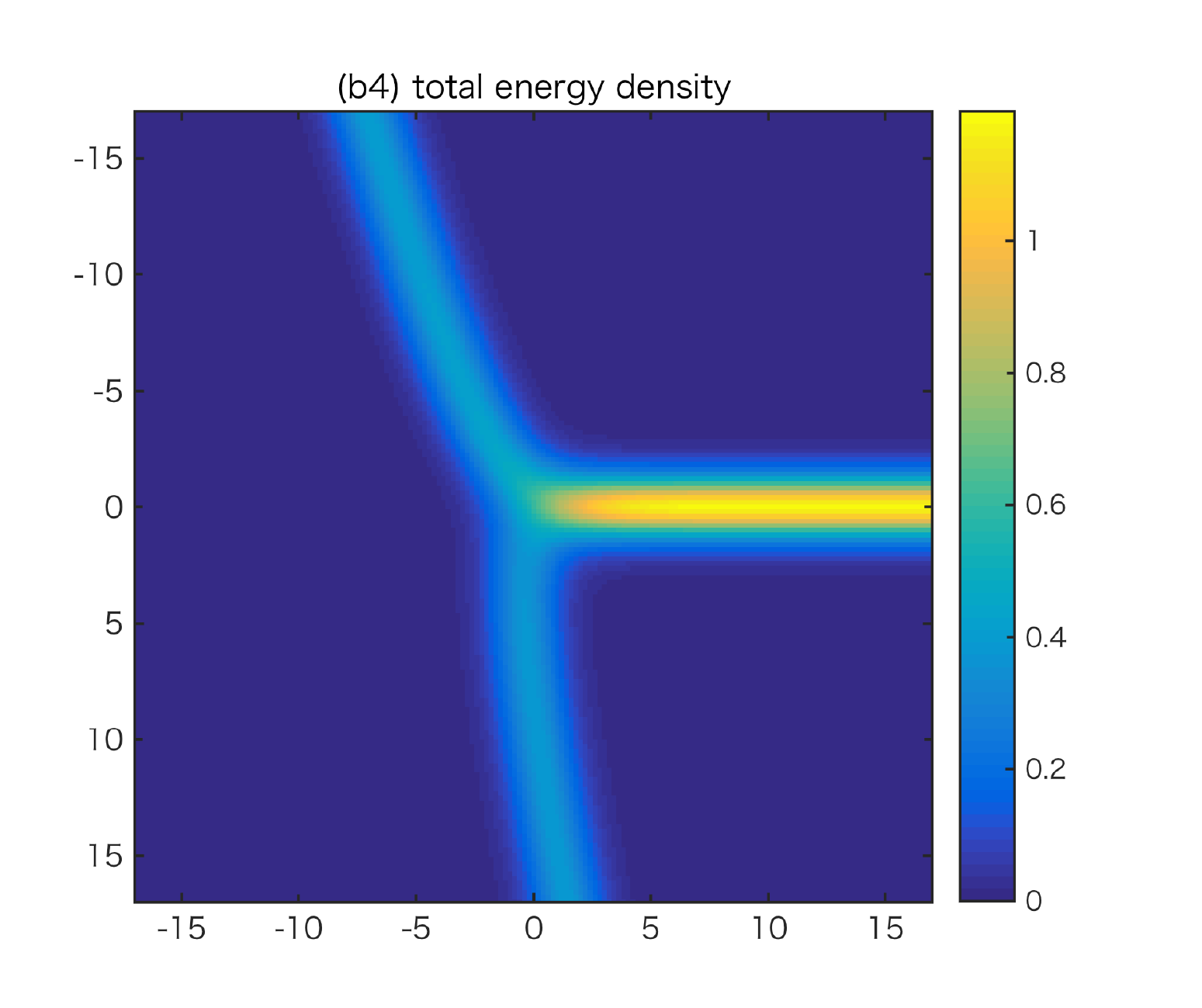}
  \end{center}
 \end{minipage}
\caption{The plots show the energy density isosurfaces of one vortex ending on one slant wall (a1, a2), where the blue and the red curves show magnetic fluxes, the wall energy density (b1), the vortex energy density (b2), the boojum energy density (b3) and the total energy density (b4).}
\label{fig:slant_1vor}
\end{figure}

Next, we put a single vortex string in the first vacuum $\left<1\right>$. The corresponding
moduli matrix is given by
\begin{eqnarray}
H_0 = \left(z,\ -e^{2\eta z}\right).
\end{eqnarray}
The master equation for this can be expressed as follows
\begin{eqnarray}
\p_k^2 u - 1 + \left(e^{\tilde m x^3-2\eta x^1+\log|z|} + e^{-\tilde mx^3 + 2\eta x^1-\log|z|}\right) e^{-u+2\eta x^1+\log|z|} = 0.
\end{eqnarray} 
An appropriate initial configuration for the gradient flow equation to this is
\begin{eqnarray}
{\mathcal U}(x^k) = u_W\left(x^3 - \frac{2\eta}{\tilde m}x^1 + \frac{1}{2\tilde m}u_S(x^a)\right) + 2\eta x^1 + \frac{u_S(x^a)}{2}.
\label{eq:U0_slant_wall_vor}
\end{eqnarray}
We show a numerical solution for $\tilde m = 1$ and $\eta = 1/4$ in Fig.~\ref{fig:slant_1vor} which
clearly demonstrates the vortex string parallel to the $x^3$ axis ends on the slanting and logarithmically bending domain wall.
The junction point is accompanied with the boojum which is also sheared as shown in the panel (b3) 
of Fig.~\ref{fig:slant_1vor}. Interestingly, the magnetic force lines supplied by the vortex string do not  
spread out in the domain wall but flow toward a direction as forming a stringy flux in $2+1$ dimensions, 
see the panel (a1) and (a2) in Fig.~\ref{fig:slant_1vor}.
This squeezing of the magnetic flux inside the domain wall occurs  
because the magnetic force lines from the vortex string repel with those of the background
magnetic flux on the slanting domain wall.

\begin{figure}[t]
\begin{center}
\includegraphics[width=15cm]{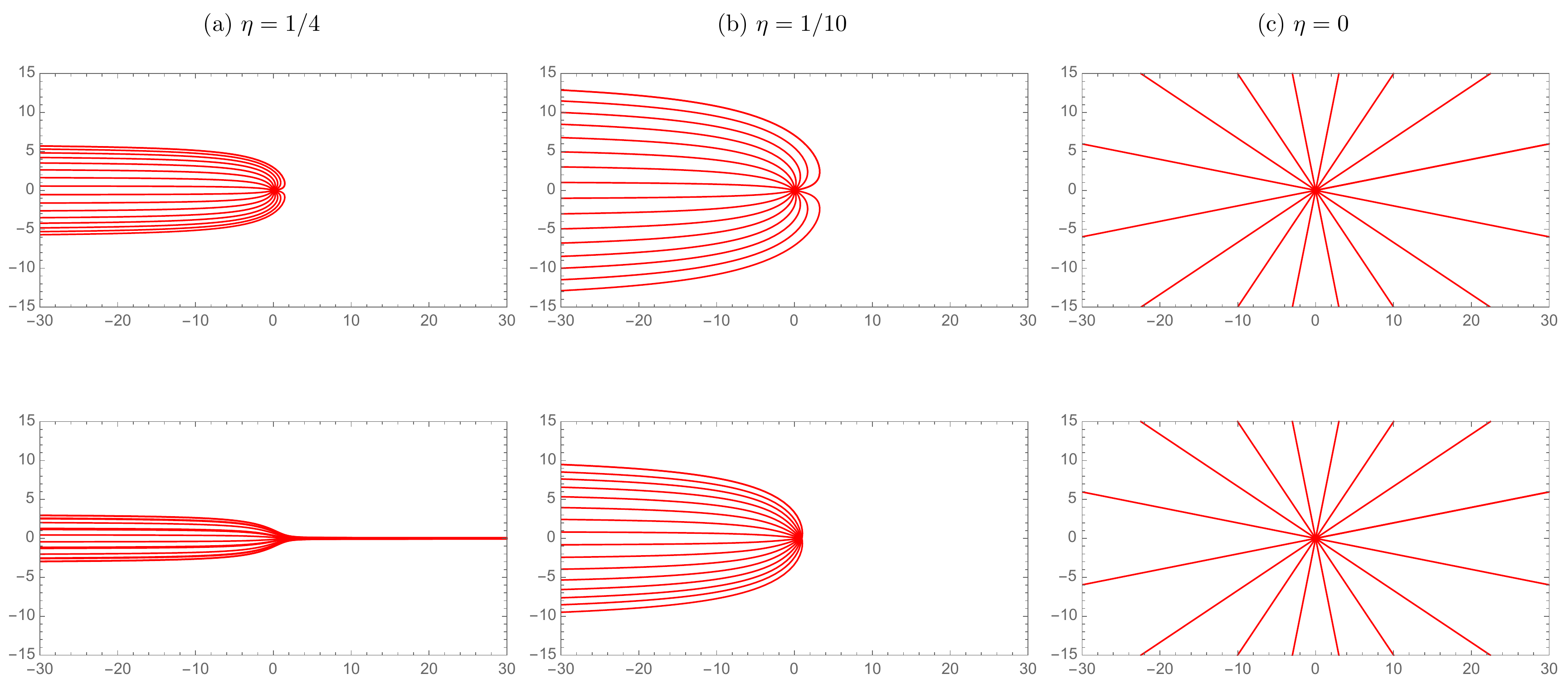}
\caption{The stream lines of the magnetic field $\tilde B_a = - \p_a\tilde \varphi$. We plot the lines which pass
the points on the unit circle surrounding the origin. The figures in the first row are for the point charge 
with $u_S = \log \rho^2$. The figures in the second row are for the finite size source with $u_S$ 
for the finite gauge coupling.}
\label{fig:flux_lines}
\end{center}
\end{figure}

This magnetic scalar potential can be read from Eq.~(\ref{eq:U0_slant_wall_vor}) as 
\begin{eqnarray}
\tilde \varphi 
= -\frac{1}{2\tilde d_W}u_S + \frac{2\eta}{\tilde d_W}x^1
\end{eqnarray}
and again correctly capture these features.
The first term corresponds to the potential generated by the endpoint of the vortex string and the 
second one expresses the potential for the background magnetic field.
We plot stream lines of the magnetic fields $\tilde B_a = - \p_a \tilde \varphi$ for $\eta = 0,1/10,1/4$ in
Fig.~\ref{fig:flux_lines} where we compare two cases: the strong gauge coupling limit with $u_s = \log \rho^2$ 
(the first row) and the finite gauge coupling case (the second row). 
The flux lines emitted from the positive magnetic source are absorbed 
into the negative magnetic charges aligned periodically at $x^1 \to -\infty$, so that they are squeezed.
This situation is quite similar to the squeezing of the magnetic
fluxes by the Higgs mechanism but it is not the case because no further symmetries are broken in the domain wall.

\begin{figure}[h]
 \begin{minipage}{0.5\hsize}
  \begin{center}
   \includegraphics[width=7cm]{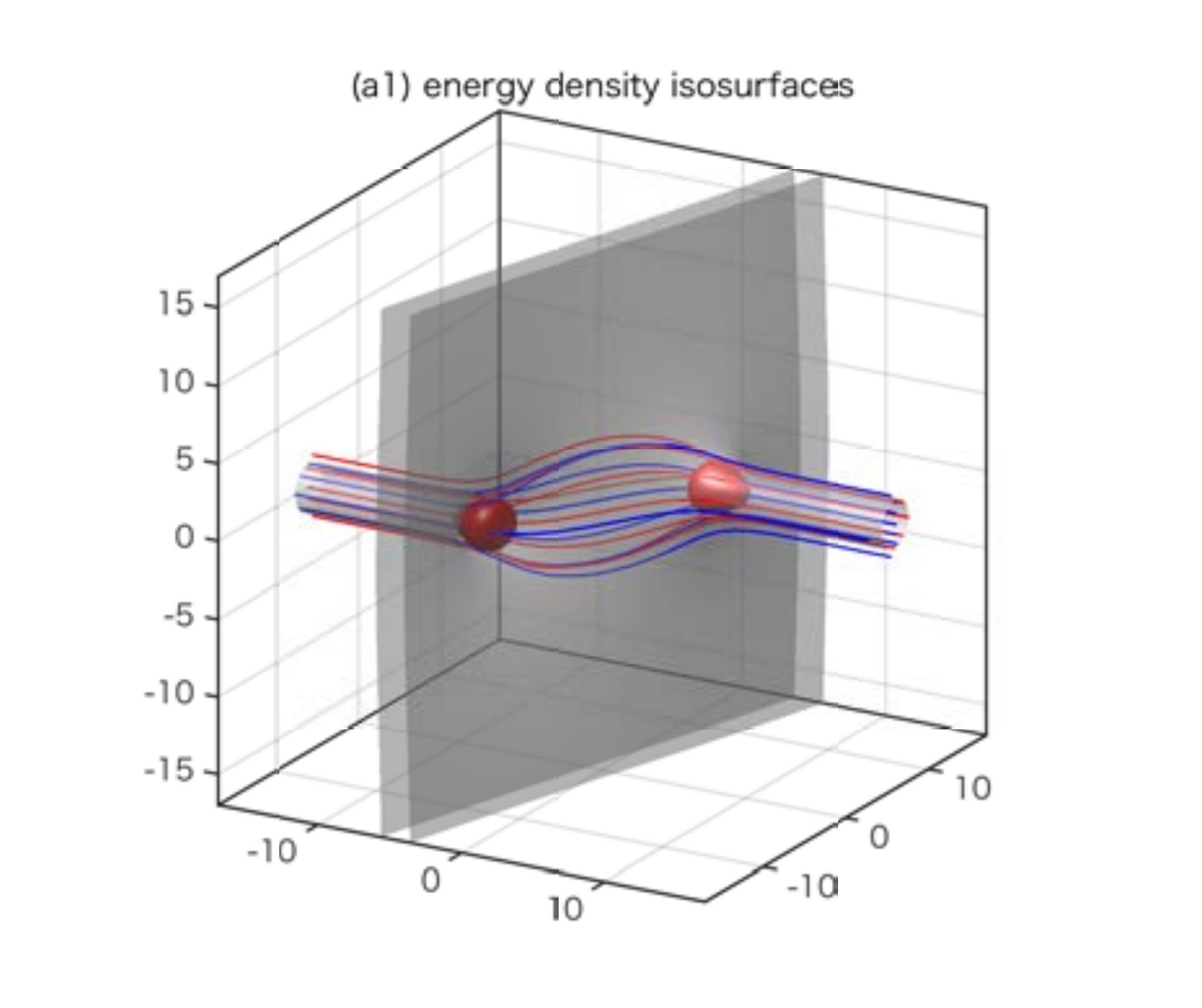}
  \end{center}
 \end{minipage}
  \begin{minipage}{0.5\hsize}
  \begin{center}
   \includegraphics[width=7cm]{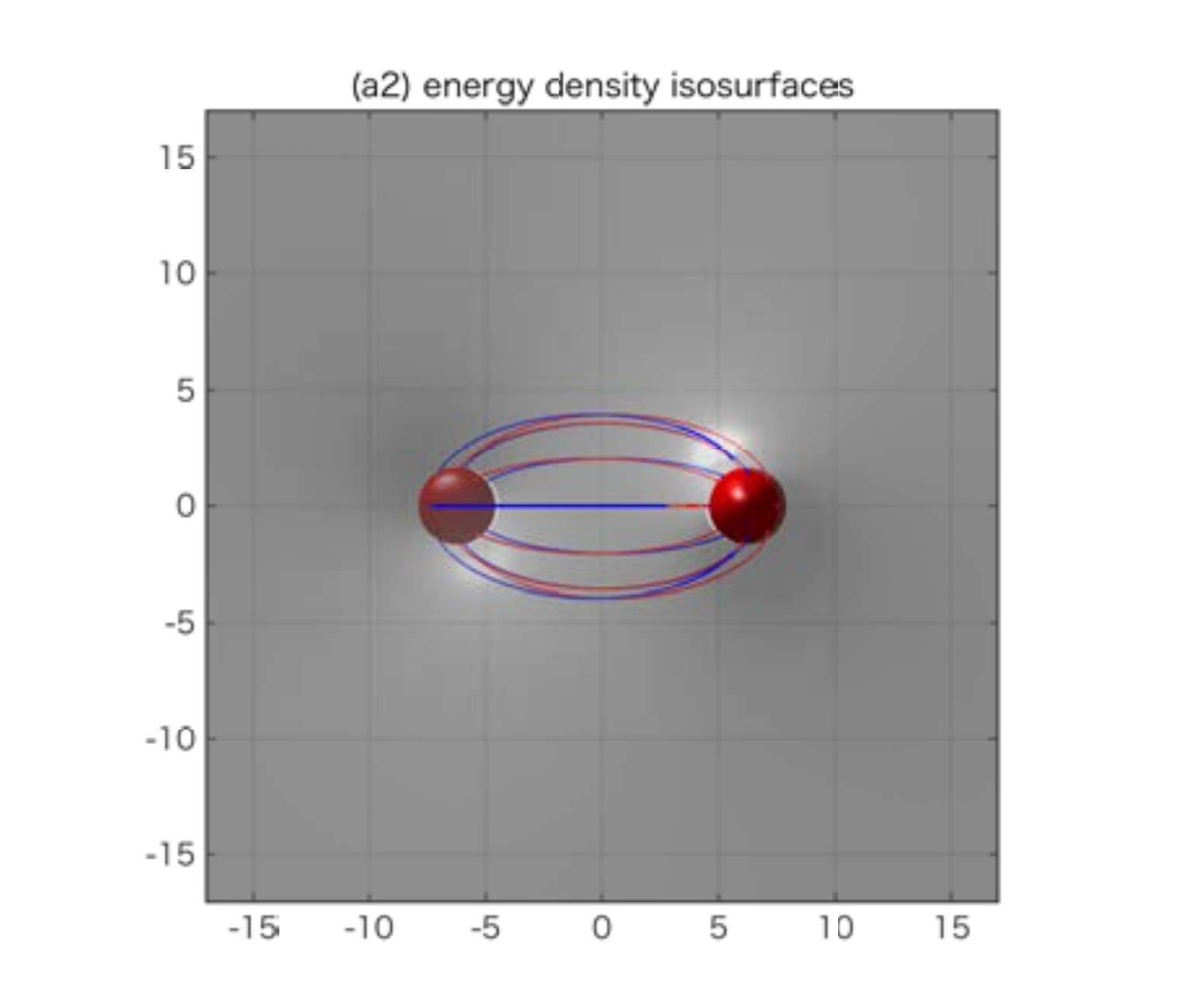}
  \end{center}
 \end{minipage}\\
 \begin{minipage}{0.5\hsize}
  \begin{center}
   \includegraphics[width=7cm]{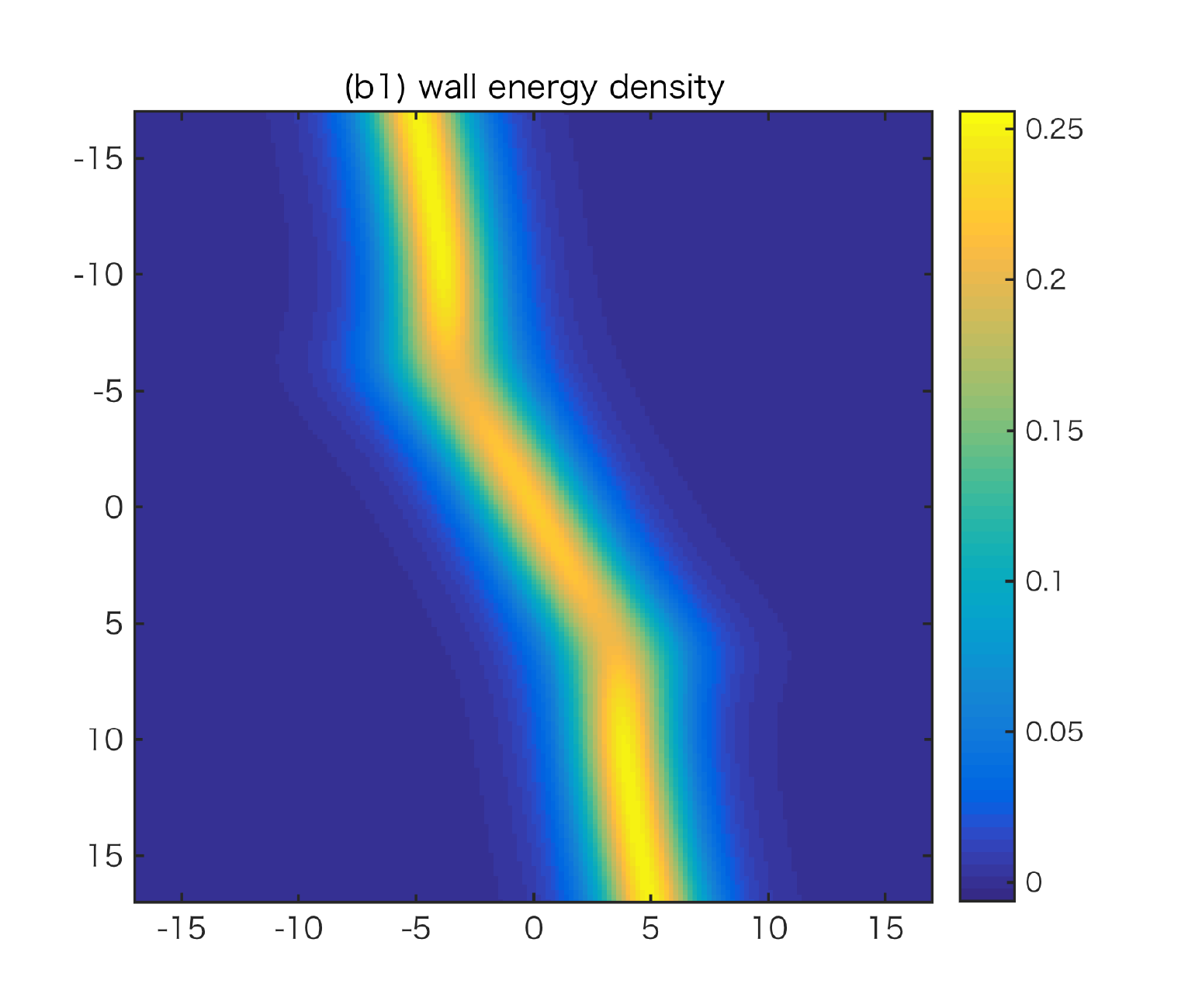}
  \end{center}
 \end{minipage}
 \begin{minipage}{0.5\hsize}
  \begin{center}
   \includegraphics[width=7cm]{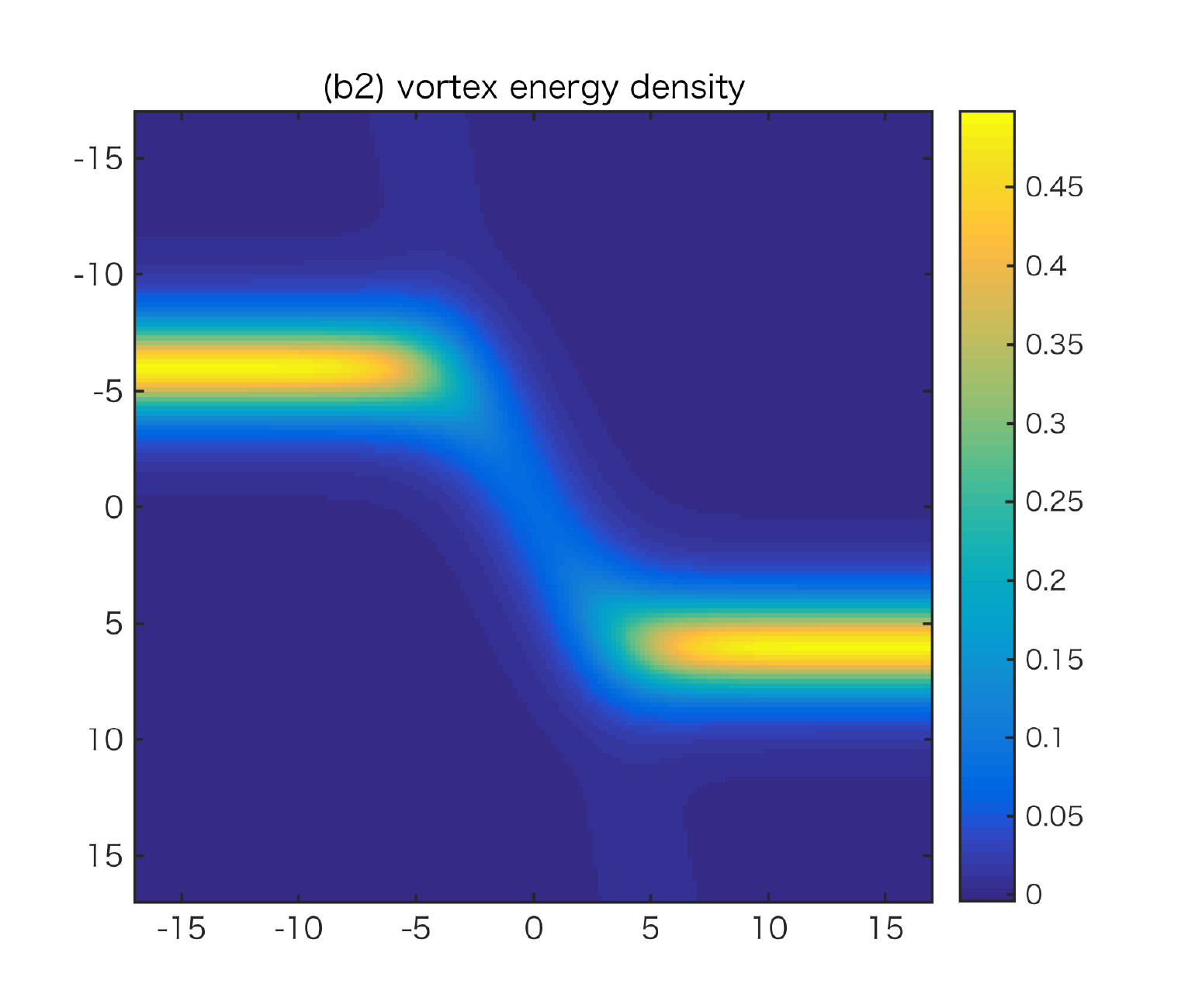}
  \end{center}
 \end{minipage}\\
  \begin{minipage}{0.5\hsize}
  \begin{center}
   \includegraphics[width=7cm]{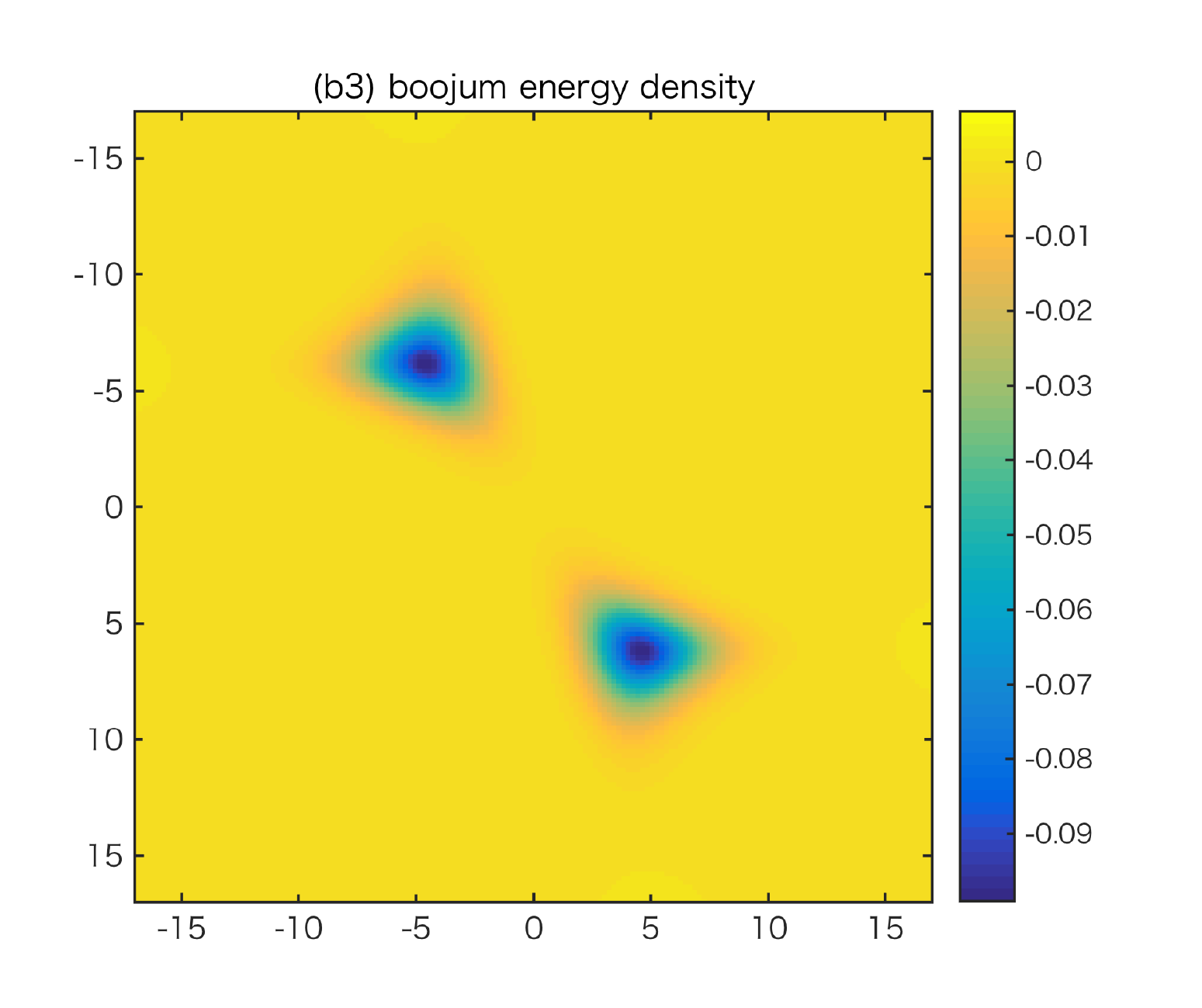}
  \end{center}
 \end{minipage}
 \begin{minipage}{0.5\hsize}
  \begin{center}
   \includegraphics[width=7cm]{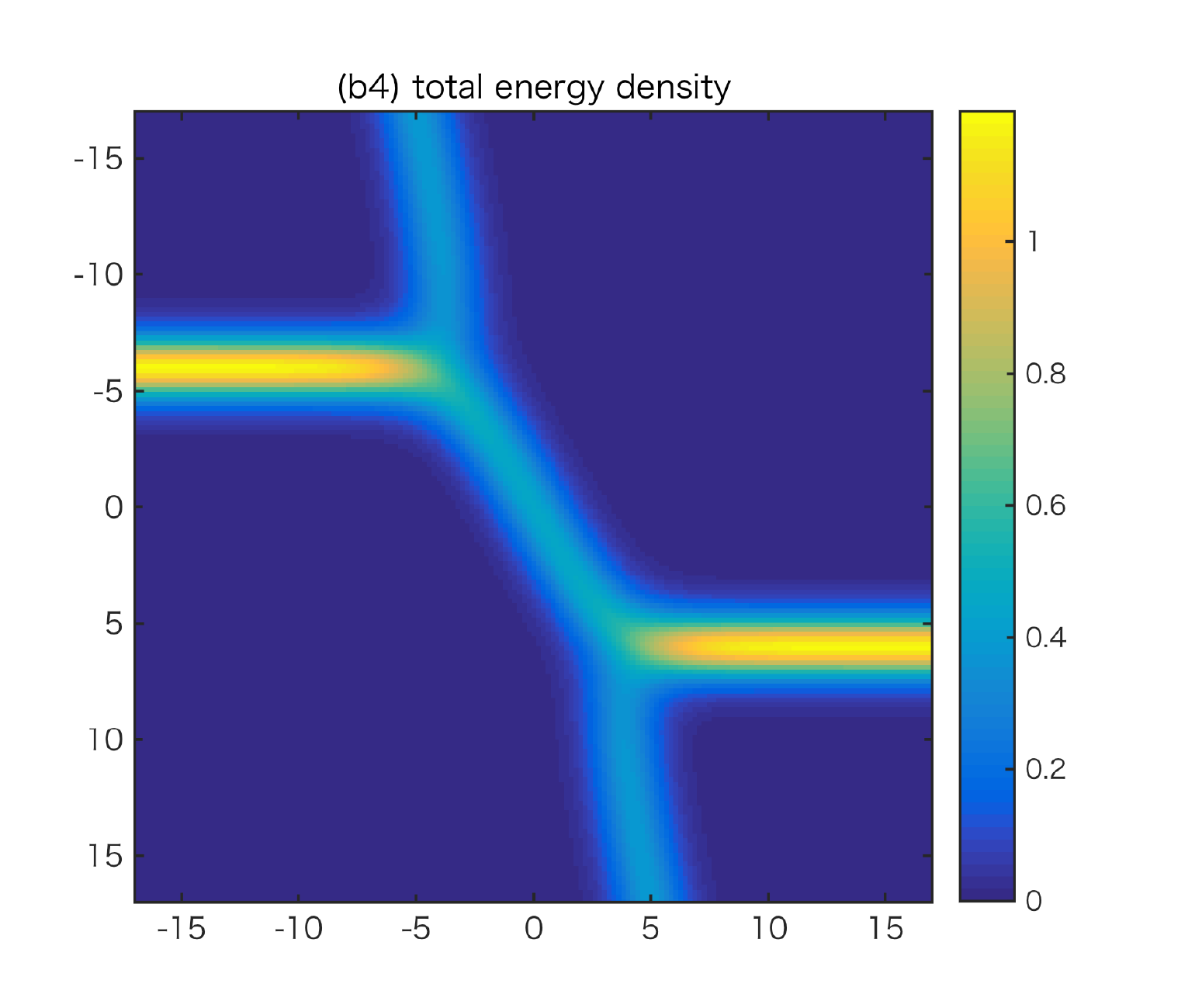}
  \end{center}
 \end{minipage}
\caption{The plots show the energy density isosurface of two vortices ending on one slant wall (a1, a2), where the blue and the red curves show magnetic fluxes, the wall energy density (b1), the vortex energy density (b2), the boojum energy density (b3) and the total energy density (b4). The distance of two vortices is taken to be $Z=6$.}
\label{fig:slant_2side_a}
\end{figure}

\begin{figure}[h]
 \begin{minipage}{0.5\hsize}
  \begin{center}
   \includegraphics[width=7cm]{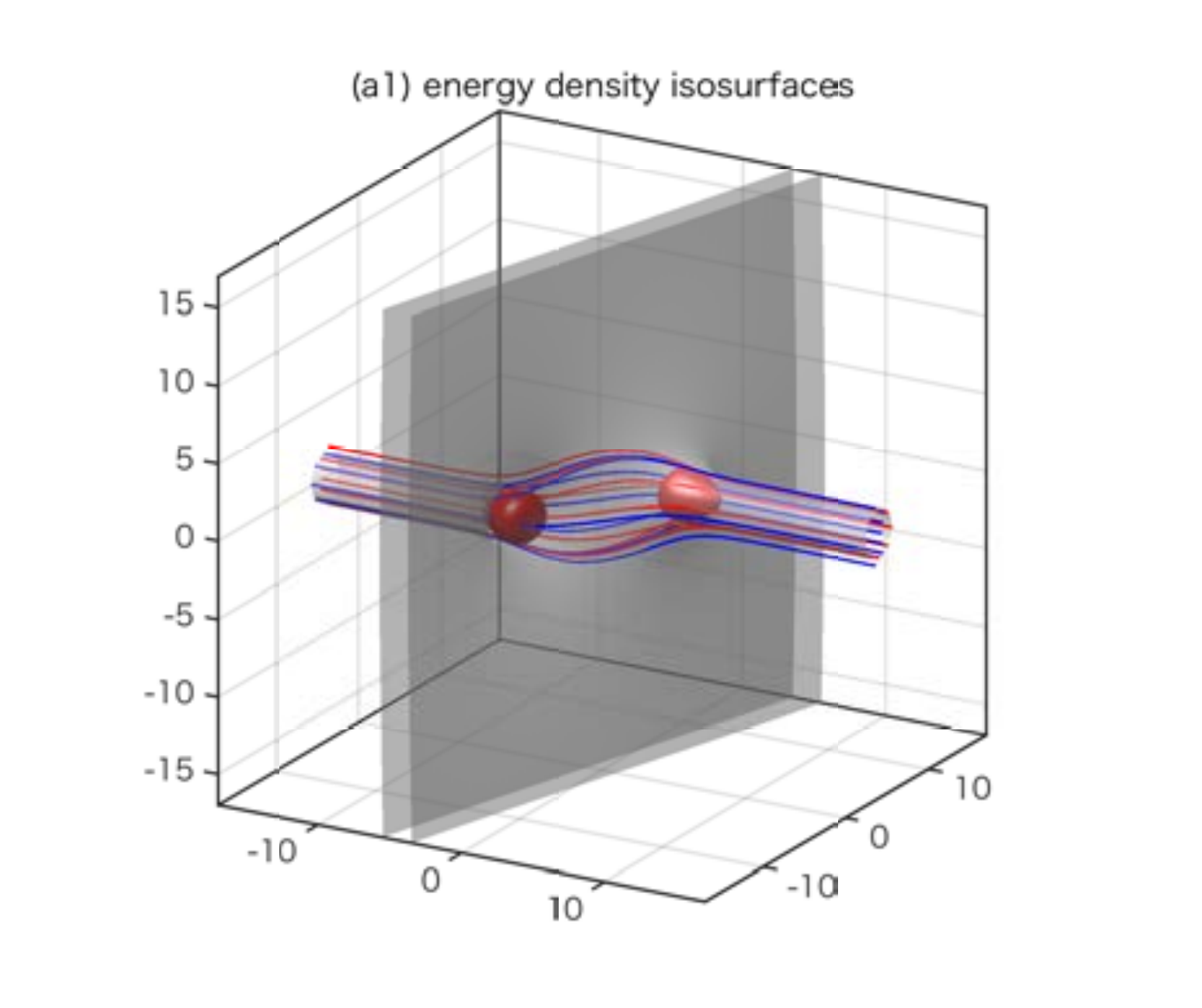}
  \end{center}
 \end{minipage}
  \begin{minipage}{0.5\hsize}
  \begin{center}
   \includegraphics[width=7cm]{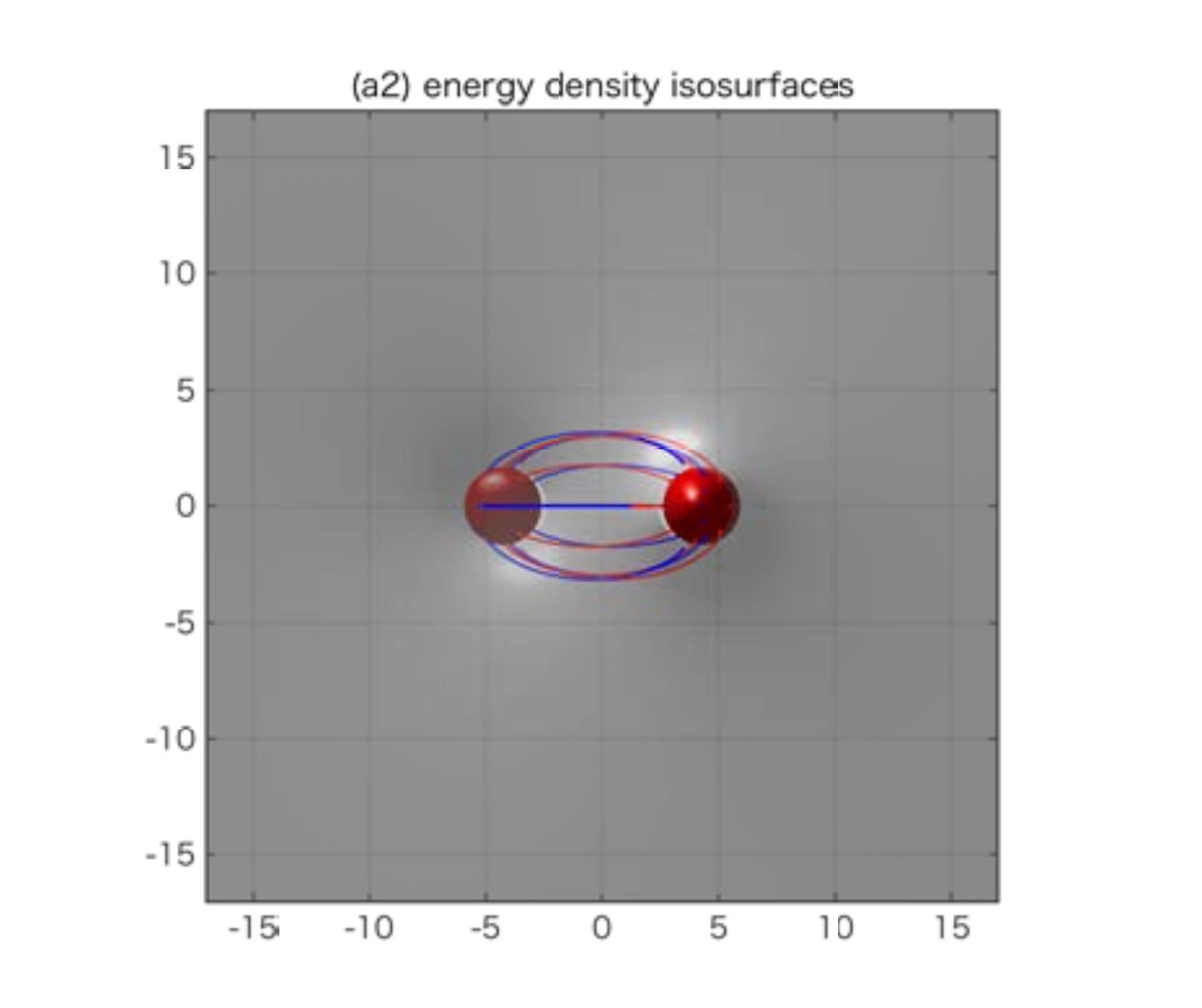}
  \end{center}
 \end{minipage}\\
 \begin{minipage}{0.5\hsize}
  \begin{center}
   \includegraphics[width=7cm]{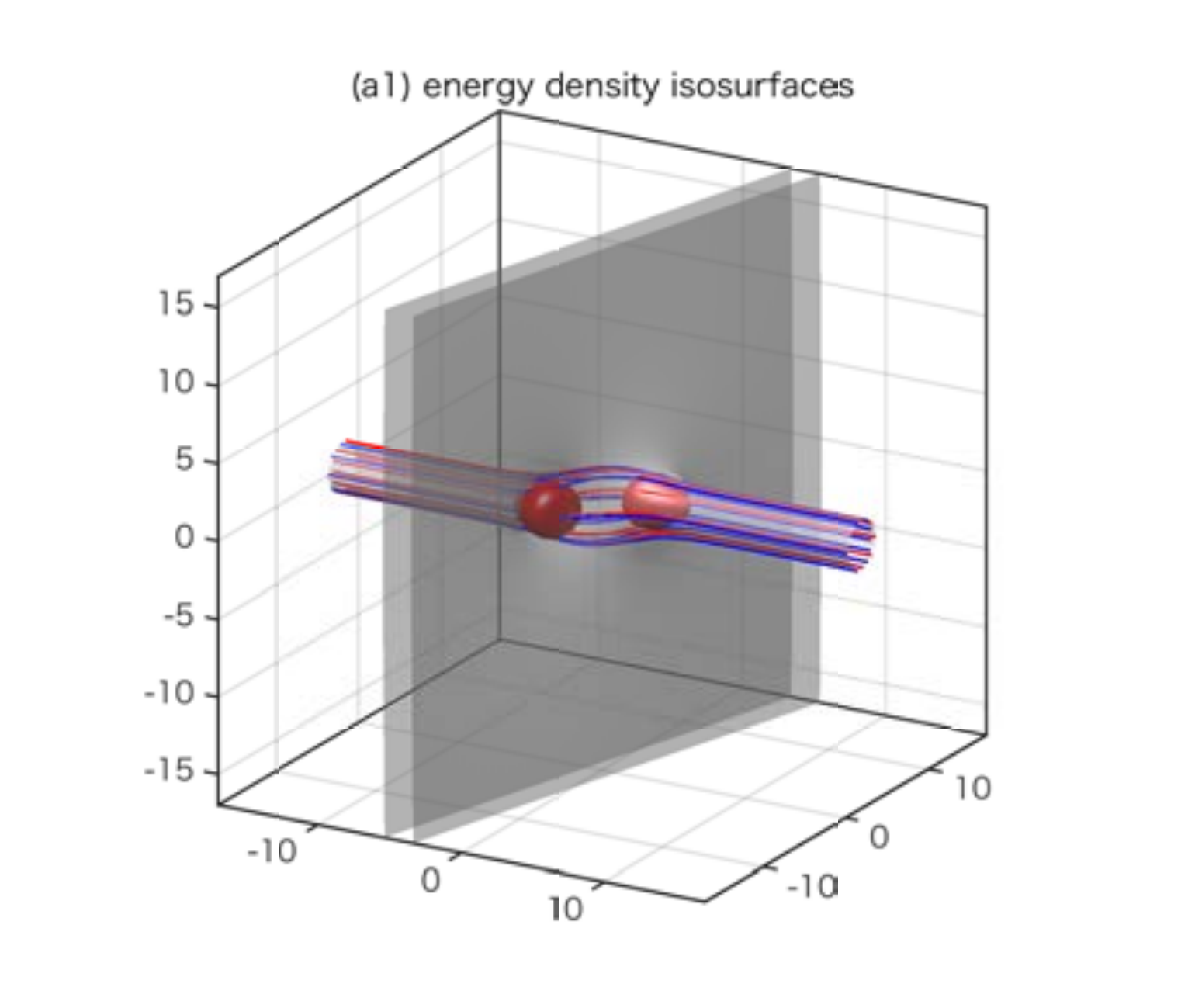}
  \end{center}
 \end{minipage}
 \begin{minipage}{0.5\hsize}
  \begin{center}
   \includegraphics[width=7cm]{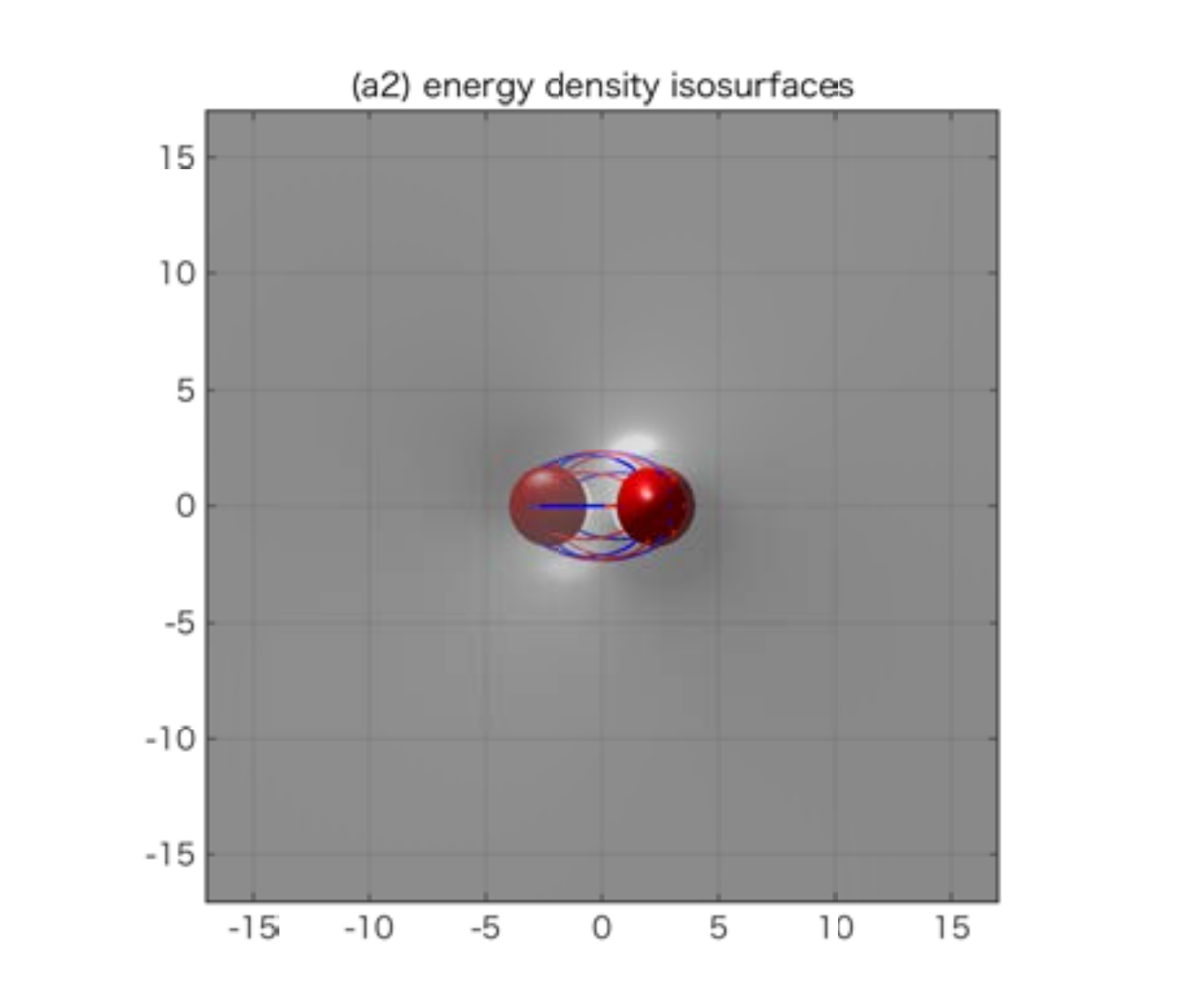}
  \end{center}
 \end{minipage}\\
  \begin{minipage}{0.5\hsize}
  \begin{center}
   \includegraphics[width=7cm]{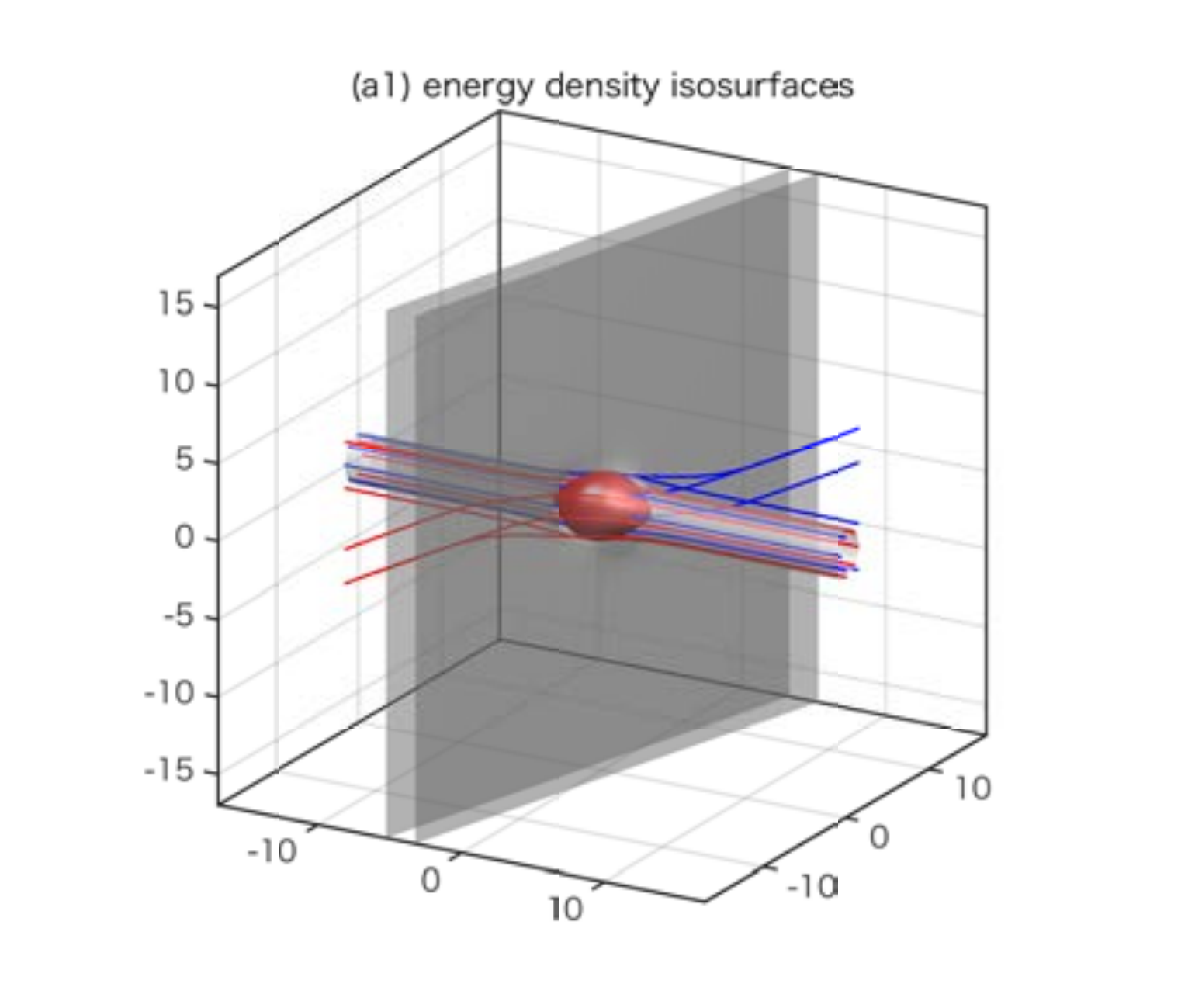}
  \end{center}
 \end{minipage}
 \begin{minipage}{0.5\hsize}
  \begin{center}
   \includegraphics[width=7cm]{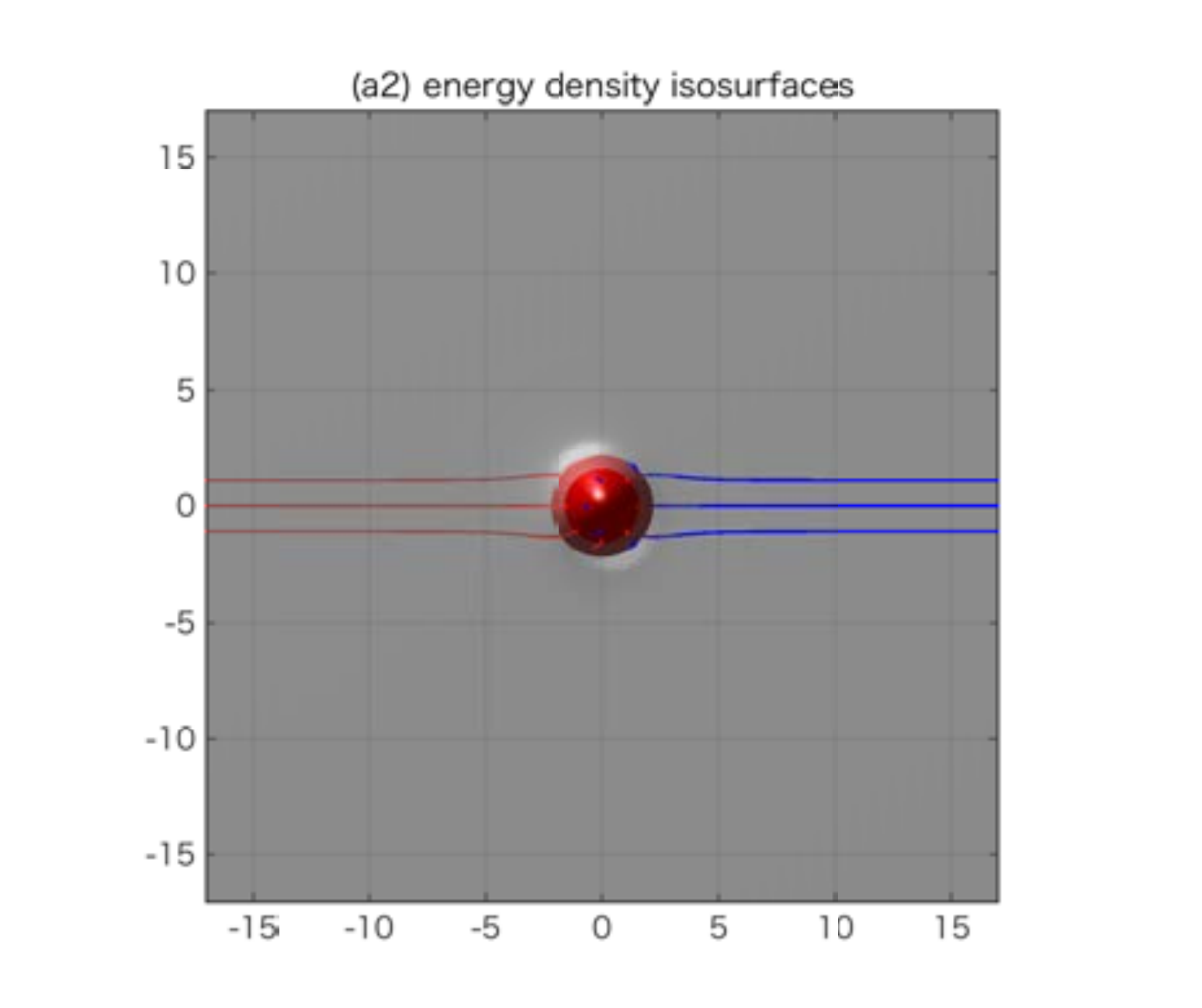}
  \end{center}
 \end{minipage}
\caption{The plots show the energy density isosurfaces of two vortices ending on one slant wall from two sides. The distance between two vortices is taken to be $Z=4, 2, 0$ from top to bottom.}
\label{fig:slant_2side_b}
\end{figure}

Finally, we put another vortex string from the other side of the domain wall. The moduli matrix is
\begin{eqnarray}
H_0 = \left((z-Z) e^{\eta z},\ -(z+Z) e^{-\eta z}\right).
\end{eqnarray}
The vortex string on the positive (negative) $x^3$ side is at $z =Z$ ($z=-Z$), and the domain wall
is asymptotically flat but slanting as 
\begin{eqnarray}
x^3 = - \frac{1}{2\tilde m}u_S\big|_{z=+Z}  + \frac{1}{2\tilde m}u_S\big|_{z=-Z} + \frac{2\eta}{\tilde m}x^1
\to \frac{2\eta}{\tilde m}x^1,\quad (\rho \to \infty).
\end{eqnarray}
We show several numerical solutions in Figs.~\ref{fig:slant_2side_a} and \ref{fig:slant_2side_b}
for $\tilde m =1$ and $\eta = 1/4$. 
We set $Z = 6$ in Fig.~\ref{fig:slant_2side_a} and $Z=4,2,0$ in Fig.~\ref{fig:slant_2side_b}.
A remarkable difference between non-slant and slant configurations can be found in the distribution of the magnetic
force lines inside the domain wall. 
The flux lines are quite similar to those around an ordinary magnetic dipole in the non-slanting domain wall.
On the other hand, they are squeezed in the slanting domain wall, so if we arrange the vortex strings in such
a way that the line segment connecting two endpoints is exactly parallel to the steepest direction of the slanting 
domain wall (the injecting vortex string is on the upper side and the ejecting one is on the lower side), the flux lines
are as if confined, see Fig.~\ref{fig:slant_2side_b}.

Now we can naturally generalize the configuration to have any slanting angle and any number of vortex strings
from both sides. The magnetic scalar potential is the most useful tool to describe it by
\begin{eqnarray}
\tilde d_W \tilde \varphi = - \sum_{k_1=1}^{n_1} \frac{1}{2} u_S\big|_{z=Z_{k_1}}
+ \sum_{k_2=1}^{n_2} \frac{1}{2} u_S\big|_{z=Z_{k_2}} + \tilde B_a^{\rm (bg)}x^a,
\end{eqnarray}
where $u_S\big|_{z=Z_k}$ stands for the solution of vortex string at $z=Z_k$,
and $\tilde B_a^{\rm (bg)}$ is the background magnetic field.

\begin{figure}[t]
\begin{center}
\includegraphics[width=15cm]{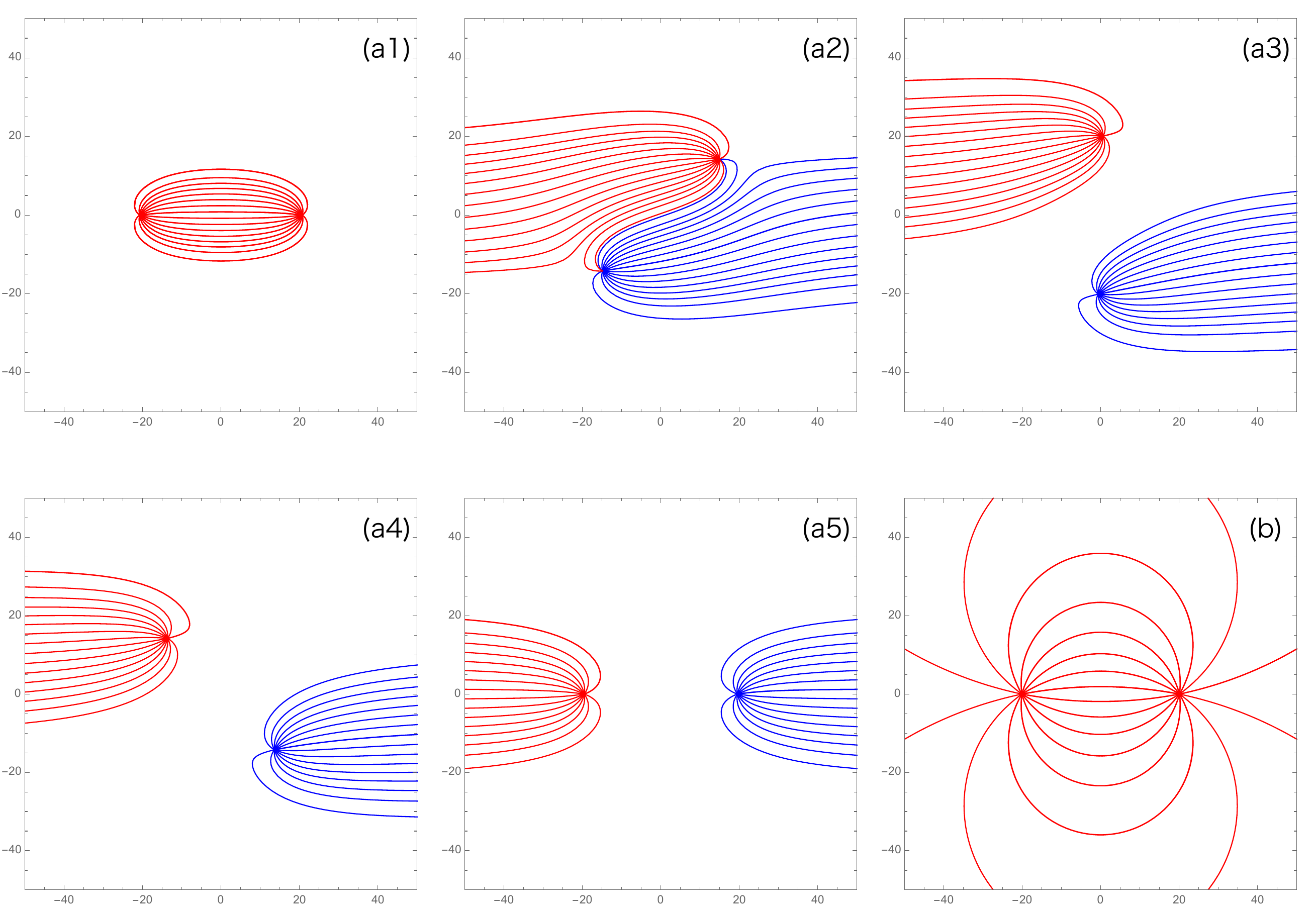}
\caption{The stream lines of the magnetic field $\tilde B_a = - \p_a\tilde \varphi$. The red lines are
fluxes from the positive charge and the blue ones are those going into the negative charge.
The panels (a1) -- (a5) are with the background magnetic field while the panel (b) is without it.}
\label{fig:flux_lines_2}
\end{center}
\end{figure}

The magnetic flux lines for $(n_1,n_2) = (1,1)$ are shown in Fig.~\ref{fig:flux_lines_2}. We put the vortex string
at $Z_{n_1=1} = - Z_{n_2=1} = 20 e^{i\theta}$ with $\theta = 0,\frac{\pi}{4},\frac{\pi}{2},\frac{3\pi}{4},\pi$
for $\tilde B_a^{\rm (bg)}  = 2\eta \delta_{1a}$ with $\eta = \frac{1}{15}$ and $\tilde m = 1$ ($\tilde d_W = 2$).
As shown in (a1) of Fig.~\ref{fig:flux_lines_2}, the magnetic sources are confined only when $\theta = 0$, where 
the magnetic flux lines from the positive source go into
the negative magnetic source. When we rotate the sources, a part of flux lines run toward the boundaries,
see (a2) -- (a5) of Fig.~\ref{fig:flux_lines_2}. When we turn off the background magnetic field, we have the
magnetic dipole regardless of the rotating angle as (b) of Fig.~\ref{fig:flux_lines_2}.

\clearpage

\clearpage

\clearpage


\section{Dyonic extension}\label{sec:dyonic}

\subsection{Basic formulae}

In this section we will study a dyonic extension of the purely magnetic 
1/4 BPS equations (\ref{eq:bpsws1})--(\ref{eq:bpsws4}) \cite{Lee:2005sv,Eto:2005sw}. 
A perfect square of the energy density including time dependence is given by
\begin{align}
{\mathcal E} & = \frac{1}{2g^2}{\Big \{}
                             \left(\xi F_{12}-\eta\cos\alpha\, \partial_3\sigma-g^2(HH^\dagger-v^2)\right)^2
                             +(F_{0k}-\sin\alpha\, \partial_k\sigma)^2 + \left(\partial_0\sigma\right)^2\nonumber \\
                    & ~~~~~~~ +(\xi F_{23}-\eta \cos\alpha\, \partial_1\, \sigma)^2+(\xi F_{31}-\eta \cos\alpha\, \partial_2\sigma)^2
                             {\Big \}} \nonumber \\
                    & ~~ + |(D_1+i\xi D_2)H|^2 
                             +|D_0 H+i \sin\alpha (\sigma H-HM)|^2 \nonumber\\
                    & ~~  + |D_3H +\eta \cos\alpha (\sigma H-HM)|^2 \nonumber \\
                    & ~~ +  v^2 \eta \cos\alpha\, \partial_3\sigma-\xi v^2 F_{12}+\xi\eta\cos\alpha\, \partial_i\left(\frac{\sigma}{g^2} \epsilon_{ijk} F_{jk}\right)  \nonumber\\
                    & ~~ +i \sin\alpha(HMD_0H^\dagger-D_0HMH^\dagger)\nonumber \\
                    & ~~ +\partial_k j_k +  \sin\alpha\, \partial_k \left(\frac{\sigma}{g^2} F_{0k}\right) \nonumber\\
                    & ~~ - \sin\alpha\left\{\frac{1}{g^2}\partial_k F_{0k} + i \left(HD_0H^\dagger - D_0HH^\dagger\right)\right\}\,,
                    \label{d-ec}     
\end{align}
where we restrict $\alpha$ to satisfy $\alpha \in (-\pi/2,\pi/2)$ because $\cos\alpha$ always appears
accompanied with $\eta = \pm 1$. The non-topological currents
$j_{a=1,2}$ are the same as in (\ref{ja}) while $j_{k=3}$ is given by 
\begin{eqnarray}
j_3 = -\eta \cos \alpha \left(\sigma H - HM\right)H^\dagger.
\end{eqnarray}
Vanishing of the squared terms  leads to the dyonic extension to 1/4 BPS equations
\begin{gather}
\label{eq:dyonicquater1} D_0 H +i \sin\alpha \bigl(\sigma H-H M\bigr) = 0\,, \\
\label{eq:dyonicquater1a} D_3 H+\eta\cos\alpha \bigl(\sigma H-HM\bigr) = 0\,, \\ 
\label{eq:dyonicquarter15} (D_1+i \xi D_2) H = 0\,, \\
\label{eq:dyonicquater2}  \eta\cos\alpha\, \partial_1\sigma = \xi F_{23}\,,
\hspace{2mm} \eta \cos\alpha\, \partial_2 \sigma = \xi F_{31}\,, \\
\label{eq:dyonicquarter25} \sin\alpha\, \partial_k\sigma = F_{0k}\,, \\
\label{eq:dyonicquarter3} \xi F_{12}-\eta\cos\alpha\, \partial_3 \sigma -g^2\bigl(\abs{H}^2-v^2\bigr) = 0\,,\\
\label{eq:dyonicquarter4} \partial_0 \sigma = 0.
\end{gather}
Also, one has to include the Gauss's law
\begin{equation}
\label{eq:lastgauss} \frac{1}{g^2}\partial_{k}F_{0k} +i \Bigl(H D_0H^{\dagger}-D_0 H H^{\dagger}\Bigr) = 0\,.
\end{equation}
The parameters $\eta^2 = \xi^2 = 1$ 
labels (anti-)vortices $\xi = (-1)1$ and (anti-)walls $\eta = (-1)1$.
In the strong gauge coupling limit, our Abelian-Higgs model reduces to the massive nonlinear 
sigma model whose target space is $\mathbb{C}P^{N_F-1}$, and the above dyonic extension
reduces to the Q-kink lump configuration without the boojums, first studied in Ref.~\cite{Portugues}. 

When the BPS equations (\ref{eq:dyonicquater1})--(\ref{eq:dyonicquarter4}) and the
Gauss's law are satisfied, 
the total energy density saturates the Bogomol'nyi bound
\begin{eqnarray}
{\cal E} \ge \left({\cal T}_W  \cos \alpha + {\cal Q}_W \sin \alpha\right) + {\cal T}_S    
+ \left({\cal T}_B  \cos\alpha + {\cal Q}_B  \sin\alpha \right)+ \partial_k j_k,
\end{eqnarray}
where ${\cal T}_{W,S,B}$ is defined in Eq.~(\ref{TB}), and the Noether charge density and the
electric Boojum charge density are defined by
\begin{eqnarray}
{\cal Q}_W &=&  i  \left(H M D_0 H^{\dagger} - D_0 H M H^{\dagger}\right),\\
{\cal Q}_B &=& \partial_k\left(\frac{\sigma}{g^2}F_{0k}\right).
\label{eq:QB}
\end{eqnarray}

The set of the BPS equations \refer{eq:dyonicquater1}--\refer{eq:dyonicquarter25} 
are solved via the moduli matrix formalism
\begin{gather}
H = v e^{-\frac{u}{2}}H_0(z) e^{M\left(x^3 \eta \cos\alpha + i x^0 \sin\alpha \right)}\,, 
\label{eq:dy_mm_1}\\
a_1+i\xi a_2=-i\partial_{\bar{z}}u\,, \\
\sigma \eta \cos\alpha + i a_3 = \frac{1}{2}\partial_3 u\,, \label{eq:dy_mm_3}\\ 
\sigma \sin\alpha +a_0 = -\frac{i}{2}\partial_0 u\,.\label{eq:dy_mm_4}
\end{gather}
We demand $u$ to be real by fixing the gauge freedom. Thus, the equation 
(\ref{eq:dy_mm_3}) gives us
\begin{eqnarray}
a_3 = 0,\qquad \sigma = \frac{\eta}{2\cos\alpha} \p_3u.
\end{eqnarray}
From Eq.~(\ref{eq:dyonicquarter4}) $\sigma$ is independent of $t$. Then,
Eq.~(\ref{eq:dy_mm_4}) gives 
\begin{eqnarray}
a_0 = - \sigma\sin\alpha =  - \frac{\eta}{2}\tan\alpha\, \p_3u.
\end{eqnarray} 
Note that this also solves the Gauss's law \refer{eq:lastgauss}. 
Now, we can express the electric and magnetic field in terms of the single real function $u$ as
\begin{eqnarray}
E_k =  \frac{\eta \tan\alpha}{2} \p_k \p_3 u,\quad
(B_1,B_2,B_3) = \frac{\xi}{2}\left(\p_3\p_1 u,\ \p_2\p_3 u,\ 
 - (\p_1^2 + \p_2^2) u\right).
\label{eq:dyonic_EB}
\end{eqnarray}
Similarly, the topological charge densities are also expressed as
\begin{eqnarray}
{\mathcal T}_W &=& \eta v^2 \p_3 \sigma = \frac{v^2}{2\cos\alpha}\p_3^2 u, \label{eq:tension-q}\\
{\mathcal T}_S &=& - \xi v^2 F_{12} = \frac{v^2}{2}\left(\p_1^2 + \p_2^2\right)u, \label{eq:tension-s}\\ 
{\mathcal T}_B &=& \frac{\eta\, \xi}{g^2} \epsilon_{klm}\partial_k (F_{lm}\sigma) \nonumber\\
&=& \frac{1}{2g^2\cos\alpha}\left\{\left(\p_1\p_3 u\right)^2 + \left(\p_2\p_3u\right)^2 - \left(\p_1^2+\p_2^2\right)\! u\  \p_3^2 u\right\}.\label{eq:tension-b}
\end{eqnarray}
Finally, we are left with the equation \refer{eq:dyonicquarter3} which turns into the master equation 
\begin{eqnarray}
\frac{1}{2g^2v^2}\partial_k^2 u = 1-\Omega_0 e^{-u},\quad
\Omega_0 = H_0
 e^{2\eta\cos\alpha\, M x^3 } H_0^{\dagger}\,. 
\label{eq:master_dyonic}
\end{eqnarray}
Comparing this with the master equation (\ref{eq:quatermaster}) for the purely magnetic case, the
only difference is the replacement $M$ by $M\cos\alpha$. 

The tension of the domain wall is the same as in the purely magnetic case
\begin{eqnarray}
T_ W = \int dx^3\ {\cal T}_W = \eta v^2 \big[\sigma\big]^{x^3=\infty}_{x^3=-\infty} 
= v^2 \eta \Delta m
= v^2 |\Delta m|.\label{eq:dyonic-dw}
\end{eqnarray}
The similar holds for the Noether charge.
Combining Eqs.~(\ref{eq:dyonicquater1}) and (\ref{eq:dyonicquater1a}), we find
\begin{eqnarray}
D_0 H = i \eta \tan\alpha\, D_3 H.
\end{eqnarray}
By using this, we have the following
expression for ${\cal Q}_W$,
\begin{eqnarray}
{\cal Q}_W = \eta \tan\alpha\, \p_3\left(HMH^\dagger\right). \label{eq:q-charge-domain-wall}
\end{eqnarray}
Thus, the Noether charge density upon the integration over $x^3$ 
gives a  constant.
\begin{eqnarray}
Q_W = \int dx^3\ {\cal Q}_W =  \eta \tan\alpha\, \bigg[HMH^\dagger\bigg]^{x^3=\infty}_{x^3=-\infty}
= v^2  \Delta m  \eta \tan\alpha \label{eq:Noether-c}.
\end{eqnarray}
Hence, the Noether charge per unit area is proportional to the domain wall tension
\begin{eqnarray}
\frac{Q_W}{T_W}  = \tan\alpha. \label{qc}
\end{eqnarray}
Therefore, the volume integral of ${\cal Q}_W$ diverges as the domain wall mass which is proportional to
$A = \int dx^1dx^2$.
Now, a part of the BPS mass can be calculated as
\begin{eqnarray}
\int d^3x\ \left({\cal T}_W\cos\alpha + {\cal Q}_W \sin\alpha\right) = \frac{T_W}{\cos\alpha} A = 
\sqrt{T_W^2 + Q_W^2}\, A. \label{dwall}
\end{eqnarray}
Contribution of the vortex string to the total mass is independent of $\cos \alpha$.
Therefore, we have $T_S = 2\pi v^2 |k|$ where $k$ stands for the vortex winding number,
and then the mass of the vortex string is given by
\begin{eqnarray}
\int d^3x\ {\cal T}_S = \int dx^3\ T_S = 2\pi v^2 |k| L. \label{eq:qvs}
\end{eqnarray}
Let us next evaluate $T_B$, the boojum mass,
\begin{eqnarray}
T_B = \frac{\eta \xi}{g^2}\int d^3x\ \p_i \left(\sigma B_i\right). \label{eq:qbje}
\end{eqnarray}
This is easy to do for the case of flat domain walls since
we have the same number $k$ of straight vortex strings at both sides of the domain walls.
The magnetic flux at $x^3 \to \pm \infty$ is given by $\int dx^1dx^2\ \xi B_i = - \delta_{i3}2\pi |k|$.
Therefore, we have
\begin{eqnarray}
T_B = \frac{\eta}{g^2} \times (-2\pi |k|) \times \big[\sigma\big]^{x^3 = +\infty}_{x^3=-\infty}
= - \frac{2\pi}{g^2}|\Delta m| |k|.
\label{eq:TB_dyonic}
\end{eqnarray}
This is independent of $\alpha$ as $T_S$. For configurations including bent domain walls, 
we should repeat the same computation as we have done in \cite{Boojum1}.
But it is clear even for such cases that $T_B$ is independent of $\alpha$. Hence, the formula
Eq.~(\ref{eq:TB_dyonic}) is valid for any configurations.
Contribution of the boojum to the total mass is then found as
\begin{eqnarray}
\cos\alpha\ T_B = \frac{T_W}{\sqrt{T_W^2 + Q_W^2}}\ T_B.
\end{eqnarray}
Since $T_B$ is negative definite, this makes  the total mass  larger.
Next, we evaluate contribution from ${\cal Q}_B$ given in Eq.~(\ref{eq:QB}). 
Using the BPS equations, it can be written as
\begin{eqnarray}
{\cal Q}_B = \partial_k\left(\frac{\sigma}{g^2}F_{0k}\right) 
= \frac{\sin\alpha}{g^2} \partial_k\left(\sigma\p_k\sigma\right)
= \frac{\sin\alpha}{2g^2} \partial_k^2\sigma^2.
\end{eqnarray}
Since the electric field is
proportional to the derivative of $\sigma$, $E_k \propto \p_k \sigma$, 
it is non-zero only inside the domain wall. 
Therefore, upon integration along $x^3$, ${\cal Q}_B$ vanishes.
The contribution from the non-topological terms $\p_k j_k$ also vanishes upon integration. 
Summing up all the contributions, we conclude that the mass of the dyonic 1/4 BPS configuration is
given by
\begin{eqnarray}
E_{1/4} = \sqrt{T_W^2 + Q_W^2}\, A + T_S L + \frac{T_W}{\sqrt{T_W^2 + Q_W^2}}\,T_B. \label{teq}
\end{eqnarray}

The electric charge density appearing in the Gauss' law (\ref{eq:lastgauss}) 
can be written as
\begin{eqnarray}
{\cal Q}_E = i \left(D_0H H^\dagger - H D_0H^\dagger\right)
= - \eta  \tan\alpha\, \p_3\left(HH^\dagger\right).
\label{eq:electric_charge}
\end{eqnarray}
This is very similar to ${\cal Q}_W$. Since we have $HH^\dagger = v^2$ at any vacua, 
$\int dx^3\ {\cal Q}_E = 0$, so that net electric charge is zero. However, note that 
the electric charge density is non-zero everywhere.

As a final remark, the following relation
\begin{eqnarray}
\vec E \cdot \vec B &=& \frac{\eta\xi}{4}\tan\alpha \left\{
\left(\p_1\p_3 u\right)^2 + \left(\p_2\p_3u\right)^2 - \left(\p_1^2+\p_2^2\right)\! u\  \p_3^2 u
\right\} \nonumber\\
&=& \frac{g^2}{2} \xi\eta \sin\alpha\, {\cal T}_B,
\label{eq:EdotB}
\end{eqnarray}
implies that $E_k$ is  perpendicular to $B_k$ far from the boojums, while
in their vicinity, they are not.

\subsection{The dyonic domain wall as an electric capacitor}

The Q-extension of the domain wall was first found in nonlinear sigma models 
in Ref.~\cite{Abraham:1992vb,Abraham:1992qv}, and lots of works have followed them.
The Q-extended domain walls in gauge theories are sometimes called 
the dyonic domain walls \cite{Lee:2005sv,Eto:2005sw,Tong:2005un}.
They are characterized by the topological and the Noether charges, so it is
suitable to call them dyonic solitons.
In the previous works \cite{Lee:2005sv,Eto:2005sw,Tong:2005un},
the dyonic domain wall was not a main focus.
Some qualitative properties
such as derivation of the BPS equations, topological charges, and the BPS mass formula 
have been given. 
To the best of our knowledge, very little has been done for solving the BPS equations, especially in
the weak gauge coupling region.
Furthermore, while the Noether charge, which gives a finite contribution to the BPS mass, has been studied
very well, the electric and/or magnetic charge densities have not been discussed.
Therefore, before studying the dyonic 1/4 BPS states, we stop for a while to clarify the dyonic 
domain wall in the weak gauge coupling region.

The master equation for the dyonic domain wall in the dimensionless coordinates
is given by
\begin{eqnarray}
\p_3^2 u - 1 + \Omega_0 e^{-u} = 0,\quad
\Omega_0 = H_0^\dagger e^{2\eta \tilde M \cos\alpha\, x^3} H_0^\dagger.
\end{eqnarray}
This is formally the same equation as the master equation for the purely magnetic domain wall.
If we write $\tilde M = \tilde M'/\cos\alpha$, the solutions $u(x^3)$ are identical
to those which have already obtained. 
In order to avoid inessential complications, we will consider $\eta=+1$ and $\tilde M' = {\rm diag}(\tilde m'/2,\ -\tilde m'/2)$
with $\tilde m' > 0$ in what follows.
The tension of the domain wall becomes
\begin{eqnarray}
\tilde T_W = \frac{2 \tilde m'}{\cos\alpha},\qquad \left(T_W = \frac{gv^3}{\sqrt2} \tilde T_W = \frac{m'v^2}{\cos\alpha}\right),
\end{eqnarray}
because of $\tilde m = \tilde m'/\cos\alpha$.

Since $u = u(x^3)$ and from Eq.~(\ref{eq:dyonic_EB}), no magnetic fields  are involved. On the other hand,
the third component of the electric field does appear
\begin{eqnarray}
\tilde E_3 = 2\tan\alpha\, \p_3\tilde\sigma = \tan\alpha\,\p_3^2 u.
\end{eqnarray}
Remember, $\p_3$ means the derivative in terms of $\tilde x^3$ and $\tilde E_k = E_k/g^2v^2$.
When $N_F=2$, 
$\tilde \sigma$ is constant outside the domain wall,
so no electric fields exist there (see the details in \cite{Boojum1}). On the other hand, a constant electric $E_3$ appears inside the domain wall,
as $\tilde \sigma$ is linear in $x^3$ there. Since the width of the domain wall is $2\tilde m'$ and
$\tilde \sigma$ changes from $-\tilde m'/2\cos\alpha$ to $\tilde m'/2\cos\alpha$, 
we have $\p_3\tilde \sigma \simeq 1/2\cos\alpha$.
Therefore, the electric field inside the domain wall for the weak coupling is
\begin{eqnarray}
\tilde E_k = \frac{\sin \alpha}{\cos^2\alpha}\, \delta_{k3}.
\end{eqnarray}
The induced electric charges which generate this electric fields
can be found from  Eq.~(\ref{eq:electric_charge}).
In terms of the dimensionless coordinates, the electric charge density is rewritten as follows
\begin{eqnarray}
\tilde {\cal Q}_E = \frac{{\cal Q}_E}{\sqrt2 gv^3} 
= -  \tan\alpha\, \p_3\left(\tilde H \tilde H^\dagger\right)
= - \tan\alpha\, \p_3 m_{\rm v}^2,
\label{eq:QE_DDW}
\end{eqnarray}
where $m_{\rm v}^2 = \tilde H \tilde H^\dagger$ is 1 in the vacua, while $m_{\rm v}^2 = 0$ inside the domain wall.
\begin{figure}[t]
\begin{center}
\includegraphics[height=7cm]{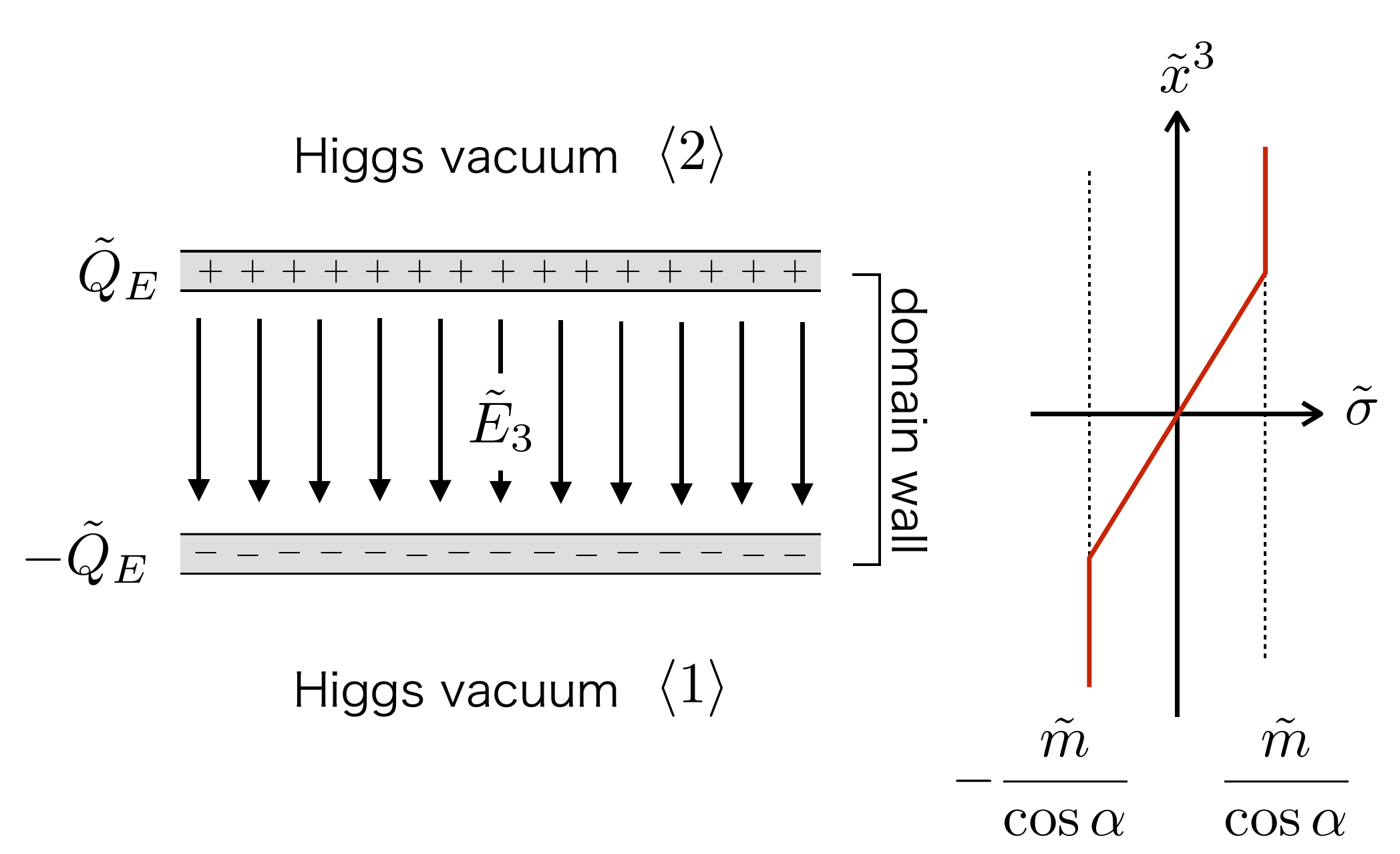}
\caption{The dyonic domain wall as an electric capacitor.}
\label{fig:capacitor}
\end{center}
\end{figure}
Therefore,  electric charges are induced on the outer layers, see Fig.~\ref{fig:capacitor}: 
positive (negative) electric charges on the left outer skin and negative (positive) charges 
on the right outer skin for $\tan \alpha > 0$ ($\tan\alpha < 0$).
Then the electric charge per unit area is given by
\begin{eqnarray}
\tilde Q_E = \pm \big[\tilde H\tilde H^\dagger \tan\alpha \big]^{\text{inside}}_{\text{ouside}}
= \pm \tan\alpha.
\end{eqnarray}
Since the distance between two outer layers is $2\tilde m'$, the difference of electric potential 
is
\begin{eqnarray}
\tilde V = 2\tilde m' \tilde E_3 =  \frac{2 \tilde m'}{\cos\alpha} \,\tan\alpha
= \frac{2\tilde m'}{\cos\alpha}  \,|\tilde Q_E|.
\end{eqnarray}
Hence, the electric capacitance per unit area is given by
\begin{eqnarray}
\tilde c = \frac{\cos\alpha}{2\tilde m'} = \frac{1}{\tilde T_W}.
\end{eqnarray}
Note that the electric capacitance, in the usual sense, is infinity because the domain wall has infinite area.
The energy stored in the capacitor is
\begin{eqnarray}
\frac{1}{2} \tilde c \tilde V^2 = \frac{1}{2}\tilde T_W \tan^2\alpha,
\end{eqnarray}
which is the excess of the domain wall's tension for small $\alpha$
\begin{eqnarray}
\sqrt{\tilde T_W^2 + \tilde Q_W^2} - \tilde T_W \simeq \frac{1}{2}\frac{\tilde Q_W^2}{\tilde T_W}
= \frac{1}{2}\tilde T_W \frac{\tilde Q_W^2}{\tilde T_W^2} = \frac{1}{2}\tilde T_W \tan^2\alpha.
\end{eqnarray}

Note that the dyonic domain wall behaves as the electric capacitor only in the weak gauge coupling
region. This is because $HH^\dagger \simeq v^2$ holds everywhere in the strong gauge coupling
region so that no electric charge can be stored on the outer skins, see Eq.~(\ref{eq:QE_DDW}).

\subsection{1/4 BPS dyonic configurations}

Let us next consider the simplest 1/4 BPS dyonic solution $H_0 = (z,1)$ in the $N_F = 2$ case.
As mentioned below Eq.~(\ref{eq:master_dyonic}), the difference between the master equation 
for the purely magnetic and the dyonic cases amounts to the replacement of $M$ by $M\cos\alpha$.
Therefore, as in the case of dyonic domain walls, all the numerical solutions which we have obtained previously \cite{Boojum1} are still valid
for the dyonic configurations. 
Indeed, the master equation \refer{eq:master_dyonic} in terms of the dimensionless parameters given in Eq.~(\ref{eq:dimless}) reduces to 
\begin{equation}
\partial_{\rho}^2u+\frac{1}{\rho}\partial_\rho u+\partial_3^2 u 
= 1-\bigl(\rho^2 e^{\eta \tilde m'\, x^3}+e^{-\eta \tilde m'\, x^3}\bigr)e^{-u}\,,
\end{equation}
where we have written $\tilde m = \tilde m'/\cos\alpha$.

The energy density consists of six parts; the domain wall ${\cal T}_W$, vortex string ${\cal T}_S$, boojum
${\cal T}_B$, ${\cal T}_4 = \p_k j_k$, ${\cal Q}_W$ and ${\cal Q}_B$. 
The first four contributions have no changes
from the purely magnetic case because of cancellation of $\cos\alpha$
\begin{eqnarray}
\tilde {\cal T}_{W;\alpha} &=& \frac{1}{g^2v^4} {\cal T}_W \cos\alpha   
=  \p_3^2 u,\\
\tilde {\cal T}_{S;\alpha} &=& \frac{1}{g^2v^4} {\cal T}_S 
= \left( \p_1^2 +  \p_2^2\right)u,\\
\tilde {\cal T}_{B;\alpha} &=& \frac{1}{g^2v^4}{\cal T}_B \cos\alpha 
= 2\left\{\left(\p_1 \p_3 u\right)^2 + \left( \p_2 \p_3u\right)^2 
- \left( \p_1^2+ \p_2^2\right)\! u\   \p_3^2 u\right\},\\
\tilde {\cal T}_{4;\alpha} &=& \frac{1}{g^2v^4}{\cal T}_4  
= - \p_k^2 \p_k^2 u.
\end{eqnarray}
The remaining quantities depend on $\alpha$ as
\begin{eqnarray}
\tilde {\cal Q}_{W;\alpha} &=& \frac{1}{g^2v^4}{\cal Q}_W \sin \alpha
= \eta\,\p_3 \left(\tilde H \tilde M' \tilde H^\dagger \right) \tan^2\alpha,\\
\tilde {\cal Q}_{B;\alpha} &=& \frac{1}{g^2v^4}{\cal Q}_B\sin\alpha
=  \partial_k^2\left[\left( \p_3u\right)^2\right] \tan^2\alpha.
\end{eqnarray}
The electric and magnetic fields are given by
\begin{eqnarray}
\tilde E_k &=& \frac{1}{g^2v^2}E_k 
=  \eta \tan\alpha\,  \p_k  \p_3 u,\\
(\tilde B_1,\tilde B_2,\tilde B_3) &=& \frac{1}{g^2v^2} (B_1,B_2,B_3) 
= \xi \left( \p_3 \p_1 u,\,  \p_2\p_3 u,\, - (\p_1^2 + \p_2^2) u\right).
\end{eqnarray}
The electric charge density is 
\begin{eqnarray}
\tilde {\cal Q}_{E;\alpha} = \frac{{\cal Q}_E}{\sqrt2 gv^3} 
= -  \eta \tan\alpha\,  \p_3\left(\tilde H \tilde H^\dagger\right).
\end{eqnarray}
Remember that the derivatives are with respect to the rescaled variables $\tilde x^k$.

In the following, we will set $\alpha = \pi/4$ and consider the masses 
$\tilde m' = 1/5, 1, 10$ ($\tilde m = \sqrt 2/5, \sqrt 2, 10\sqrt 2$), 
as examples for the strong, intermediate and weak gauge couplings, respectively.

\begin{figure}[h]
  \begin{center}
   \includegraphics[width=15cm]{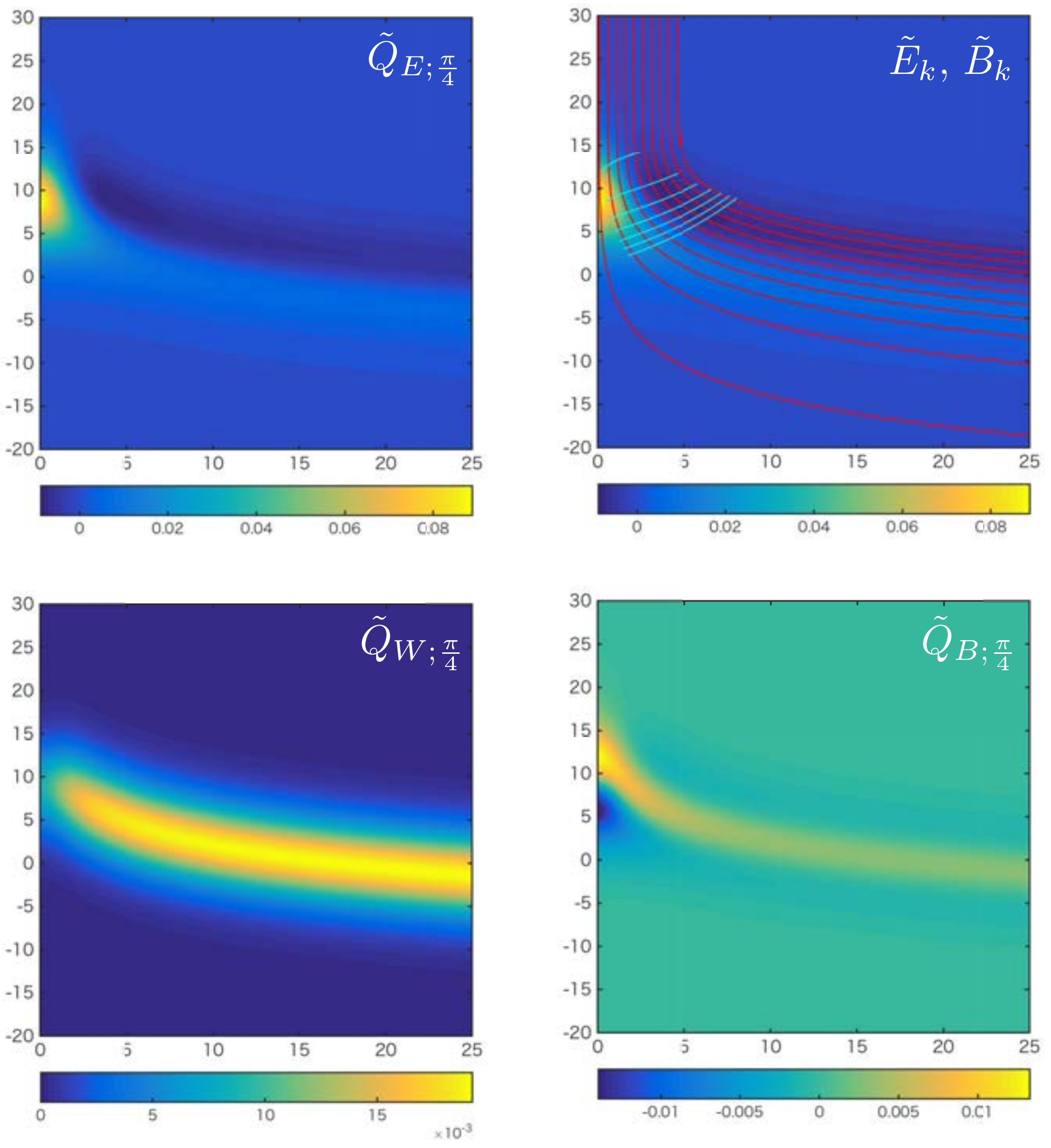}
   \caption{The dyonic charges $\tilde Q_{E,W,B;\alpha=\pi/4}$ are shown for $\tilde m' = 1/5$
   (strong gauge coupling region). The topological charge densities are given in Fig.~5. The red curves
   show the magnetic force lines and the cyan ones correspond to the electric force lines.}
   \label{fig:dyonic_m1o5}
  \end{center}
\end{figure}

Let us first look at Fig.~\ref{fig:dyonic_m1o5} in which the dyonic charge densities 
for $\tilde m'=1/5$ are shown.
The distributions are quite different from those in the weak coupling solution.
The domain wall steeply bends.
Since $HH^\dagger \simeq v^2$ holds everywhere for 
the strong gauge coupling, the induced electric charge is tiny ($\p_3(HH^\dagger) \simeq 0)$,
so that it is no longer suitable to regard it as an electric capacitor,
see the top-left panel of Fig.~\ref{fig:dyonic_m1o5}. 
Only the region near the junction point is evidently charged positively, whereas the
electric charge densities become diluted far away from the junction
point. The electric and magnetic force lines are shown in the top-right panel of Fig.~\ref{fig:dyonic_m1o5}.

\begin{figure}
  \begin{center}
   \includegraphics[width=15cm]{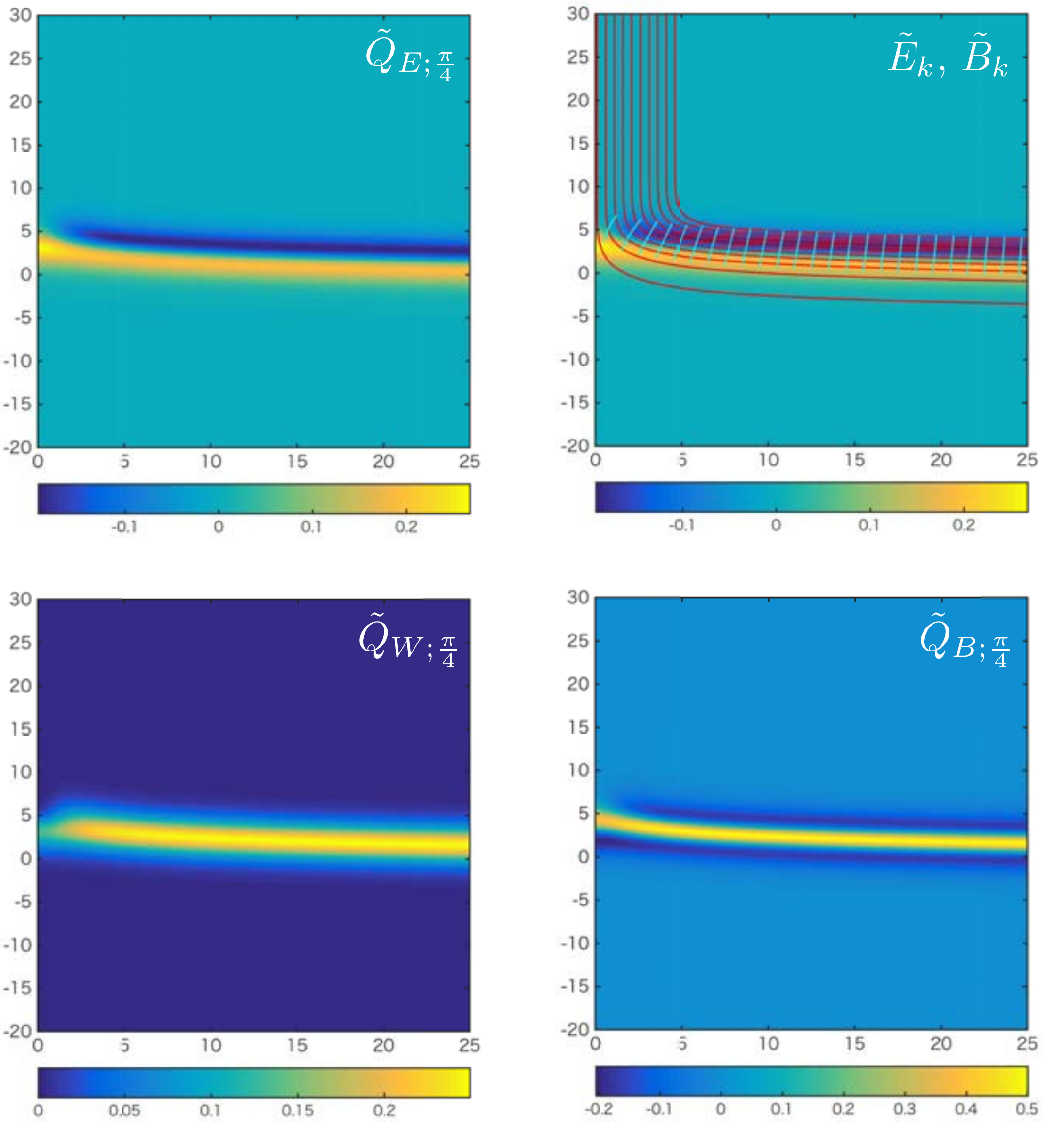}
   \caption{The dyonic charges $\tilde Q_{E,W,B;\alpha=\pi/4}$ are shown for $\tilde m' = 1$
   (strong gauge coupling region). The topological charge densities are given in Fig.~6. The red curves
   show the magnetic force lines and the cyan ones correspond to the electric force lines.}
   \label{fig:dyonic_m1}
  \end{center}
\end{figure}

Fig.~\ref{fig:dyonic_m1} shows the electric charge densities for the intermediate mass $\tilde m'=1$.
Since the curvature of the domain wall is now smaller, the separation between the positive and negative 
electric charges is visible. Mean distance is of the same order as the domain wall width $2\tilde m' = 2$.
Unlike the strong coupling case, both the positive and negative electric charge densities are not localized
near the junction point, but they extend along the domain wall. The positive charges are distributed
across the whole domain wall, whereas the negative charges have no support around the junction point.
Therefore, the electric force lines bend near the junction point and asymptotically becomes vertical 
far from the boojum (see the top-right panel of Fig.~\ref{fig:dyonic_m1}).

\begin{figure}[b]
  \begin{center}
   \includegraphics[width=15cm]{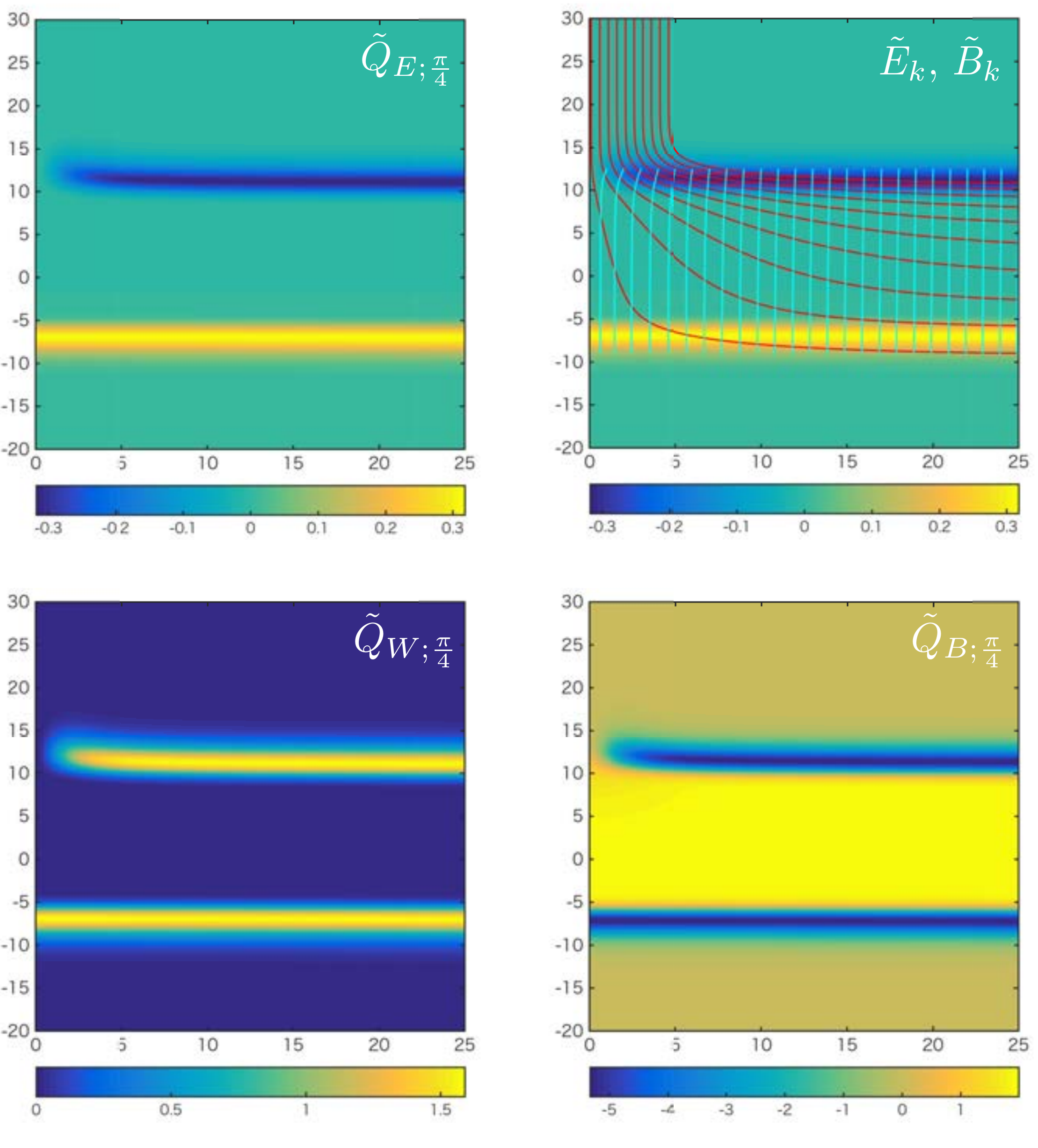}
   \caption{The dyonic charges $\tilde Q_{E,W,B;\alpha=\pi/4}$ are shown for $\tilde m' = 10$
   (strong gauge coupling region). The topological charge densities are given in Fig.~7. The red curves
   show the magnetic force lines and the cyan ones correspond to the electric force lines.}
   \label{fig:dyonic_m10}
  \end{center}
\end{figure}

Finally, we show the dyonic solution for $\tilde m' = 10$ in Fig.~\ref{fig:dyonic_m10}.
As expected, it is clearly similar to an electric capacitor with a large distance $2\tilde m'=20$
between positive and negative charges. The electric force lines are vertical except for
the region near the boojum. From Fig.~\ref{fig:dyonic_m10} one clearly sees that the electric 
charge and Noether charge appear on the outer skins of the domain wall in the weak coupling region.
The Noether charge densities on the two outer skins have the same sign so that the total Noether charge
does not vanish. The charge $\tilde Q_B$ is negative on the outer skins but it  is positive inside the domain wall, which is consistent with the fact that $\int^\infty_{-\infty}dx^3\ \tilde Q_B = 0$.

\clearpage

\clearpage

\clearpage

\clearpage


\section{Low energy effective theory and Nambu-Goto/DBI action}
\label{sec:DBI}
In this section, we study 1/2 and 1/4 BPS configurations from the viewpoint of the low energy effective actions, namely the Nambu-Goto (NG) action and the Dirac-Born-Infeld (DBI) action for the domain wall. As is well known, a low energy effective theory 
of a simple domain wall with translational zero modes is the NG action.  The low energy effective action for the domain wall
with not only the translational zero modes but also the internal moduli have been found to be the NG type \cite{Portugues,Eto3,Eto:2015wsa} by regarding the internal space as extra dimensions.
It is also known that the DBI action is dual to the NG action. In this section, we show that the domain wall, the vortex string ending on the domain wall and their dyonic extension are reproduced in the NG action when the gauge coupling constant is taken to the infinity. In the strong gauge coupling limit, the vortex string asymptotically becomes the singular lump string attached to the domain wall. We call this configuration the spike domain wall. The dyonic extension of this configuration has been already studied in the massive nonlinear sigma model on $T^*{\mathbb C}P^1$, and it was shown that the configuration is realized as BIon
in the DBI action \cite{Portugues}. We review the dyonic extension of the spike domain wall from the viewpoint of the NG action. Finally, we discuss whether the non-singular lump string with the size moduli, the semi-local boojums studied in subsection \ref{sec:semilocal}, can be realized in the DBI action.

%
%
\subsection{Nambu-Goto action and Hamiltonian}
We start with the NG action in $(2+1)$-space-time dimensions \cite{Portugues}:
\begin{align}
{\mathcal L}_{\rm NG} & = -\hat{T}_W \sqrt{\mbox{det}\left(\eta_{\alpha\beta}-\partial_\alpha X\partial_\beta X-\partial_\alpha\phi\partial_\beta\phi\right)}\,, \quad \alpha,~\beta=0,1,2\,,
 \label{eq:lagng} 
\end{align}
where $X$ and $\phi$ are scalar fields, 
which will be identified with the position and the phase moduli of the domain wall solution 
and $\hat{T}_W$ is the membrane tension. Here we have used the so-called physical gauge where the induced metric on the world-volume of the brane is flat (i.e. $\eta = \mbox{diag}(1,-1,-1)$). We can explicitly write this as
\begin{align}
{\mathcal L}_{\rm NG}= -\hat{T}_W \sqrt{D_{\rm NG}}\,, \label{eq:NGaction}
\end{align} 
where
\begin{eqnarray}
 D_{\rm NG}=1-(\partial_\alpha X \partial^\alpha X)-(\partial_\alpha \phi \partial^\alpha \phi)+(\partial_\alpha X \partial^\alpha X)(\partial_\beta \phi \partial^\beta \phi)-(\partial_\alpha X \partial^\alpha \phi)^2\,.
\end{eqnarray}
The canonical momenta for $X$ and $\phi$ are given by
\begin{eqnarray}
 P_X={\partial {\cal L} \over \partial \dot{X}}&=& -\hat{T}_W D_{\rm NG}^{-1/2} \left\{\dot{X}-\dot{X}(\partial_\alpha \phi \partial^\alpha \phi)+\dot{\phi}(\partial_\alpha X \partial ^\alpha \phi)\right\}\,, \\
 P_{\phi}={\partial {\cal L} \over \partial \dot{\phi}}&=& \hat{T}_W D_{\rm NG}^{-1/2} \left\{\dot{\phi}-\dot{\phi}(\partial_\alpha X \partial^\alpha X)+\dot{X}(\partial_\alpha X \partial ^\alpha \phi)\right\}\,,
\end{eqnarray}
so that the Hamiltonian is obtained as
\begin{eqnarray}
 {\cal H}_{\rm NG}&=&P_X \dot{X}+P_\phi \dot{\phi}-{\cal L}_{\rm NG} \nonumber \\
   &=& \hat{T}_W D_{\rm NG}^{-1/2}
   \left\{
    1+(\partial_i X)^2+(\partial_i \phi)^2+(\partial_i X)^2(\partial_j \phi)^2-(\partial_i X \partial_i \phi)^2
   \right\}\,, \label{eq:H-NG}
\end{eqnarray}
where the index $i, j=1,2$ are summed over.
%
%
\subsection{Domain wall, its dyonic extension and the NG action}
Let us recall the domain wall solution discussed in section \ref{model}. The master equation for the flat domain wall is given in (\ref{eq:quatermaster}) when $u$ is restricted to depend on $x^3$ coordinate only. In the strong gauge coupling limit, the master equation can be solved to give 
\begin{eqnarray}
 u_{g\rightarrow \infty}=\log \Omega_0\,, \label{eq:ud}
\end{eqnarray}
where $\Omega_0$ is given in (\ref{eq:quatermaster}). For simplicity, let us consider $N_F=2$ case. In this case, the moduli matrix $H_0(z)$ in (\ref{a3}) is just a constant. We choose
\begin{eqnarray}
 H_0(z)=\left(e^{-\frac{m}{2} (X + i \phi)},e^{\frac{m}{2} (X + i \phi)}\right)\,, \quad M={\rm diag}(m/2, -m/2)\,.\label{eq:mmdw}
\end{eqnarray}
with $\xi=\eta=1$.
Then (\ref{eq:ud}) gives
\begin{eqnarray}
 u_{g\rightarrow \infty}(x^3)=\log(e^{m(x^3-X)}+e^{-m(x^3-X)})\,.
\end{eqnarray}
This shows that the constant parameter $X$ corresponds to the position moduli. The other constant parameter $\phi$
is the internal moduli which is the Nambu-Goldstone mode associated with the spontaneously broken $U(1)_F$ symmetry.
The energy of the domain wall is readily calculated by integrating (\ref{eq:tw}) over all the space-directions
\begin{eqnarray}
 E_{\rm wall}=\int d^3x\ {\mathcal T}_W=T_W A\,, \label{eq:dwb}
\end{eqnarray}
where $T_W=mv^2$ is the domain wall's tension and $A$ is the area of the domain wall.

Now let us study the flat domain wall solution in the NG action. It is just given by considering constants for $X$ and $\phi$. 
The NG Hamiltonian (\ref{eq:H-NG}) reduces to
\begin{eqnarray}
 {\cal H}_{\rm NG}= \hat{T}_W\,.
\end{eqnarray}
The energy is obtained by integrating along the membrane directions:
\begin{eqnarray}
 E_{\rm NG}=\int d^2x\, {\cal H}_{\rm NG}=\hat{T}_WA\,. \label{eq:dwd}
\end{eqnarray}
The energy (\ref{eq:dwb}) and (\ref{eq:dwd}) completely coincide if the domain wall tension $T_W$ is identified with the membrane tension $\hat{T}_W$. Therefore, as expected, the NG action with constant $X$ and $\phi$ realizes the domain wall in the field theoretical model.

Next we consider the Q-extension (dyonic-extension) of domain wall. Let us first recall the field theory solution
given in Eqs.~(\ref{eq:dy_mm_1}) -- (\ref{eq:dy_mm_4}).
The dyonic extension of the master equation is given in (\ref{eq:master_dyonic}). Substituting (\ref{eq:mmdw}) with 
$X$ and $\phi$ being constants into (\ref{eq:master_dyonic}), we obtain the solution in the strong gauge coupling limit as
\begin{eqnarray}
 u_{g\rightarrow \infty}(x^3)=\log(e^{m\cos\alpha\, x^3 -mX}+e^{-m\cos\alpha\, x^3 + mX})\,.
\end{eqnarray}
The energy of the system is obtained from (\ref{dwall}) with (\ref{eq:tension-q}) and (\ref{eq:q-charge-domain-wall}):
\begin{eqnarray}
 E_{\text{Q-wall}}=\int d^3x\, ({\cal T}_W\cos\alpha+{\cal Q}_W\sin\alpha)=\sqrt{T_W^2+Q_W^2}A\,, \label{eq:dwall-a}
\end{eqnarray}
where we have used (\ref{qc}). The domain wall tension $T_W$ and the Noether charge $Q_W$ are 
calculated by (\ref{eq:dyonic-dw}) and (\ref{eq:Noether-c}): $T_W=mv^2$ and $Q_W=mv^2\tan\alpha$. 

Let us consider the corresponding configuration in the NG action. We take time-dependent phase 
\begin{eqnarray}
 X=\text{const.}\,,\quad \phi=\frac{\omega}{\hat m} t\,, \label{q-domain-wall-NG}
\end{eqnarray}
 where $\omega$ is a constant angular velocity and we have introduced a constant parameter $\hat m$ of mass dimension one
 in order to make $\phi$ have mass dimension $-1$. In this case, the energy is obtained as
\begin{eqnarray}
 E_{\rm NG}={\hat{T}_W \over \sqrt{1-(\omega/\hat m)^2}}A\,. \label{eq:dwall-NG}
\end{eqnarray}
The conserved momentum $P_\phi$ for this solution is given by
\begin{eqnarray}
P_\phi = \hat T_W \frac{\omega/\hat m}{\sqrt{1-(\omega/\hat m)^2}}.
\end{eqnarray}
Then the energy can be written as
\begin{eqnarray}
 E_{\rm NG}=\sqrt{{\hat T}_W^2+P_\phi^2}A\,, \label{eng-qd}
\end{eqnarray}
We identify $T_W=\hat{T}_W$ as before. Furthermore, we should identify $\omega/\hat m = \sin\alpha$ from
Eq.~(\ref{eq:dy_mm_1}), which tells us that the Q-charge (\ref{qc}) in the field theory is understood as
\begin{eqnarray}
P_\phi = \hat T_W \tan \alpha = Q_W 
 \,. \label{qcharge-NG}
\end{eqnarray}
Thus, Eq.~(\ref{eng-qd}) coincides with (\ref{eq:dwall-a}). We conclude that the configuration (\ref{q-domain-wall-NG}) in the NG action realizes the Q-extension of the domain wall in the field theory.

%
%
\subsection{Dyonic extension of spike domain wall and NG action}
In this subsection, we study the 1/4 BPS dyonic extension of the spike domain wall that the lump vortex attaches on the domain wall. The master equation for this configuration is given in (\ref{eq:master_dyonic}) where $u$ depends on all the space coordinates. Considering $N_F=2$ case, the moduli matrix and the mass matrix are given by
\begin{eqnarray}
 H_0(z)=(z,1),\quad M={\rm diag}(m/2,-m/2)\,. \label{eq:nf2}
\end{eqnarray}
In the strong gauge coupling limit the master equation (\ref{eq:master_dyonic}) is solved as
\begin{eqnarray}
 u_{g\rightarrow \infty}(z,\bar{z}, x^3)=\log(\rho^2 e^{m\cos\alpha\, x^3}+e^{-m\cos\alpha\, x^3})\,,
\end{eqnarray}
where $|z|^2=\rho^2$. The total energy of the configuration is
\begin{eqnarray}
 E_{1/4}=\int d^3x ({\cal T}_W\cos\alpha+{\cal Q}_W\sin\alpha+{\cal T}_S+{\cal T}_B)\,,
\end{eqnarray}
where ${\cal T}_S$ and ${\cal T}_B$ are defined in (\ref{eq:tension-s}) and (\ref{eq:tension-b}), respectively. Taking into account the fact that in the strong gauge coupling limit (\ref{eq:tension-b}) is vanishing, the energy is given as
\begin{eqnarray}
E_{1/4}&=&\sqrt{T_W^2+Q^2_W}A+T_S L\,. \label{eq:dyonic-spike}
\end{eqnarray}
Here we have used (\ref{qc}) and $T_S$ is the string tension given in (\ref{eq:qvs}), $T_S=2\pi v^2 |k|$ where $k$ is the vortex winding number and $L$ is length of the vortex string.

We consider the corresponding configuration in the NG action. We take the following configuration:
\begin{eqnarray}
 X\rightarrow X(x^1,x^2), \quad \phi \rightarrow \frac{\omega}{\hat m} t+\phi(x^1,x^2)\,. \label{eq:spiky-kink-conf}
\end{eqnarray}
To simplify notation, let us introduce
\begin{eqnarray}
a_i \equiv  \partial_i X\,,\quad b_i \equiv \partial_i \phi\,.
\end{eqnarray}
Then the Hamiltonian (\ref{eq:H-NG}) is expressed as
\begin{eqnarray}
 {\cal H}_{\rm NG}= \hat T_W\frac{1+a_i^2 + b_i^2 + (\epsilon_{ij}a_ib_j)^2}{\sqrt{
 1+a_i^2 + b_i^2 + (\epsilon_{ij}a_ib_j)^2 - (\omega/\hat m)^2 (1+a_i^2)
 }}\,,
\end{eqnarray}
where $\epsilon_{12}=1$ and we have used $a_i^2 b_j^2 - (a_ib_i)^2 = (\epsilon_{ij}a_ib_j)^2$.
Let us further rewrite this in terms of $ \tilde{b}_i \equiv b_i/ \sqrt{1-(\omega/\hat m)^2}$, 
\begin{eqnarray}
 {\cal H}_{\rm NG}={\hat{T}_W \over \sqrt{1-(\omega/\hat m)^2}} \left( {\mathcal H}^{1/2}-{(\omega/\hat m)^2 (\tilde{b}_i^2+(\epsilon_{ij}a_i\tilde{b}_j)^2) \over {\mathcal H}^{1/2}} \right)\,, \label{eq:H-NG-R}
\end{eqnarray}
where
\begin{eqnarray}
 {\cal H}=1+(a_i)^2+(\tilde{b}_i)^2+(\epsilon_{ij}a_i\tilde{b}_j)^2\,.
\end{eqnarray}
Now we are ready to minimize the Hamiltonian ${\cal H}_{\rm NG}$. To this end, we first minimize ${\cal H}$ as
\begin{eqnarray}
 {\cal H}=(a_i\pm \epsilon_{ij}\tilde{b}_j)^2+(1\mp \epsilon_{ij}a_i\tilde{b}_j)^2 
 \ge  (1\mp \epsilon_{ij}a_i \tilde{b}_j)^2\,. \label{eq:NG-KL-H}
\end{eqnarray}
The last inequality is saturated when the equation
\begin{eqnarray}
 a_i \pm \epsilon_{ij}\tilde{b}_j=0 \label{eq:NG-KL}
\end{eqnarray}
is satisfied. The key observation is that when the equation (\ref{eq:NG-KL}) holds the first term in (\ref{eq:H-NG-R}) is minimized while the second term is maximized. Indeed, the numerator in the second term is maximized because 
$(\epsilon_{ij}a_i\tilde b_j)^2$ becomes maximum when $a_i$ is orthogonal to $b_i$, 
and at the same time the denominator is minimized. 
Taking account of the minus sign in front of the second term of Eq.~(\ref{eq:H-NG-R}), it is found that the Hamiltonian  is minimized
when the BPS equation (\ref{eq:NG-KL}) is satisfied.  The BPS energy in terms of the original variable $X$
is given by
\begin{eqnarray}
 {\cal H}_{\rm NG}=\hat{T}_W\left\{{1 \over \sqrt{1-(\omega/\hat m)^2}}+\sqrt{1-(\omega/\hat m)^2}(\partial_i X)^2\right\}\,. \label{eq:KL-H2}
\end{eqnarray}

We now consider that a point particle corresponding to the endpoint of the lump string is placed on the membrane. Comparing Eqs.~(\ref{eq:mmdw}) and (\ref{eq:nf2}), one is naturally lead to the following identification 
\begin{eqnarray}
 \phi=\pm \frac{1}{\hat m} \arctan{x^2 \over x^1}\,. \label{eq:KL-P}
\end{eqnarray}
Again, we have introduced a certain parameter $\hat m$  of mass dimension one.
Combining (\ref{eq:NG-KL}) with (\ref{eq:KL-P}) gives 
\begin{eqnarray}
 X(\rho)=-{1 \over \sqrt{\hat m^2-\omega^2}}\log\rho\,, \label{eq:KL-sol2}
\end{eqnarray}
where $\rho=\sqrt{(x^1)^2+(x^2)^2}$. With the solution (\ref{eq:KL-sol2}), the Hamiltonian (\ref{eq:KL-H2}) turns out to be
\begin{eqnarray}
  {\cal H}_{\rm NG}={\hat{T}_W \over \sqrt{1-(\omega/\hat m)^2}}\left(1+{1 \over \hat{m}^2 \rho^2}\right)\,.
\end{eqnarray}
The energy is
\begin{eqnarray}
 E_{\rm NG}=\int d^2x\ {\cal H}_{\rm NG}={\hat{T}_W \over \sqrt{1-(\omega/\hat m)^2}} A+{2\pi \hat{T}_W \over \hat{m}\sqrt{\hat m^2-\omega^2}}(\log R-\log \delta)\,,
\end{eqnarray}
where we have introduced the ultraviolet cutoff $\rho=\delta$ and the infrared cutoff $\rho=R$. With the use of (\ref{qcharge-NG}) and (\ref{eq:KL-sol2}), the energy is rewritten as
\begin{eqnarray}
 E_{\rm NG}=\sqrt{\hat{T}_W^2 + P_\phi^2} A+{2\pi \hat{T}_W \over \hat{m}}\hat{L}\,, \label{eq:KL-E}
\end{eqnarray}
where $\hat{L}\equiv X(\delta)-X(R)$. Identifying $T_W=\hat{T}_W$ and $m=\hat{m}$, it is found that $2\pi \hat{T}_W/\hat{m}$ 
($T_W = mv^2$) coincides with the string tension $T_S = 2\pi v^2$ in the field theory.  Respecting $\hat{L}$ with the length of the vortex string, the energy (\ref{eq:KL-E}) coincides with (\ref{eq:dyonic-spike}) in the field theoretical model.

%
%
\subsection{Relation between solutions of NG action and DBI action}
In this subsection, we show that the dyonic extension of the spike domain wall (\ref{eq:nf2}) in the field theory is also realized in the DBI action \cite{Portugues}. Rather than minimizing the Hamiltonian of the DBI action we derive the BPS equations in the DBI action by transforming Eq.~(\ref{eq:NG-KL}).

First we derive $(2+1)$-dimensional DBI action from the NG action (\ref{eq:lagng}) by dualization
\begin{equation}
{\mathcal L}_{\rm DBI} = {\mathcal L}_{\rm NG}+\frac{\kappa}{2} \hat{T}_W\varepsilon_{\alpha \beta \gamma}\hat{F}^{\alpha\beta}\partial^\gamma\phi\,, \label{dual}
\end{equation}
where the last term is called the BF term consisting of $\hat{F}_{\alpha\beta} = \partial_\alpha \hat{A}_\beta-\partial_\beta \hat{A}_\alpha$ being an abelian field strength and $\kappa$ an arbitrary constant of mass dimension $-2$. Notice that the term we added is a total divergence with no effect on dynamics. Let us eliminate $\phi$ by using its equation of motion. Variation of the above Lagrangian with respect to $\partial_\alpha \phi$ leads to the condition
\begin{equation}\label{eq:condition}
\kappa \hat{F}_{\alpha}^{*} = \frac{-1}{\sqrt{D_{\rm NG}}}\Bigl[\left\{1-(\partial_\beta X \partial^\beta X)\right\}\partial_\alpha\phi+\bigl(\partial_\beta X \partial^\beta \phi\bigr)\partial_\alpha X\Bigr]\,,
\end{equation}
where $\hat{F}_\alpha^*= \tfrac{1}{2}\varepsilon_{\alpha\beta\gamma}\hat{F}^{\beta\gamma}~(\epsilon_{123}=1)$. Contracting the above with $\partial^\alpha \phi$ we obtain
\begin{equation}\label{eq:dbidev}
\kappa (\hat{F}_\alpha^* \partial^\alpha\phi) = \sqrt{D_{\rm NG}}-\frac{1-(\partial_\alpha X \partial^\alpha X)}{\sqrt{D_{\rm NG}}}\,.
\end{equation}
Substituting this into (\ref{dual}), we have
\begin{equation}
{\mathcal L}_{\rm DBI} = - \hat{T}_W\frac{1-(\partial_\alpha X \partial^\alpha X)}{\sqrt{D_{\rm NG}}}\,.
\end{equation}
In order to eliminate $\phi$ in $D_{\rm NG}$, we consider contractions of \refer{eq:condition} with $\partial^\alpha X$ and $\hat{F}^{*\alpha}$ as
\begin{align}
&\kappa (\hat{F}_\alpha^*\partial^\alpha \phi)={-1 \over \sqrt{D_{\rm NG}}}(\partial_\alpha \phi \partial^\alpha X)\,, \\
&\kappa(\partial_\alpha \phi \partial^\alpha\phi)(1-(\partial_\beta X \partial^\beta X))^2 \nonumber \\
&\qquad ={-1 \over \sqrt{D_{\rm NG}}}
\left\{(1-(\partial_\alpha X \partial^\alpha X))(\partial_\beta \phi \hat{F}^{\beta*})+(\partial_\alpha X \partial^\alpha \phi)(\partial_\beta X \hat{F}^{\beta*})
\right\}\,. \label{dbi2}
\end{align} 
Eqs. \refer{eq:dbidev}--\refer{dbi2} can be combined together to solve for $D_{\rm NG}$:
\begin{equation}
D_{\rm NG} = \frac{\bigl(1-(\partial_\alpha X \partial^\alpha X)\bigr)^2}{1-(\partial_\beta X \partial^\beta X)+\kappa^2 (\hat{F}_\beta^*\hat{F}^{\beta *})-\kappa^2 \bigl(\partial_\beta X \hat{F}^{\beta *}\bigr)^2}\,.
\end{equation}
With this expression at hand we use Eq.~\refer{eq:dbidev} to obtain the DBI action 
\begin{align}
{\mathcal L}_{\rm DBI} & = -\hat{T}_W \sqrt{1-(\partial_\alpha X \partial^\alpha X)+\kappa^2 (\hat{F}_\alpha^*\hat{F}^{\alpha *})-\kappa^2 \bigl(\partial_\alpha X \hat{F}^{\alpha*}\bigl)^2} \nonumber \\ \label{eq:lagdbi}
& = -\hat{T}_W \sqrt{\mbox{det}\bigl(\eta_{\alpha\beta}-\partial_\alpha X\partial_\beta X+ \kappa \hat{F}_{\alpha \beta}\bigr)}\,,
\end{align}
where the validity of the last equality can be checked by direct evaluation of the determinant.

Next, we derive the Hamiltonian. In the following, we set $\dot{X}=0$ since we are not interested in the configuration where $X$ depends on time. First we shall write the Lagrangian (\ref{eq:lagdbi}) in terms of the electric and magnetic fields defined by $\hat{F}_{0i}=\hat{E}_i$ $(i=1,2)$ and $\hat{F}_{12}=\hat{B}$:
\begin{eqnarray}
{\mathcal L}_{\rm DBI} = -\hat{T}_W \sqrt{D_{\rm DBI}}\,,\label{dbi1}
\end{eqnarray}
where
\begin{eqnarray}
D_{\rm DBI}=1+(\partial_i X)^2-\kappa^2 \hat{E}_i^2 +\kappa^2 \hat{B}^2 - \kappa^2 \left(\epsilon_{ij}\hat{E}_i \partial_j X\right)^2\,. \label{eq:def-DBID}
\end{eqnarray}
Here the index $i, j$ are summed over. In the following we rescale
\begin{eqnarray}
 \kappa\hat{E}_i\rightarrow \hat{E}_i, \quad  \kappa\hat{B}_i\rightarrow \hat{B}_i\,.
\end{eqnarray}
A canonical momentum is obtained by differentiating (\ref{dbi1}) with respect to $\hat{E}_i$:
\begin{eqnarray}
\Pi_i&=&{\partial {\cal L}_{\rm DBI} \over \partial \hat{E}_i} 
          = -\hat{T}_W D_{\rm DBI}^{-1/2}(\hat{E}_i-\epsilon_{ij}\partial_j A)\,, \label{eq:pi-define}
\end{eqnarray}
where
\begin{eqnarray}
 A=\epsilon_{ij}\partial_i X \hat{E}_j\,. \label{eq:a-define}
\end{eqnarray}
The Hamiltonian of the DBI action is then obtained as
\begin{eqnarray}
{\mathcal H}_{\rm DBI} &=& \Pi_i\hat{E}_i - {\cal L}_{\rm DBI} \nonumber \\
 &=&\hat{T}_W D_{\rm DBI}^{-1/2}(1+(\partial_i X)^2+\hat{B}^2)\,. \label{eq:dbiham}
\end{eqnarray}
Remaining task for the derivation of the Hamiltonian is to write $D_{\rm DBI}$ in terms of $X, \Pi_i$ and $\hat{B}$.
To this end, we calculate $\Pi_i^2$ and $\epsilon_{ij}\partial_i X \Pi_j$:
\begin{eqnarray}
 \Pi_i^2&=&\hat{T}_W^2D_{\rm DBI}^{-1}\left\{\hat{E}_i^2+A^2+(1+(\partial_i X)^2)A^2\right\}\,, \label{eq:pi2}\\
 \epsilon_{ij}\partial_i X \Pi_j&=&\hat{T}_WD_{\rm DBI}^{-1/2}(1+(\partial_i X)^2)A\,. \label{eq:pir}
\end{eqnarray}
From (\ref{eq:pir}) we have
\begin{eqnarray}
 A={D_{\rm DBI}^{1/2} \over \hat{T}_W(1+(\partial_i X)^2)}(\epsilon_{kl}\partial_k\Pi_l)\,. \label{eq:A}
\end{eqnarray}
From (\ref{eq:def-DBID}) we find
\begin{eqnarray}
 \hat{E}_i^2+A^2=1+(\partial_i X)^2+\hat{B}^2-D_{\rm DBI}\,. \label{eq:ea}
\end{eqnarray}
Substituting (\ref{eq:A}) and (\ref{eq:ea}) into (\ref{eq:pi2}), we reach the following equation:
\begin{eqnarray}
 \Pi_i^2= \hat{T}_W^2 D_{\rm DBI}^{-1}
 \left\{
  1+(\partial_i X)^2+\hat{B}^2-D_{\rm DBI}+{D_{\rm DBI}(\epsilon_{ij}\partial_i X \Pi_j)^2 \over 1+(\partial_k X)^2}
 \right\}\,.
\end{eqnarray}
Solving this equation with respect to $D_{\rm DBI}$, we have
\begin{eqnarray}
 D_{\rm DBI}={\hat{T}_W^2(1+(\partial_iX)^2+\hat{B}^2)(1+(\partial_iX)^2) \over \hat{T}_W^2(1+(\partial_i X)^2)+\Pi_i^2 + (\partial_iX \Pi_i)^2}\,. \label{eq:ddbi}
\end{eqnarray}
We substitute (\ref{eq:ddbi}) into the Hamiltonian (\ref{eq:dbiham}) and obtain the final expression for the Hamiltonian
\begin{eqnarray}
 {\mathcal H}_{\rm DBI}=\sqrt{\left\{\hat{T}_W^2(1+(\partial_i X)^2)+\Pi_i^2 + (\partial_iX \Pi_i)^2\right\}\left\{1+(\partial_jX)^2+\hat{B}^2\right\} \over 1+(\partial_kX)^2}\,. \label{eq:dbiham-can}
\end{eqnarray}
Note that this is different from the DBI Hamiltonian obtained in Ref.~\cite{Portugues}.

Let us next rewrite the BPS equation (\ref{eq:NG-KL}) in terms of the DBI variables. To this end, we first 
show how $\hat{B}$ and $\hat{E}_i$ are written in terms of the fields $X$ and $\phi$ in the NG action. 
For $X \to X(x^1,x^2)$ and $\phi \to (\omega/\hat m) t + \phi(x^1,x^2)$ as is given in (\ref{eq:spiky-kink-conf}), 
from (\ref{eq:condition}) we have
\begin{eqnarray}
 \hat{B}&=&-{(\omega/\hat{m}) \over \sqrt{D_{\rm NG}}}(1+(\partial_i X)^2)\,, \label{eq:b-define}\\
 \hat{E}_i&=&{1 \over \sqrt{D_{\rm NG}}}\left\{(1+(\partial_k X)^2)\epsilon_{ij}\partial_j\phi-(\partial_k X \partial_k \phi)(\epsilon_{ij}\partial_j X) \right\}\,,
 \label{eq:e-define}
\end{eqnarray}
from which we find
\begin{eqnarray}
 \hat{B}^2&=& {(\omega/\hat{m})^2 \over D_{\rm NG}}(1+(\partial_i X)^2)^2\,, \label{eq:b1} \\
 \hat{E}_i^2&=&{1 \over D_{\rm NG}}
  \left\{
   (1+(\partial_i X)^2)^2(\partial_j \phi)^2-(\partial_i X \partial_i \phi)^2(\partial_j X)^2-2(\partial_i X \partial_i \phi)^2
  \right\}\,. \label{eq:e2}
\end{eqnarray}
Combining (\ref{eq:a-define}) with (\ref{eq:e2}) we obtain
\begin{eqnarray}
  A^2 = (\partial_i X)^2 \hat{E}_j^2-(\partial_i X \hat{E}_i)^2 = {(\partial_i X \partial_i \phi)^2 \over D_{\rm NG}}\,. \label{eq:a2}
\end{eqnarray}
Substituting (\ref{eq:b1})--(\ref{eq:a2}) into $D_{\rm DBI}$ given in (\ref{eq:def-DBID}), it can be shown that
\begin{eqnarray}
 D_{\rm DBI}={(1+(\partial_i X)^2)^2 \over D_{\rm NG}}\,.
\label{eq:DDBI_DNG}
\end{eqnarray}

Now we are ready to rewrite the BPS equation (\ref{eq:NG-KL}) in terms of the DBI language. 
From (\ref{eq:NG-KL}) we have
\begin{eqnarray}
\p_i \phi \p_i X = 0,
\end{eqnarray}
and
\begin{eqnarray}
D_{\rm NG} &=& \left(1 +(\p_iX)^2\right)\left(1-(\omega/\hat m)^2+(\p_j\phi)^2\right) \nonumber\\
&=& (1-(\omega/\hat m)^2)\left(1+(\p_iX)^2\right)^2.
\end{eqnarray}
Substituting this into (\ref{eq:DDBI_DNG}), we have
\begin{eqnarray}
D_{\rm DBI} = \frac{1}{1-(\omega/\hat m)^2}.
\end{eqnarray}
Furthermore, (\ref{eq:e-define}) gives us the relation
\begin{eqnarray}
 \hat{E}_i=\sqrt{D_{\rm DBI}}\,\epsilon_{ij}\partial_j\phi\,. \label{eq:e4}
\end{eqnarray}
Since $A=0$, (\ref{eq:pi-define}) is simplified as
\begin{eqnarray}
 \Pi_i=\hat{T}_W{\hat{E}_i \over \sqrt{D_{\rm DBI}}}\,. \label{eq:e-pi-relation}
\end{eqnarray}
Combining (\ref{eq:e4}) and (\ref{eq:e-pi-relation}),  tell us that
\begin{eqnarray}
 \epsilon_{ij}\partial_j\phi=\hat{T}_W^{-1}\Pi_i\,. \label{eq:NG-RHS}
\end{eqnarray}
Substituting (\ref{eq:NG-KL}) into  (\ref{eq:b-define}), we find
\begin{eqnarray}
 \hat{B}=- \frac{\omega}{\hat m}\sqrt{\frac{1+(\p_iX)^2}{1-\omega^2 + (\p_i\phi)^2}} = -{\omega \over \sqrt{\hat{m}^2-\omega^2}}\,.  \label{eq:B-NG}
\end{eqnarray}
Using (\ref{eq:NG-RHS}) and (\ref{eq:B-NG}), the BPS equation (\ref{eq:NG-KL}) in the NG action is rewritten as
\begin{eqnarray}
 \partial_i X =\pm\sqrt{1+\hat{B}^2}\,\hat{T}_W^{-1}\Pi_i\,. \label{eq:DBI-BPS}
\end{eqnarray}
This is the BPS equation for the dyonic extension of the spike domain wall in terms of the DBI variables. 
Note that we eventually arrive 
at the same BPS equation given in \cite{Portugues}, although our Hamiltonian (\ref{eq:dbiham-can}) is different from
one in \cite{Portugues}. 

In order to check that this equation leads to the desired result, we substitute (\ref{eq:DBI-BPS}) into the Hamiltonian (\ref{eq:dbiham-can}). It yields
\begin{eqnarray}
 {\cal H}_{\rm DBI}=\hat{T}_W\sqrt{1+\hat{B}^2}\left(1+{\hat{T}_W}^{-1}\Pi_i^2\right)\,. \label{eq:ham-DBI-spiky}
\end{eqnarray}
We shall consider the following configuration:
\begin{eqnarray}
\Pi_i={\hat{T}_W\over \hat{m}}{x_i \over \rho^2}\,. \label{eq:DBI-source}
\end{eqnarray}
This configuration represents that the unit electric charge is placed on the membrane. This fact is understood from the relation (\ref{eq:e-pi-relation}), which gives 
\begin{eqnarray}
 \hat{E}_i=\hat{T}_W^{-1}\Pi_i={1\over \hat{m}}{x_i \over \rho^2}\,.
\label{eq:EPI}
\end{eqnarray}
Thus, the factor $1/\hat{m}$ is interpreted as an electric charge. 
A solution of the BPS equation (\ref{eq:DBI-BPS}) which is called the BIon is in this case
\begin{eqnarray}
 X=-{1 \over \hat{m}}\sqrt{1+\hat{B}^2}\log\rho\,. \label{eq:sol-DBI-final}
\end{eqnarray}
Substituting (\ref{eq:DBI-source}) into (\ref{eq:ham-DBI-spiky}) and integrating over the membrane directions, we find the energy of the configuration as
\begin{eqnarray}
 E_{\rm DBI}=\hat{T}_W\sqrt{1+\hat{B}^2}A+{2\pi \hat{T}_W \over \hat{m}}\sqrt{1+\hat{B}^2}(\log R-\log \delta)\,,
\end{eqnarray}
where we have introduced the infrared and ultraviolet cutoff $\rho=R$ and $\rho=\delta$, respectively. The energy is rewritten as
\begin{eqnarray}
 E_{\rm DBI}=\hat{T}_W\sqrt{1+\hat{B}^2}A+{2\pi \hat{T}_W \over \hat{m}}\hat{L}\,,
\end{eqnarray}
where $\hat{L}=X(\delta)-X(R)$. Taking (\ref{eq:B-NG}) 
with the choice $\omega/\hat m=\sin\alpha$, $\hat T_W =T_W$,
and $\hat m = m$ into account, it is found that this expression is the same with (\ref{eq:KL-E}) obtained in the NG action. Therefore it is concluded that the BPS equation (\ref{eq:DBI-BPS}) in the DBI action reproduces the dyonic extension of the spike domain wall (\ref{eq:nf2}) in the field theory.

Before closing this subsection, let us make a comment on the relation between the magnetic scalar potential $\varphi$ introduced
by  (\ref{eq:def_msp}) in Sec. \ref{msp} and $X$ given in (\ref{eq:sol-DBI-final}). $\varphi$ is the scalar potential for the
magnetic field $B_i(x^1,x^2)$ in the original gauge theory. We have found the relation 
$\varphi = m x^3(x^1,x^2)/d_W$ and $x^3(x^1,x^2) = - u_S(x^1,x^2)/2m$.  In the strong gauge coupling limit, we have
$u_S = \log \rho^2$, so that $x^3(x^1,x^2) = - (\log \rho)/m$, which precisely coincides with (\ref{eq:sol-DBI-final})
in the case of $\hat B = 0$. On the other hand, 
from (\ref{eq:EPI}) and (\ref{eq:sol-DBI-final}), $X$ is the scalar potential for the dual electric field $\hat E_i(x^1,x^2)$.
Thus, the magnetic field $B_i$ in the original gauge theory and the dual electric field $\hat E_i$ in the effective
theory are generated by the same static potential.

%
%
\subsection{The semi-local BIon: Round spike configuration and DBI action}
\label{sec:semilocal_BIon}
In this subsection, we discuss the semi-local BIon -- the round spike domain wall configuration, where the spike's tip is smoothed out by introducing a lump size moduli. This configuration should be a DBI counterpart to the semi-local boojum which we studied 
for the finite gauge coupling in subsection \ref{sec:semilocal}. 
Our purpose in this subsection is to investigate whether the semi-local boojum in the strong gauge coupling limit in the field theory is realized in the DBI action.

\begin{figure}[t]
\begin{center}
$
\begin{array}{c}
\includegraphics[width=10cm]{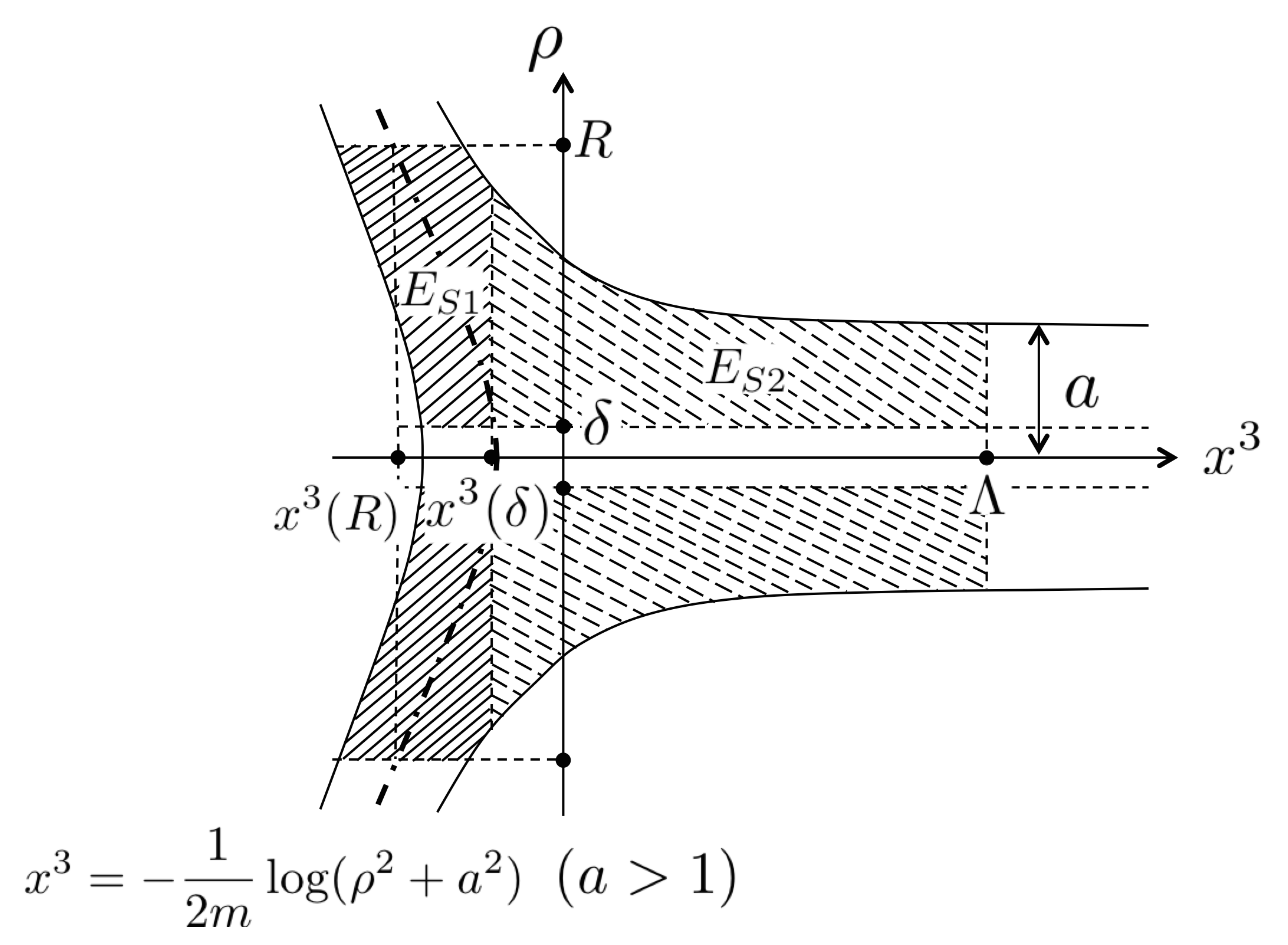} \\
\includegraphics[width=10cm]{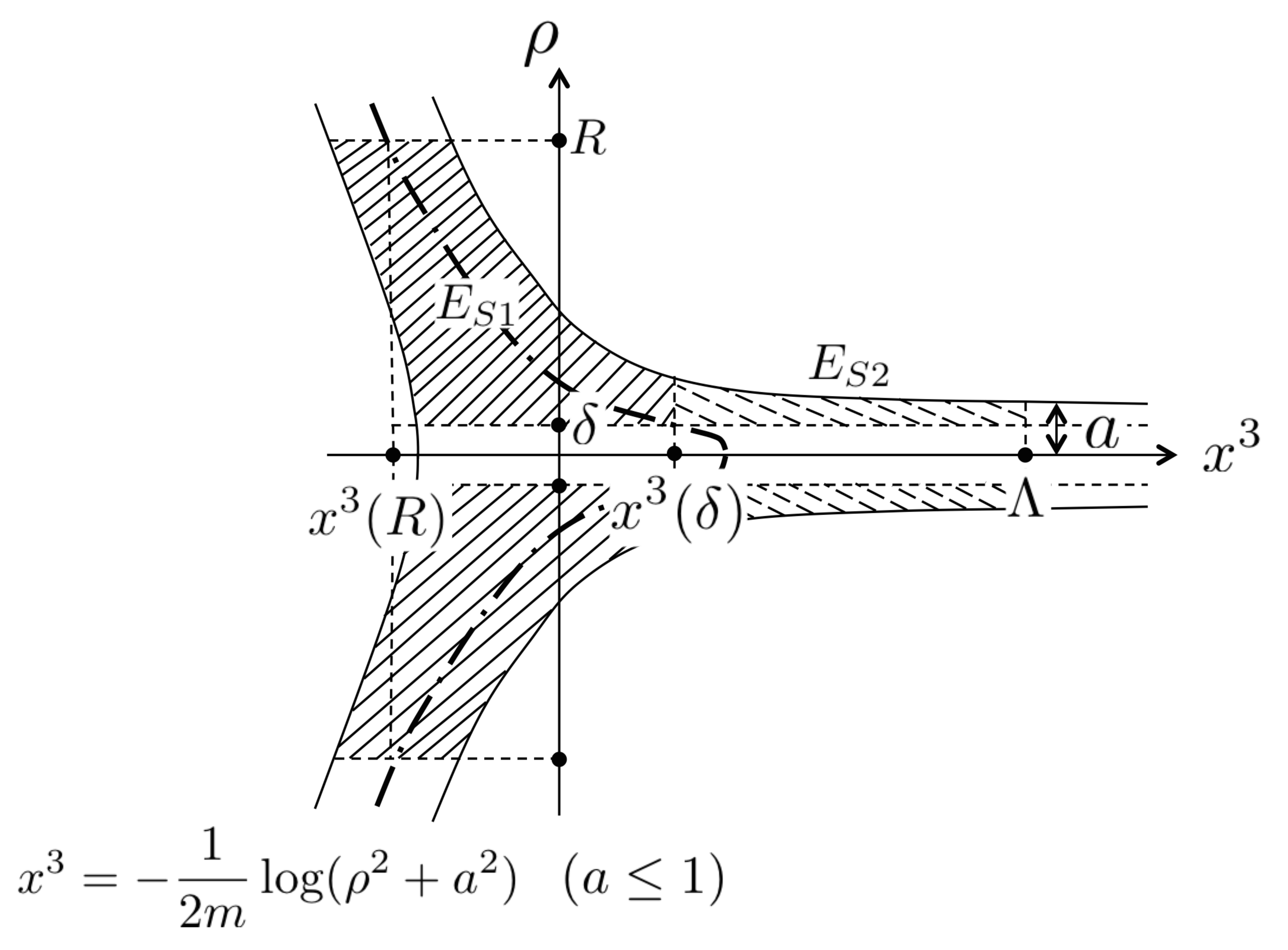}
\end{array}
$
\caption{Schematic picture of the semi-local vortex ending on the domain wall for $a>1$(upper) and $a\le 1$(lower). The dashed-dotted curve shows the position of the domain wall described by $x^3=-1/(2m)\log(\rho^2+a^2)$. The solid shaded part of the configuration contributes to the vortex string energy $E_{S1}$ while the dashed shaded part contributes to $E_{S2}$.}
\label{fig:slv}
\end{center}
\end{figure}

As the simplest example of the semi-local vortex, let us consider $N_F=3$ case with vanishing Q-charge. The moduli matrix of the configuration is
\begin{eqnarray}
H_0(z)=(z,a,1)\,, \quad  M={\rm diag}(m/2,m/2,-m/2). \label{mm}
\end{eqnarray}
where $a \in \mathbb{R}$ is a size moduli. Supposing that the $u$ depends on all the space-directions, the master equation (\ref{eq:quatermaster}) is solved as
\begin{eqnarray}
u_{g\rightarrow \infty}=\log\left\{(\rho^2+a^2)e^{mx^3}+e^{-m x^3}\right\}\,.
\end{eqnarray}
The domain wall position is read off from the condition $(\rho^2+a^2)e^{mx^3}=e^{-m x^3}$, which gives
\begin{eqnarray}
 x^3(\rho)=-{1 \over 2m}\log(\rho^2+a^2)\,. \label{pos}
\end{eqnarray}
In the strong gauge coupling limit, since the boojum energy density (\ref{eq:tb}) is vanishing, the total energy (\ref{evw}) is just a sum of the domain wall tension and the vortex string tension:
\begin{eqnarray}
E_{1/4}=\int d^3x\ ({\mathcal T}_W+ {\mathcal T}_S)\,,
\end{eqnarray}
where ${\mathcal T}_W$  and ${\mathcal T}_S$ are given in (\ref{eq:tw}) and (\ref{eq:ts}). We perform the integral along only $x^3$-direction:
\begin{eqnarray}
 E_{1/4}&=& \displaystyle  T_W A +\int d^2x\ {\cal E}_S(x^2,x^2)\,, \label{eq:energy-vsd}
 \label{te}
\end{eqnarray}
where ${\mathcal E}_S$ is given by
\begin{eqnarray}
 {\mathcal E}_S(x_1,x_2)={v^2 \over m} \left[{\rho^2 \over (\rho^2+a^2)^2}+{a^2 \over (\rho^2+a^2)^2}\log\left\{1+(\rho^2+a^2)e^{2m\Lambda}\right\}\right]\,. \label{e2d} 
\end{eqnarray}
Here we have introduced the infrared cutoff $x_3=\Lambda$. 
In the following, we focus on the contribution of the energy from the vortex string in (\ref{e2d}). We separate it into two parts as
\begin{eqnarray}
 E_{S} = \int d^2 x\ {\mathcal E}_S(x_1,x_2)=E_{S1}+E_{S2}\,,
\end{eqnarray}
where
\begin{eqnarray}
E_{S1}&=& {v^2 \over m} \int d^2 x\ {\rho^2 \over (\rho^2+a^2)^2}\,, \\
E_{S2}&=& {v^2 \over m} \int d^2 x\ {a^2 \over (\rho^2+a^2)^2}\log\left\{1+(\rho^2+a^2)e^{2m\Lambda}\right\}\,. \label{eq:dominant}
\end{eqnarray}
Integration over $\rho$ from a UV cutoff $\rho =\delta$ ($\delta \ll a$) to an IR cutoff $\rho = R$ ($R\gg a$), see Fig.~\ref{fig:slv}, we find
\begin{eqnarray}
 E_{S1}&=&{\pi v^2 \over m}\left\{\log(R^2+a^2)-\left({a^2 \over \delta^2 +a^2}+\log(\delta^2+a^2)\right) \right\}\,, \label{es1} \\
 E_{S2}&\simeq&{\pi v^2 a^2\over m}\left\{{1+\log(\delta^2+a^2) \over \delta^2+a^2}-2m\Lambda \left({1 \over R^2+a^2}-{1 \over \delta^2+a^2}\right)\right\}\,, \label{es2}
\end{eqnarray}
where we have assumed that the cutoff $\Lambda$ is sufficiently large and that the second term in the logarithm in (\ref{eq:dominant}) is dominant. 
Taking account of (\ref{pos}) and assuming $a>\delta$,  (\ref{es1}) and (\ref{es2}) are rewritten as
\begin{eqnarray}
 E_{S1} &\simeq & -2\pi v^2 \left(x^3(R)-x^3(\delta)+{1 \over 2m} \right)+{\mathcal O}\left({\delta^2 \over a^2}\right)\,, \label{es1a} \\
 E_{S2} &\simeq& -2\pi v^2 (x^3(\delta)-\Lambda)+{\mathcal O}\left({\delta^2 \over a^2}\right)\,. \label{es2a}
\end{eqnarray}
These results tell us that $E_{S1}$ is the vortex string energy between $x^3(R)$ and $x^3(\delta)$ while $E_{S2}$ is one between $x^3(\delta)$ and $\Lambda$. We show each contribution schematically in Fig.~\ref{fig:slv}. 
As can be seen from Fig.~\ref{fig:slv}, $E_{S2}$ corresponds to the energy contributed by the lump string which is
perpendicular to the domain wall. On the other hand, $E_{S1}$ is contribution from a part of lump string $(\rho > a)$ whose
angle from the $x^1$-$x^2$ plane is in the range of $[0,\pi/2)$. Thus, we expect that only $E_{S1}$ is reproduced within
the DBI theory.

For $a>1$ the curve (\ref{pos}) does not cross the $\rho$ axis (the upper panel in Fig. \ref{fig:slv}). As $a$ is decreased, the curve crosses the $\rho$ axis and $x^3(\delta)$ goes to right along $x^3$ axis and the contribution of $E_{S1}$ is dominant (the lower panel in Fig. \ref{fig:slv}) in $E_S$. As $a$ is decreased further, the expressions (\ref{es1a}) and (\ref{es2a}) are no longer valid. In such case, it is convenient to go back to the original expression (\ref{es1}) and (\ref{es2}). There we can safely take $a$ to be zero and find that $E_{S2}$ is vanishing while $E_{S1}$ becomes
\begin{eqnarray}
 E_{S1}={2 \pi v^2 \over m}(\log R-\log\delta)\,.
\end{eqnarray}
Using (\ref{pos}) with $a=0$, this expression is rewritten as
\begin{eqnarray}
 E_{S1}=T_S L\,,
\end{eqnarray}
where $L=x^3(\delta)-x^3(R)$ is the string length. Therefore, $E_{S1}$ at $a=0$ is nothing but the lump string energy of zero size. 
Note that the lump string with $a=0$ becomes singular at $x^3 \to \infty$. Thus, $a=0$ lump string attached to the domain wall
is regular except for $\rho=0$, namely the angle to the $x^1$-$x^2$ plane is always in the range of $[0,\pi/2)$ except for
the junction point. Therefore, the lump string of zero size $E_{S1}$ should be reproduced in the DBI theory for
the whole region on the domain wall except for $\rho = 0$. 

Now let us consider the corresponding configuration in the DBI action. The Hamiltonian and the BPS equation in the DBI action are (\ref{eq:ham-DBI-spiky}) and (\ref{eq:DBI-BPS}) with $\hat{B}=0$:
\begin{eqnarray}
& {\cal H}_{\rm DBI}=\hat{T}_W \left(1+\hat{T}_W^{-1}\Pi_i^2 \right)\,,& \\
&\partial_i X =\pm\hat{T}_W^{-1}\Pi_i\,.& \label{eq:BPS-Pi}
\end{eqnarray}
The energy of the configuration is
\begin{eqnarray}
 E_{\rm DBI}=\hat{T}_W A+\hat{T}_W^{-1}\int d^2x\ \Pi_i^2\,. \label{eq:energy-DBI}
\end{eqnarray}
We expect that the second term of (\ref{eq:energy-DBI}) coincides with $E_{S1}$ given in  (\ref{es1}). 
In order to check this under the identification $T_W=\hat{T}_W$, the following relation should hold:
\begin{eqnarray}
{v^2 \over m} {\rho^2 \over (\rho^2+a^2)^2}=\hat{T}_W^{-1}\Pi_i^2\,\quad (\hat T_W = mv^2). \label{eq:relation-DBI}
\end{eqnarray}
This is solved by
\begin{eqnarray}
 \hat{T}_W^{-1}\Pi_i={1 \over m}{x_i \over \rho^2+a^2}\,. \label{eq:pi-sol}
\end{eqnarray}
Combining this with (\ref{eq:BPS-Pi}), we find
\begin{eqnarray}
 X= -{1 \over 2 m}\log(\rho^2+a^2)\,. \label{eq:desired-sol}
\end{eqnarray}
This is precisely equal to  (\ref{pos})! Thus, we find the semi-local BIon that is 
the desired counterpart of the semi-local boojum in the original gauge theory. Note that $E_{S1}$ in (\ref{es1a})
with $\delta = 0$ does
not depend on $a$, since we can absorb it into the IR cutoff by taking $R = (1+\tilde R)a$, where $\tilde R >0$ is  a new cutoff. 
In this sense, we can interpret $a$ as a moduli parameter of the semi-local BIon.

While $E_{S1}$ is correctly reproduced in the DBI action, $E_{S2}$ is missing. As explained above, $E_{S2}$ corresponds
to a part of the lump string perpendicular to the domain wall. To reproduce this correctly, we should take additional
zero modes into account. So far, we have considered only the zero modes localized on the domain wall. However, 
there are non-normalizable zero modes in the bulk because a part of flavor symmetry remains unbroken in
the model with the partially degenerate masses (\ref{mm}). We expect that $E_{S2}$ would be correctly reproduced 
once we include coupling between the localized zero modes and non-normalizable zero modes. 
It would be interesting to figure out the interaction, but it is beyond the scope of this paper, so we leave it as a future
work.

The last comment is on the dual electric charge. The electric field (\ref{eq:EPI}) for the semi-local BIon is given by
\begin{eqnarray}
 \hat{E}_i = {1 \over \kappa \hat{m}}{x_i \over \rho^2+a^2}\,, \label{ele-dbi}
\end{eqnarray}
where we recover the dimensional parameter $\kappa$. Interestingly, the behavior of $\hat{E}_i$ is similar to the magnetic field in the domain wall discussed in subsection \ref{sec:semilocal}. There it was found that the observer in the domain wall sees that the magnetic field of the magnetic point source, which obeys the modified Coulomb's law in a region far from the magnetic source. Similarly, in the DBI theory the observer in the domain wall sees that the electric field obeys the modified Coulomb's law in a region far from the electric point source placed in the domain wall. Recalling the discussion in section \ref{msp}, we see that the electric charge placed on the membrane is given as
\begin{eqnarray}
 q_E={2 \over \hat{m}\kappa}\,.
\end{eqnarray}
%


\section{Outlook}
\label{sec:out}

In this paper, using our previous results \cite{Boojum1}, we have furthered the understanding of 1/4 BPS composite solitons in   the Abelian-Higgs theory.

We obtained the solution of the vortex strings attached to the tilted domain walls on which constant background magnetic
field is turned on. This is quite similar to D1-branes suspended between tilted D3-branes. The D1-branes can be seen
as the magnetic monopoles in non-commutative spacetime \cite{Hashimoto:1999xh,Hashimoto:1999zw}. 
As a natural analogy, the vortex strings suspended between
tilted domain walls in the gauge theory should be seen as electrically charged particles in a low energy
effective theory which is a non-commutative theory. If this is the case, the similarity between the solitons in
field theories and D-branes in string theories, which is repeated in the literature, is further reinforced.

Further studies of composite solitons, especially in non-Abelian setting, may be useful for ironing out the phenomenology of dynamically realized brane-world scenarios. In string theory, it is known that intersecting D-branes can generate Standard Model gauge group, chiral fermions, and family replication (see \cite{2005Blumenhagen} and references therein). 
Similar results were obtained within the field theory as well. Indeed, domain walls and magnetic vortices have been long since used to localize scalar and fermionic fields \cite{Akama, Rubakov}. The localization of gauge fields is achieved either by Dvali-Shifman mechanism \cite{Dvali}, where confining phase in the bulk is assumed, or dynamically using field-dependent gauge coupling constant via Ohta-Sakai mechanism \cite{Ohta}. Using the former, the Standard Model gauge group has been constructed at the junction of perpendicular domain walls in \cite{Volkas}, while the later was used to localize large gauge group on coincident domain walls \cite{Us1, Us2}. In the future, using the method developed here, we would like to construct a realistic brane-world scenario on intersecting solitons at finite gauge coupling constant and to clarify the role of negative binding energy on low-energy effective theory.

We have shown that the spike domain wall configuration in the field theory can be reproduced in the NG action and the DBI action. We have also observed that the DBI action correctly realizes the semi-local boojum as the semi-local BIon. On the other hand, 
as is addressed in Sec.~\ref{sec:semilocal_BIon}, the DBI action cannot reproduce the semi-local lump string itself ($E_{S2}$)
which is  perpendicular to the domain wall. A possible reason for this situation is that all the massless modes may not be taken into account in the low energy effective theory. The semi-local boojum is constructed by introducing a partially degenerate masses such as (\ref{mm}). If there is a mass splitting in the first two components, there are three discrete vacua and an extra domain wall appears. Associated with this, extra massless modes localize around the extra domain wall. The extra domain wall becomes increasingly broader as the mass splitting disappears. Correspondingly, the extra massless modes spread into the whole half-space and they become  non-normalizable. We expect that these non-normalizable modes would be necessary for reproducing $E_{S2}$ in the
low energy effective theory. Normally, it is not easy to deal with non-normalizable modes. We may achieve it by
considering a small mass splitting which introduces the broad extra domain wall. Further, we may consider a 
low energy effective action for the two parallel domain walls and investigate the limit when we turn on the mass degeneracy.
We leave this problem for a future work.


\section*{Acknowledgements} 
This work is supported by Grant-in Aid for Scientific Research 
No.25400280 (M.\ A.\ ) and from the Ministry of Education, 
Culture, Sports, Science and Technology  (MEXT) of Japan. 
The work of M. E. is supported in part by
JSPS Grant-in-Aid for Scientic Research (KAKENHI Grant No.~26800119)
and the MEXT-Supported Program for the Strategic
Research Foundation at Private Universities ``Topological Science''
(Grant No.~S1511006).
F. B. is an international research fellow of the Japan Society for the Promotion of Science.
This work was supported by Grant-in-Aid for JSPS Fellows, Grant Number 26004750.

\bibliographystyle{jhep}
\bibliography{references}
\end{document}